\documentclass[a4paper,fleqn]{cas-dc}
\usepackage[numbers]{natbib}

\usepackage{subcaption}
\usepackage{bm}
\usepackage{multirow} 
\usepackage{cancel}
\usepackage{makecell}
\usepackage{float}
\def\tsc#1{\csdef{#1}{\textsc{\lowercase{#1}}\xspace}}
\tsc{AM}
\tsc{TO}
\tsc{PP}
\tsc{FDM}
\tsc{MinS}
\tsc{MinV}
\tsc{MaxOA}
\newcommand{\deriv}[2]{\ensuremath{\frac{\mathrm{d}#1}{\mathrm{d}#2}}}
\newcommand{\derivp}[2]{\ensuremath{\frac{\partial#1}{\partial#2}}}
\newcommand{\minus}{\scalebox{0.75}[1.0]{$-$}}

\usepackage[dvipsnames]{xcolor}
\begin{document}
	\let\WriteBookmarks\relax
	\def\floatpagepagefraction{1}
	\def\textpagefraction{.001}
	
	\begin{center}
		\Large{\bf{Topology optimization for additive manufacturing with length scale, overhang, and building orientation constraints}}\\
		
	\end{center}
	
	\begin{center}
			\large{Prabhat Kumar\,$^{\star,}\,^{\dagger,}$\footnote[1]{Corresponding author, email: pkumar@mae.iith.ac.in,\,prabhatkumar.rns@gmail.com}, Eduardo Fern\'{a}ndez$\,^{\ddagger}$ }\\
		\vspace{4mm}
		
		\small{\textit{$^\star$Department of Mechanical and Aerospace Engineering, Indian Institute of Technology Hyderabad, 502285, India}}\\
		
		\small{\textit{$^\dagger$Department of Mechanical Engineering, Indian Institute of Science, 560012, Karantaka, India}}
		
		\small{\textit{$^\ddagger$Department of Aerospace and Mechanical Engineering, University of Liege, All\'{e}e de la D\'{e}couverte 13A, B52, 4000, Liege, Belgium.}}

	\end{center}
	
	\vspace{3mm}
	\rule{\linewidth}{.15mm}
	{\bf Abstract:}
	This paper presents a density-based topology optimization approach considering additive manufacturing limitations. The presented method considers the minimum size of parts, the minimum size of cavities, the inability of printing overhanging parts without the use of sacrificial supporting structures, and the printing directions. These constraints are geometrically addressed and implemented. The minimum size on solid and void zones is imposed through a well-known filtering technique. The sacrificial support material is reduced using a constraint that limits the maximum overhang angle of parts by comparing the structural gradient with a critical reference slope. Due to the local nature of the gradient, the chosen restriction is prone to introduce parts that meet the structural slope but that may not be self-supporting. The restriction limits the maximum overhang angle for a user-defined printing direction, which could reduce structural performance if the orientation is not properly selected. To ease these challenges, a new approach to reduce the introduction of such non-self-supporting parts and a novel method that includes different printing directions in the maximum overhang angle constraint are presented. The proposed strategy for considering the minimum size of solid and void phases, maximum overhang angle, and printing direction, is illustrated by solving a set of 2D benchmark design problems including stiff structures and compliant mechanisms. We also provide MATLAB codes in the appendix for educational purposes and for replication of the results.\\
	

	\vspace{-4mm}
	\rule{\linewidth}{.15mm}
	
	Additive manufacturing (AM), a 3D design printing technique, creates components in a layer-by-layer fashion. With the recent developments in this field, printing a complex geometry  is no longer  a challenging and costly affair. However, AM processes, e.g., stereo lithography, selective laser melting (SLM) and fused deposition melting (FDM) have certain geometrical limitations with respect to overhang angles, building orientations, minimum feature size, etc, which they can print to \citep{Gaynor2016}. Therefore, nowadays AM processes is directly being associated to topology optimization to fully capitalize on the advantages AM processes offer. Topology optimization (TO), a design technique, that provides  optimized designs by extremizing the desired objective with the given geometrical/physical constraints/limitations. These designs typically have finer details of the structures with complex geometries which can be printed by the AM processes if their limitations are taken care of. Thus, the main focus of this paper is to present a TO approach with length scale, overhang, and building orientation constraints for additive manufacturing.
	
	Till date, several authors have attempted to incorporate the AM limitations into TO. The minimum size of the parts and the minimum size of the cavities have been addressed in TO by \citet{Lazarov2016,Liu2016}.  
	The robust design approach \citep{Wang2011} has been proven effective in controlling minimum size in several design problems, e.g. compliant mechanisms \citep{daSilva2019a,daSilva2019b,kumar2021topology}, electromechanical actuators \citep{Qian2013} to name a few. The robust design approach accounts for the manufacturing errors with help of the dilated, intermediate and eroded projected fields. The approach solves a min-max optimization problem \citep{Wang2011}. We have adopted this formulation in our approach to control the minimum feature size of the optimized designs for AM.
	
\begin{figure}
    \centering
	\includegraphics[scale=0.20]{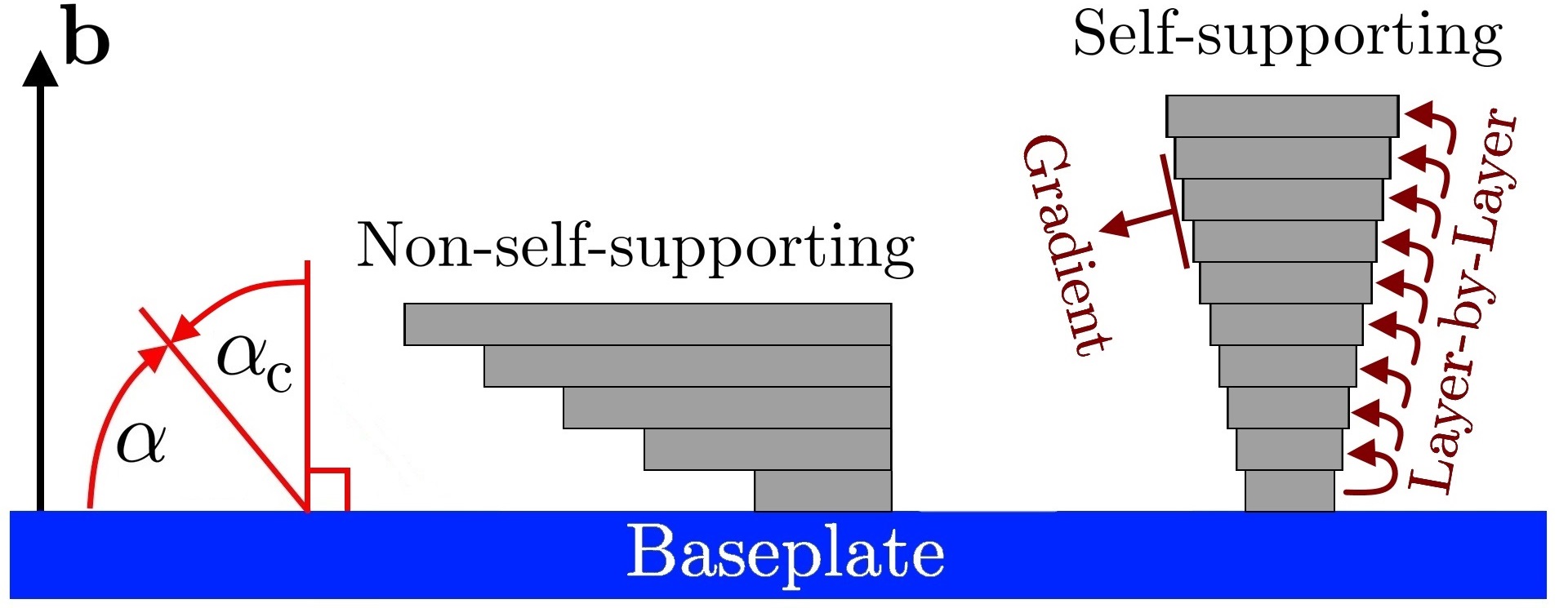}
	\caption{The critical overhanging angle $\alpha$ determines the printability of the component. $\mathbf{b}$ indicates the printing/building direction.}			
	\label{FIG:OverhangingAngle}
\end{figure}

	In general, AM processes fail to print the structural parts that overhang with respect to a critical reference value \citep{Gaynor2016,mezzadri2020second}. This is because material supports are essential: (i) to retain the new deposition of the material, (ii) for thermal conduction to avoid collapses, and (iii) to obviate distortion and failure of the newly deposited part \citep{Gaynor2016,mezzadri2020second, misiun2021topology}. Alternatively, one may decide to use sacrificial supporting structures, which have to be removed in the post processing steps. This however increases material, manufacturing and post-processing costs. TO approaches with overhang constraint is presented by \citet{Langelaar2017,Garaigordobil2019,Pellens2019,mezzadri2020second}. These methods seek to ensure no material layout beyond a maximum overhang angle (MaxOA), $\alpha_\mathrm{c}$, i.e., they ensure to prevent downward facing surfaces of the material below a critical angle $\alpha$  (see Fig.~\ref{FIG:OverhangingAngle}) in their approaches. Although this does not consider complex physics of the AM processes, it remarkably  helps improve the printability of optimized designs \citep{Langelaar2017,Liu2018,Garaigordobil2019,Pellens2019,mezzadri2020second}. One can classify the  MaxOA constraint based TO approaches into two: the local   \citep{Allaire2017,Qian2017,Garaigordobil2018,Zhang2019} and the global approaches \citep{Gaynor2016,Langelaar2016,Langelaar2017,Amir2018,vandeVen2018}. The former are based on  the structural gradient, whereas  layer-by-layer analyses are performed in the latter wherein the material deposition is done sequentially \footnote{Similar to the AM processes}. Note that the serial evaluation hinders full parallelization and thus, it increases computational cost especially for the large scale problems.

\begin{figure}
\centering
  		\begin{subfigure}{0.27\textwidth}
			\centering
				\includegraphics[scale = 0.125]{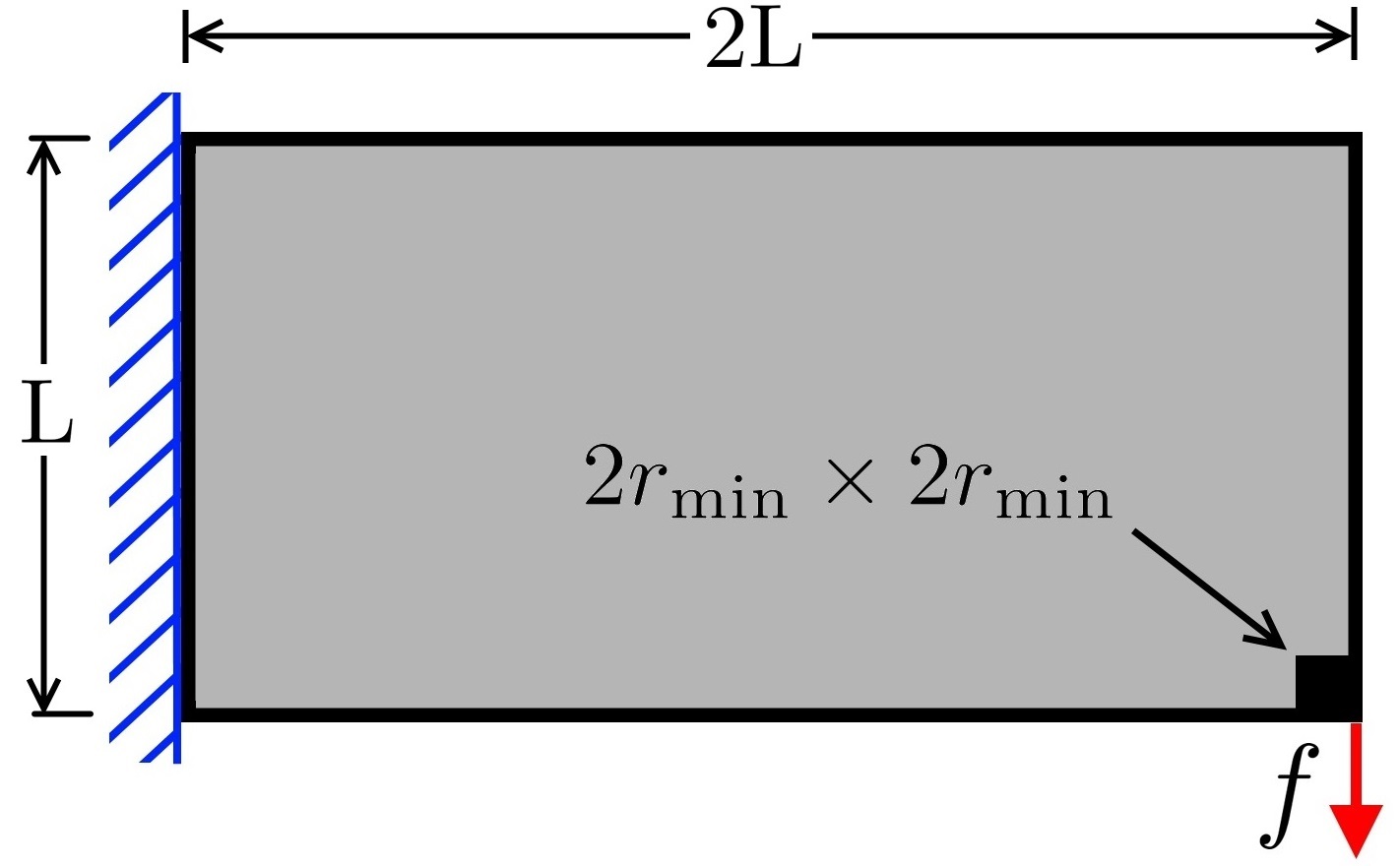}
				\caption{}			
				\label{FIG:Cone_Example_a}
		\end{subfigure}
		\begin{subfigure}{0.27\textwidth}
			\centering
				\includegraphics[scale= 0.25]{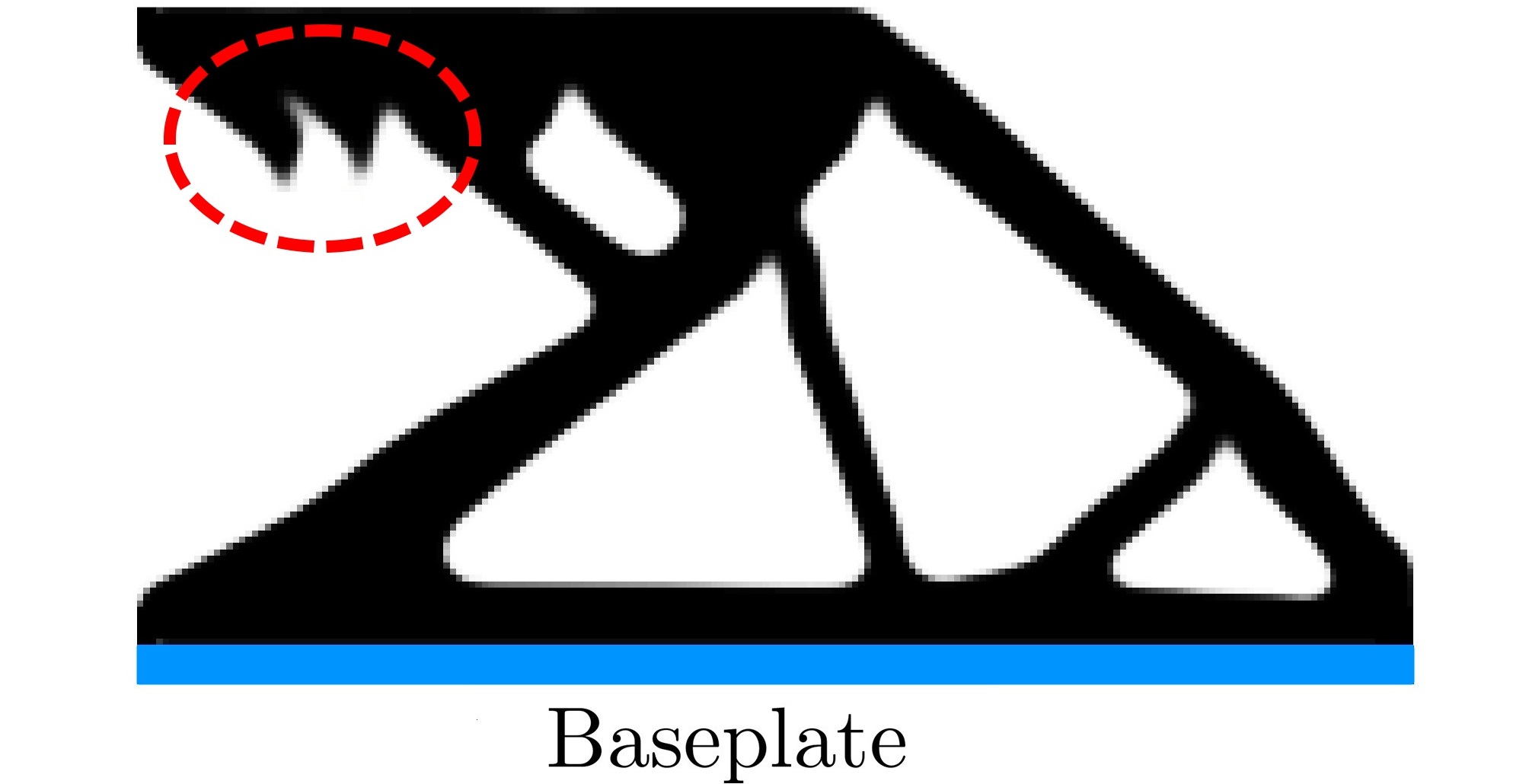}
				\caption{}
				\label{FIG:Cone_Example_b}			
		\end{subfigure}
		\begin{subfigure}{0.27\textwidth}
			\centering
				\includegraphics[scale = 0.25]{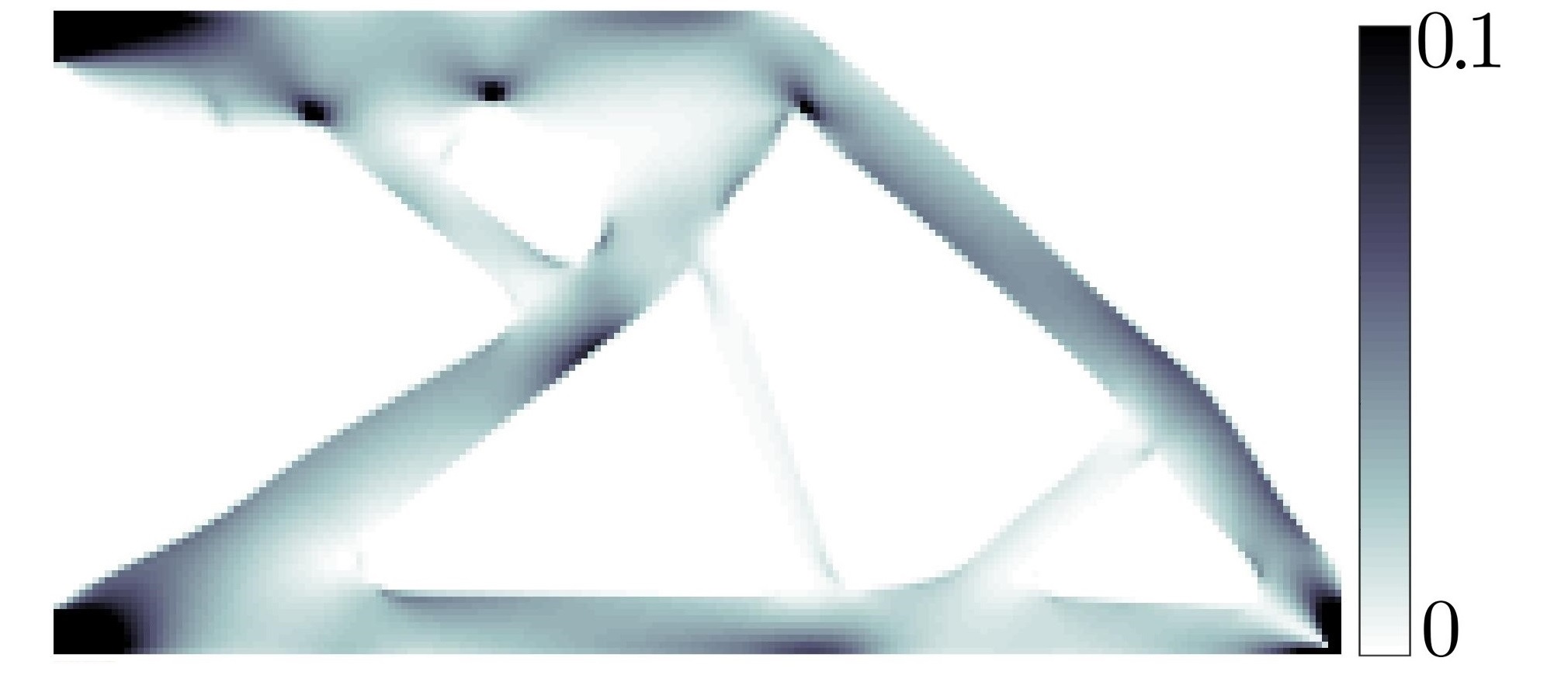}
				\caption{}
				\label{FIG:Cone_Example_c}			
		\end{subfigure} 
	\label{FIG:Cone_Example}
	\caption{(a) 2D-cantilever beam domain, (b) Solution to compliance minimization with gradient-based overhang constraint. The dashed circle highlights the non-self-supporting parts, and  (c) Elemental strain energy plot.}	
\end{figure}

	Figure~\ref{FIG:OverhangingAngle} illustrates both the  layer-by-layer and gradient-based techniques.  Due to local nature of the structure gradient, gradient-based techniques can readily be parallelized and thus, computational cost can be reduced significantly.  However,  they permit the introduction of triangular structures oriented towards the base plate \citep{Allaire2017}, which are undesirable as they meet the slope restriction but are usually not self-supporting (Fig. \ref{FIG:Cone_Example_b}). \citet{Qian2017} employs additional restriction to alleviate such down facing cones. A different constraint is used by \citet{Zhang2019} and several 2D-self-supporting designs are obtained. \citet{garaigordobil2021preventing} reduce the down facing cones using a continuation strategy over the area for gradient computation. We notice that such desirable cones have the lowest/no strain energy (Fig.~\ref{FIG:Cone_Example_c}) and based on this observation, we detect and alleviate such appendages without using any additional constraints. \citet{Pellens2019} proposed an approach with length scale and overhang angle control. Further, the above mentioned methods employ a user-defined building orientation. As per \citet{Langelaar2018}, selection of the proper printing direction  is however important to reduce deviation of the overhang-constrained components from the free-designs while printing. Therefore, considering building orientation in overhang-constrained problems is one of the important aspects for the AM processes. To the best of authors' knowledge, an approach that includes the printing direction, minimum size of member/solid phase (MinS), minimum size of cavity/void phase (MinV), MaxOA and building orientation does not exist yet in the current state-of-the-arts.

	In this paper, we present a strategy  to simultaneously control MinS, MinV,  MaxOA and the building orientation in a density-based TO setting. With a gradient-based overhang restriction, we evaluate the MaxOA constraint at different printing orientations using the Prewitt operator and select the least restrictive using aggregation functions (Sec.~\ref{Sec:Gradient-overhangcons}). In addition, we propose a new method to detect none self-supporting triangular parts from the optimized design using the strain energy information, wherein such structures have the least/no strain energy (see Fig.~\ref{FIG:Cone_Example_c} and Sec.~\ref{Sec:Strategy_to-avoid-nonSelfSupporting}) and subdue them using the free evolution and boolean operation (Sec.~\ref{Sec:PP-triangules_removal}). To impose MinS and MinV, the robust design approach \citep{Wang2011} based on eroded, intermediate and dilated projections is employed. The proposed strategy is explained on the 2D cantilever beam (Fig. \ref{FIG:Cone_Example_a}) for compliance minimization. As the strategy involves multiple computational algorithms. We also provide the associated in-house developed MATLAB codes for educational purposes.

	The reminder of this article is structured as follows. Sec.~\ref{Sec:2} describes the formulation of the topology optimization problem, the test case used to explain the strategy, description for gradient-based overhang constraint formulation, strategy to avoid non-self-supporting parts and results showing importance of the building directions. Sec.~\ref{Sec:3} provides various numerical examples including stiff structures and compliant mechanism with length scale, overhang angle and printing direction constraints. Pertinent discussions are also provided in Sec.~\ref{Sec:2} and Sec.~\ref{Sec:3}. Finally, Sec.~\ref{Sec:4} provides the concluding remarks.
	
	\section{Topology optimization framework} \label{Sec:2}
	We first demonstrate the presented concepts (MinS, MinV, MaxOA and building orientation) on the 2D cantilever beam  design (Fig.~\ref{FIG:Cone_Example_a}), and to illustrate the efficacy and versatility of the proposed approach, the MBB beam and compliant mechanism designs are also optimized, and the pertinent discussions are provided.
	
	A density-based TO formulation is adopted with three-phase (original, filtered and projected fields) technique. Rectangular tessellation is employed to represent the design domain. Each element is assigned a design variable $\rho \in[0,\,1]$ which is considered constant in the element. The original, filtered and projected design vectors are indicated via $\bm{\rho}$, $\bm{\tilde{\rho}}(\bm{\rho})$ and   $\bm{\bar{\rho}} \left( \bm{\tilde{\rho}} \left( \bm{\rho} \right) \right)$, respectively. 
	
\begin{figure}
\captionsetup[subfigure]{labelformat=empty}
\centering
\begin{tabular}{l} 
	\hspace{25mm}
  	\begin{subfigure}{1\textwidth}
		    \includegraphics[width=0.85\textwidth]{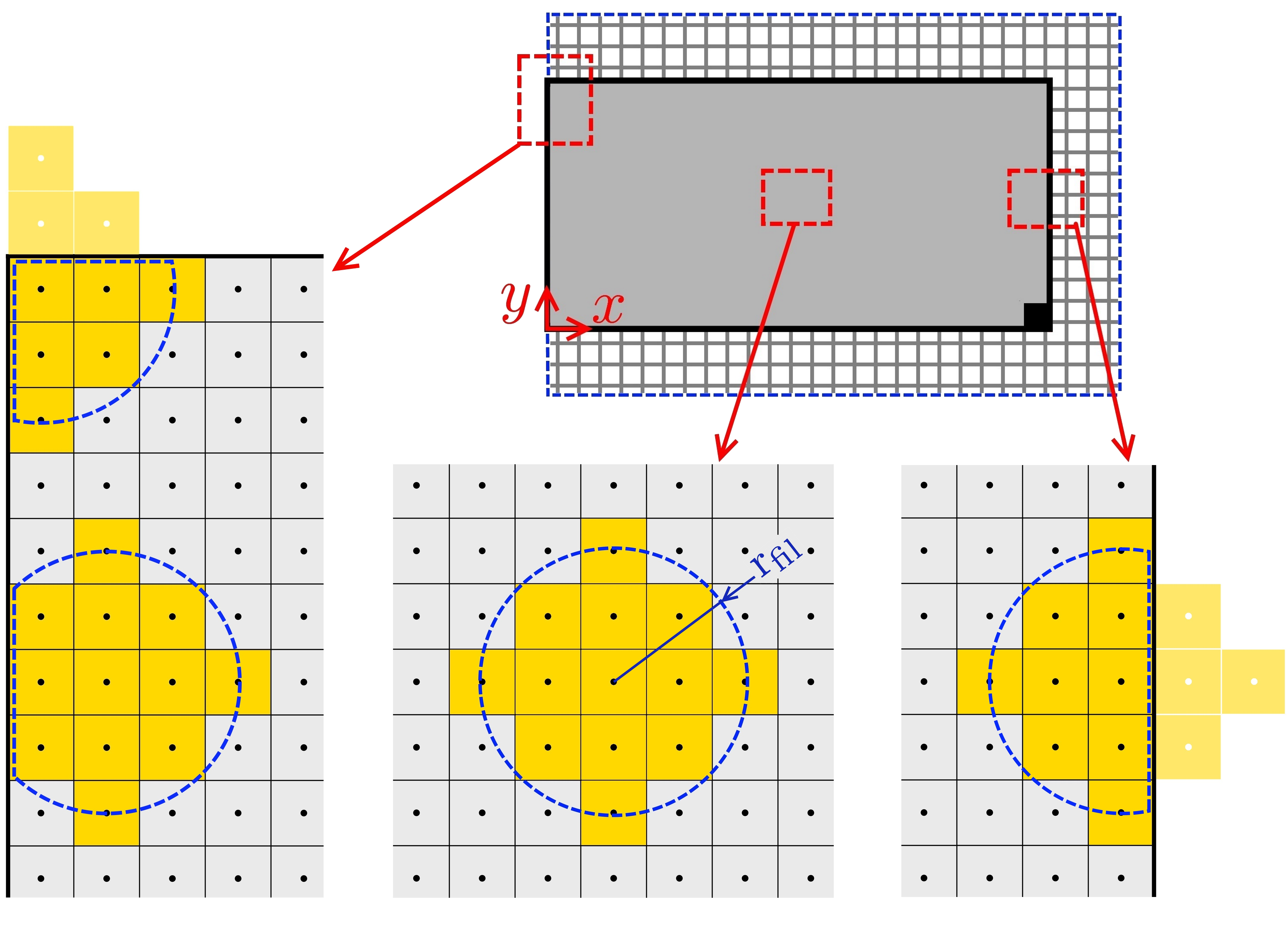}
			\caption{}
			\label{FIG:Filtering_a}					
	\end{subfigure}
			
			\begin{minipage}{0.25\textwidth}
			\subcaption{\label{FIG:Filtering_b}}
			\end{minipage}
			
			\begin{minipage}{0.25\textwidth}			
			\subcaption{\label{FIG:Filtering_c}}
			\end{minipage}			
			
			\begin{minipage}{0.25\textwidth}
			\subcaption{\label{FIG:Filtering_d}}	
			\end{minipage}
	\\  
	\vspace{-20mm}
	\\	
	\hspace{50mm}(b) \hspace{35mm} (c) \hspace{40mm} (d)
	\\	
	\vspace{-55mm} 
	\\
	\hspace{120mm} (a)
	\\
\end{tabular}
\label{FIG:Filtering}	
\vspace{54mm}
\caption{Filtering treatment. (a) Design domain including the mesh extension, (b) Fractioned filtering regions on the left side of the design domain, (c) Filtering region inside the design domain, and (d) Numerically extended fractioned filtering region as (Eq.~\ref{EQ:density_filter_corrected}).}	
\end{figure}

	The filtered design variable of element $i$, $\tilde{\rho}_i$, can evaluated as:
	\begin{equation} \label{EQ:density_filter}
	\tilde{\rho}_i = \frac{\displaystyle\sum_{j=1}^{N}\rho_j \mathrm{v}_j \mathrm{w}(\mathbf{x}_i,\mathbf{x}_j)}{\displaystyle\sum_{j=1}^{N} \mathrm{v}_j \mathrm{w}(\mathbf{x}_i,\mathbf{x}_j) } \; ,
	\end{equation}
	where  weight $\mathrm{w}(\mathbf{x}_i,\mathbf{x}_j)$ = $\mathrm{max}  \left(0 \; , \; 1-\frac{\| \mathrm{\mathbf{x}}_i - \mathrm{\mathbf{x}}_j \|}{\mathrm{r}_\mathrm{fil}} \right)$ \citep{Bruns2001}. The filter radius  is indicated via $r_\text{fill}$, $N$ indicates the total number of FEs employed to describe the design domain, and $\mathrm{v}_j$ is the volume of element $j$. We modify Eq.~\ref{EQ:density_filter} to negate boundary effects of the filtering process \citep{Andreassen2011} by extending the design domain (Fig.~\ref{FIG:Filtering_a}, cf. \citet{fernandez2020imposing}). This treatment is needed to maintain the uniformity of MinS and MinV irrespective of location within the domain \citep{fernandez2020imposing}. We use solid passive elements around the external force (Fig. \ref{FIG:Filtering_a}) to avoid the numerical issues \citep{Andreassen2011}.
	
	As per the boundary extension method \citep{fernandez2020imposing}, the elements away from the fixed boundary have same filtering regions (Figs. \ref{FIG:Filtering_c} and \ref{FIG:Filtering_d}). Therefore, for such elements the denominator of Eq.~\ref{EQ:density_filter} remains same. However, in the vicinity of the fixed boundary, filtering regions are fractionated as displayed in Fig. \ref{FIG:Filtering_b}. In this light, elements with  same $x$-coordinates will have filtering regions of the same size, and the weighted volume of the filtering region for each element $i$ is then determined as:
	\begin{equation} 
	\mathrm{V}^{\mathrm{w}}_i = {\displaystyle\sum_{j=1}^{N} \mathrm{v}_j \mathrm{w}(\mathrm{\mathbf{x}}_k,\mathrm{\mathbf{x}}_j) } \;,\; 
	\end{equation}
	\noindent where 
	\begin{equation}
	\mathrm{\mathbf{x}}_k = 
	\left\lbrace
	\begin{matrix}
	\left( \: {x}_{i} \; , \; \mathrm{L}/2 \: \right)
	\;,
	& \text{if} \; \; x_i  < \mathrm{r_{fil}} \; ,
	\vspace{2mm} 
	\\
	\left( \: 
	\mathrm{L} \; , \; \mathrm{L}/2 \: \right)
	\;
	,
	& \text{otherwise.}
	\end{matrix} 
	\right.
	\end{equation}
	Now, the filtered design variable (Eq.~\ref{EQ:density_filter}) is evaluated as 
	\begin{equation} \label{EQ:density_filter_corrected}
	\tilde{\rho}_i = \frac{\displaystyle\sum_{j=1}^{N}\rho_j \mathrm{v}_j \mathrm{w}(\mathrm{x}_i,\mathrm{x}_j)}{ \mathrm{V}^{\mathrm{w}}_i } \:,
	\end{equation}
	which is used in this paper.
	\noindent Note that $\mathrm{v}_j$, $\mathrm{w}(\mathrm{x}_i,\mathrm{x}_j)$ and $\mathrm{V}^\mathrm{w}_i$ do not vary as TO progresses, thus these are stored in a matrix $\mathbf{H}$ as:
	\begin{equation}
	\mathrm{H}_{i,j} = \frac{\mathrm{v}_j \: \mathrm{w}(\mathrm{\mathbf{x}}_i,\mathrm{\mathbf{x}}_j)}{\mathrm{V}^{\mathrm{w}}_i} \: .
	\end{equation} 
	Finally, the filtered field with boundary treatment is obtained as $\bm{\tilde{\rho}}=\mathbf{H}\bm{\rho}$ and its derivative as $\deriv{\bm{\tilde{\rho}}}{\bm{\rho}}=\mathbf{H}^\intercal$.
	
\begin{figure}
    \centering
	\begin{subfigure}{0.93\linewidth}		
		\includegraphics[width=0.75\linewidth]{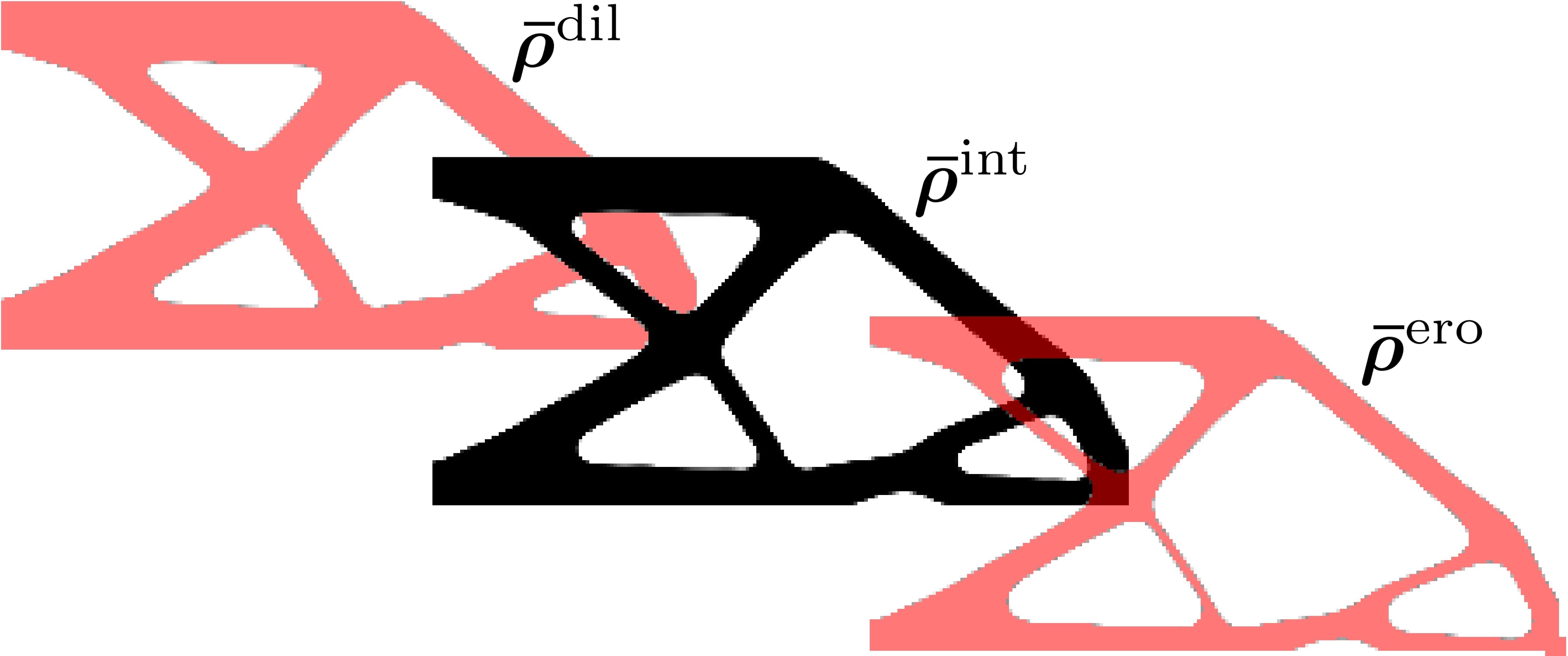}
		\caption{}
		\label{FIG:Ero_Int_Dil_a}
	\end{subfigure}
	\begin{subfigure}{0.90\linewidth}
		\centering	
		\includegraphics[width=0.65\linewidth]{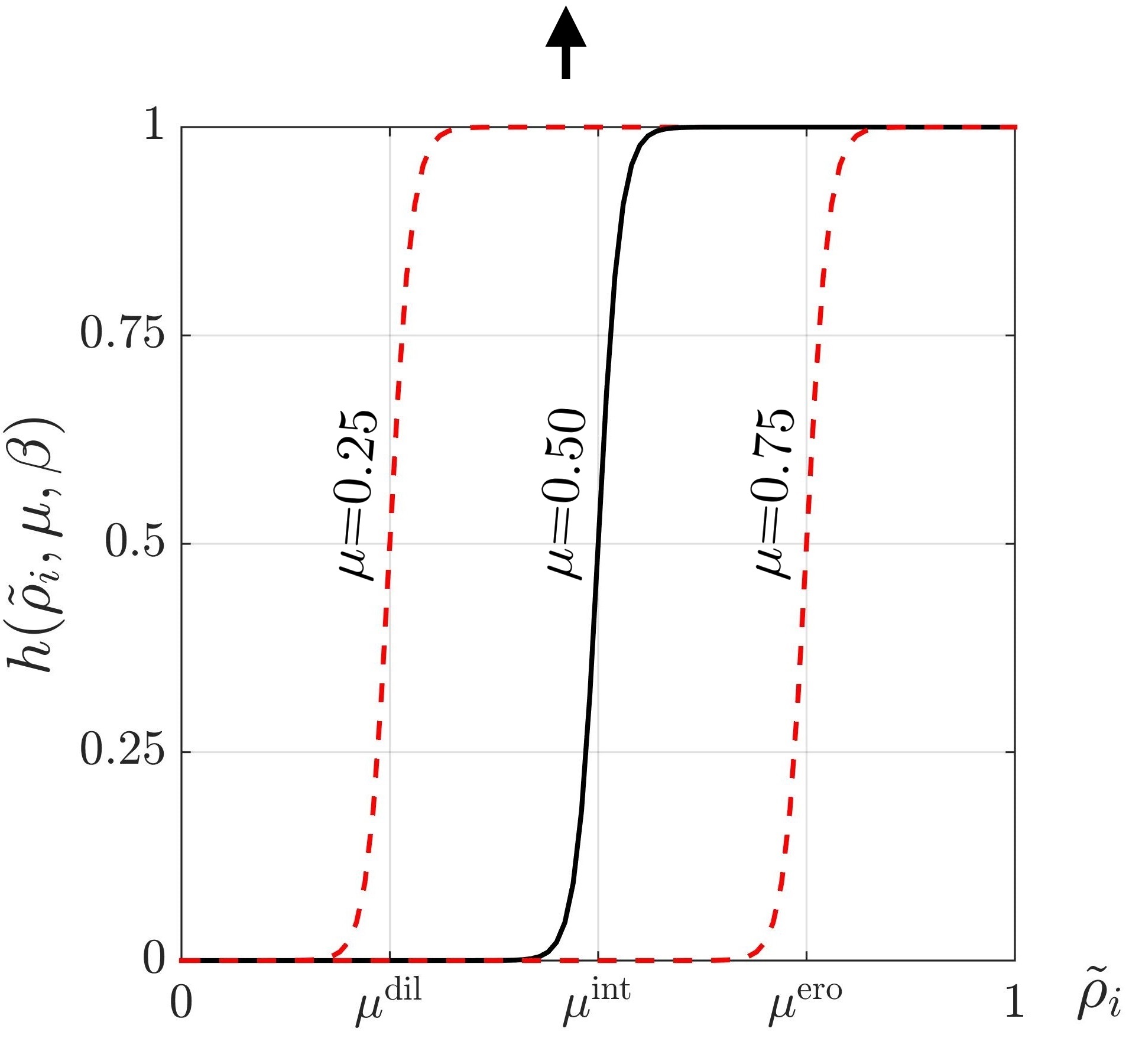}
		\caption{}
		\label{FIG:Ero_Int_Dil_b}
	\end{subfigure}	
	\begin{subfigure}{0.90\linewidth}
		\centering			
		\includegraphics[width=0.40\linewidth]{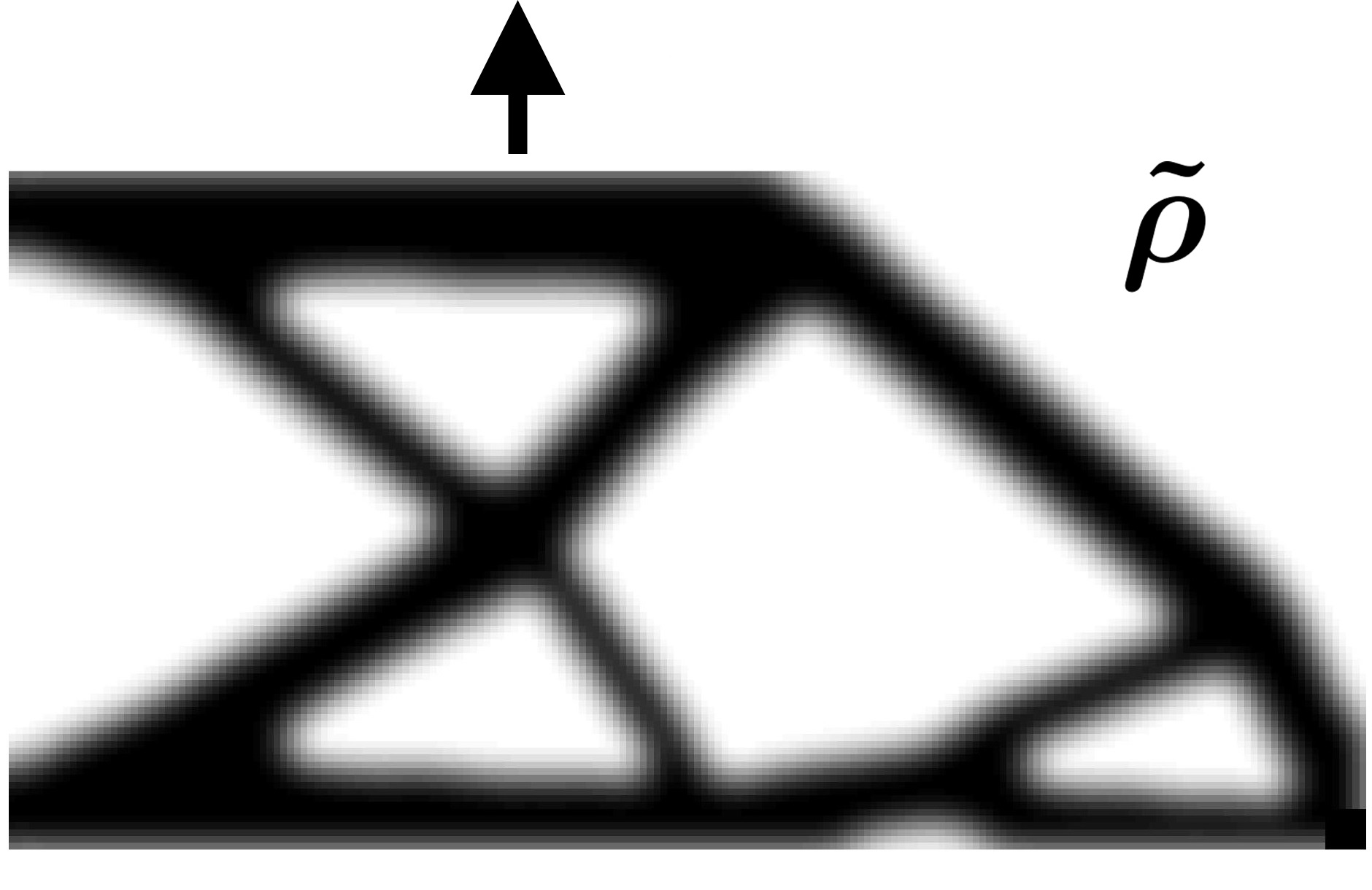}   
		\caption{}
		\label{FIG:Ero_Int_Dil_c}
	\end{subfigure}	 
	\caption{(a) Optimized designs corresponding to dilated, intermediate and eroded fields, (b) Plots for smoothed Heaviside function at 3 different thresholds and $\beta=38$, and (c) The final filtered material field.}			
	\label{FIG:Ero_Int_Dil}
\end{figure}

	As mentioned before, to impose MinS and MinV, we adopt the robust design optimization approach \citep{Wang2011}. The physical design variable of element $i$ is obtained as:
	\begin{equation} \label{eq:Heaviside}
	\bar{\rho}_i = {h}(\tilde{\rho}_i,\beta,\mu) = 
	\frac
	{\mathrm{tanh}(\beta\mu)+\mathrm{tanh}(\beta\:(\tilde{\rho}_i-\mu))}
	{\mathrm{tanh}(\beta\mu)+\mathrm{tanh}(\beta\:(        1     -\mu))},
	\end{equation}
	where $\beta$ and $\mu$ control the steepness and the threshold of the projection, respectively. ${h}(\tilde{\rho}_i,\beta,\mu)$ is displayed in Fig. \ref{FIG:Ero_Int_Dil_b} for three different thresholds at $\beta=38$.  $\mu$ simulates the manufacturing error. The subscripts $\mathrm{ero}$, $\mathrm{int}$ and $\mathrm{dil}$ are used for the parameters pertaining to eroded, intermediate and dilated deigns, respectively in this paper.
	
	The minimum size is defined by a circular region of radius $r_\mathrm{min}$ and that for the intermediate design depends on the filter radius $r_\mathrm{fil}$ and on the thresholds $\mu_\mathrm{ero}$, $\mu_\mathrm{int}$ and $\mu_\mathrm{dil}$ \citep{Wang2011,Qian2013,trillet2021analytical}. In this work, we set  $[\mu_\mathrm{ero},\mu_\mathrm{int},\mu_\mathrm{dil}]=[0.75,0.50,0.25]$ in order to obtain MinS equal to MinV and $r_\mathrm{min}=0.5r_\mathrm{fil}$. This indicates that if the desired minimum size is $r_\mathrm{min}=3s$, then $r_\mathrm{fil}=6s$, where $s$ is the size of an element.

	\subsection{Reference topology optimization formulation} \label{Sec:2.TOProblem}
	
	The reference optimization problem is considered \citep{Amir2018b} that impose MinS and MinV as:
	
	\begin{align} \label{EQ:OPTI}
	\begin{split}
	{\min_{\bm{\rho}}} & \quad C_\mathrm{ero}=\mathbf{f}^{\intercal} \mathbf{u}_\mathrm{ero} \\
	\mathrm{s.t.:} &\quad  \mathbf{v}^{\intercal} \bm{\bar{\rho}}_\mathrm{dil} \leq V^*_\mathrm{dil} \left( V^*_\mathrm{int} \right) 	\\
	&\quad 0 \leq {\rho_i} \leq1  \;\;,\;\; i=1,...\:,N \;\;, 
	\end{split}
	\end{align}
	\noindent where $\mathbf{f}$ is the external force, and $C_\mathrm{ero}$ represents compliance of the eroded design. $\mathbf{u}_\mathrm{ero}$ contains the nodal displacements of the eroded design, which is obtained by solving  $\mathbf{K}(\bm{\bar{\rho}}_\mathrm{ero})\:\mathbf{u}_\mathrm{ero}=\mathbf{f}$. $\mathbf{K}(\bm{\bar{\rho}}_\mathrm{ero})$ is the global stiffness matrix and  $\mathbf{k}(\bm{\bar{\rho}}_\mathrm{ero})$ is the stiffness matrix of an element. $ \mathbf{k}_i(\bar{\rho}_{\mathrm{ero}(i)}) = \mathrm{E}_i(\bar{\rho}_{\mathrm{ero}(i)}) \mathbf{k_0}$.  $\mathrm{E}_i(\bar{\rho}_{\mathrm{ero}(i)})$ is the element Young's modulus and $\mathbf{k_0}$ denotes the element stiffness matrix at unit Young's modulus.  To interpolate $\mathrm{E}_i$ between solid and void states of an element, the modified SIMP interpolation scheme is used \citep{Sigmund2007}. Mathematically, one writes
	\begin{equation}
	\mathrm{E}_i = \mathrm{E}_\mathrm{min} + \bar{\rho}_{\mathrm{ero}(i)}^\eta(\mathrm{E}_0-\mathrm{E}_\mathrm{min}),
	\end{equation}
	where $\eta$ is a penalization parameter, and $E_0$ and $E_\text{min}$ are Young's moduli of solid and void elements respectively.
	
	As done in \citep{Amir2018b}, the volume constraint is applied on the dilated design through the intended volume restriction $V^*_\mathrm{int}$ by scaling $V^*_\mathrm{dil}$ every 10 iterations as follows:
	\begin{equation}
	V^*_\mathrm{dil} = \frac{\mathbf{v}^{\intercal} \bm{\bar{\rho}}_\mathrm{dil}}{\mathbf{v}^{\intercal} \bm{\bar{\rho}}_\mathrm{int}} V^*_\mathrm{int} \; .
	\end{equation}
	
\begin{figure}
\centering   	
			\includegraphics[width=0.600\linewidth]{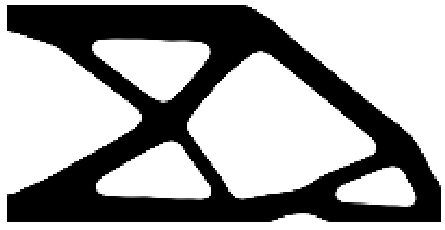}	
\caption{Optimized intermediate design obtained by running code as \texttt{topCbeam(200,100,0.4,1.0,6.1)}.}	\label{FIG:Filtering_2_a}	
\end{figure}

	The optimization problem is solved using the Method of Moving Asymptotes (MMA, cf. \cite{Svanberg1987}) with the above mentioned filtering and projection filters. The MATLAB code \texttt{topCbeam} provided in Appendix~\ref{APP:1} adopts the element connectivity matrix for the rectangular tessellation and stiffness evaluation from \citet{Andreassen2011}. $\mathrm{E}_0=1$ and  $\mathrm{E}_\mathrm{min}=10^{-9}$ are set. Continuation schemes are used for the SIMP and Heaviside parameters. $\eta$ is initialized at $1.0$ and is increased by $0.125$ in every $40$ iterations. Initial value of $\beta$ is set to 1.0, which is increased by  $1.5$ times in every 40 iterations. The maximum number of iterations is set to $340$. To ease convergence, we restrict the maximum change of design variables as 
	\begin{equation}
	\mathrm{max}(0,\rho_i-\mathrm{m_L}) \leq \rho_i \leq \mathrm{min}(1,\rho_i+\mathrm{m_L}),
	\end{equation}
	\noindent where $\mathrm{m_L}$ is the move limit. As the $\eta$ and $\beta$ parameters increase, so does the non-linearity of the optimization problem \citep{fernandez2020imposing}. To reduced the effects of non-linearity,  we define the move limit $\mathrm{m_L}$ using a continuation approach as
	\begin{equation}
	\mathrm{m_L} = \frac{0.7-0.1}{1-2}(\eta-2)+0.1 \;.
	\end{equation}
	Using the above mentioned parameters, the reference optimization problem (Eq.~\ref{EQ:OPTI}) is solved for the cantilever beam design (Fig.~\ref{FIG:Cone_Example_a}). The optimized design obtained by calling \texttt{topCbeam(200,100, 0.4,1.0,6.1)} is displayed in  
	Fig.~\ref{FIG:Filtering_2_a}.

	\subsection{Gradient-based overhang constraint}\label{Sec:Gradient-overhangcons}
	
\begin{figure}
	\centering
	\begin{subfigure}{0.36\linewidth}
		    \includegraphics[width=1\linewidth]{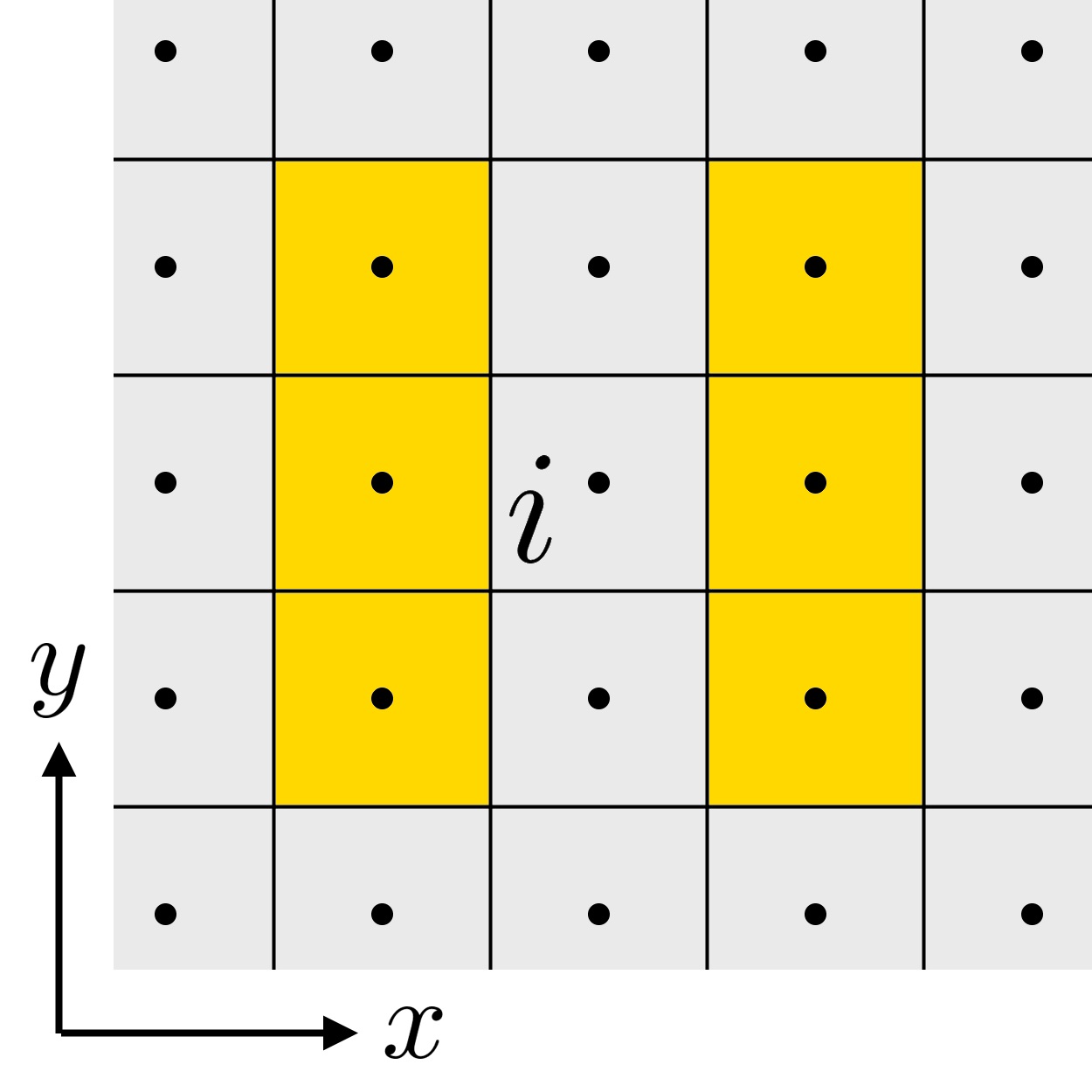}
			\caption{}			
			\label{FIG:Gradient_Computation_a}	
	\end{subfigure}
	\begin{subfigure}{0.36\linewidth}
		    \includegraphics[width=1\linewidth]{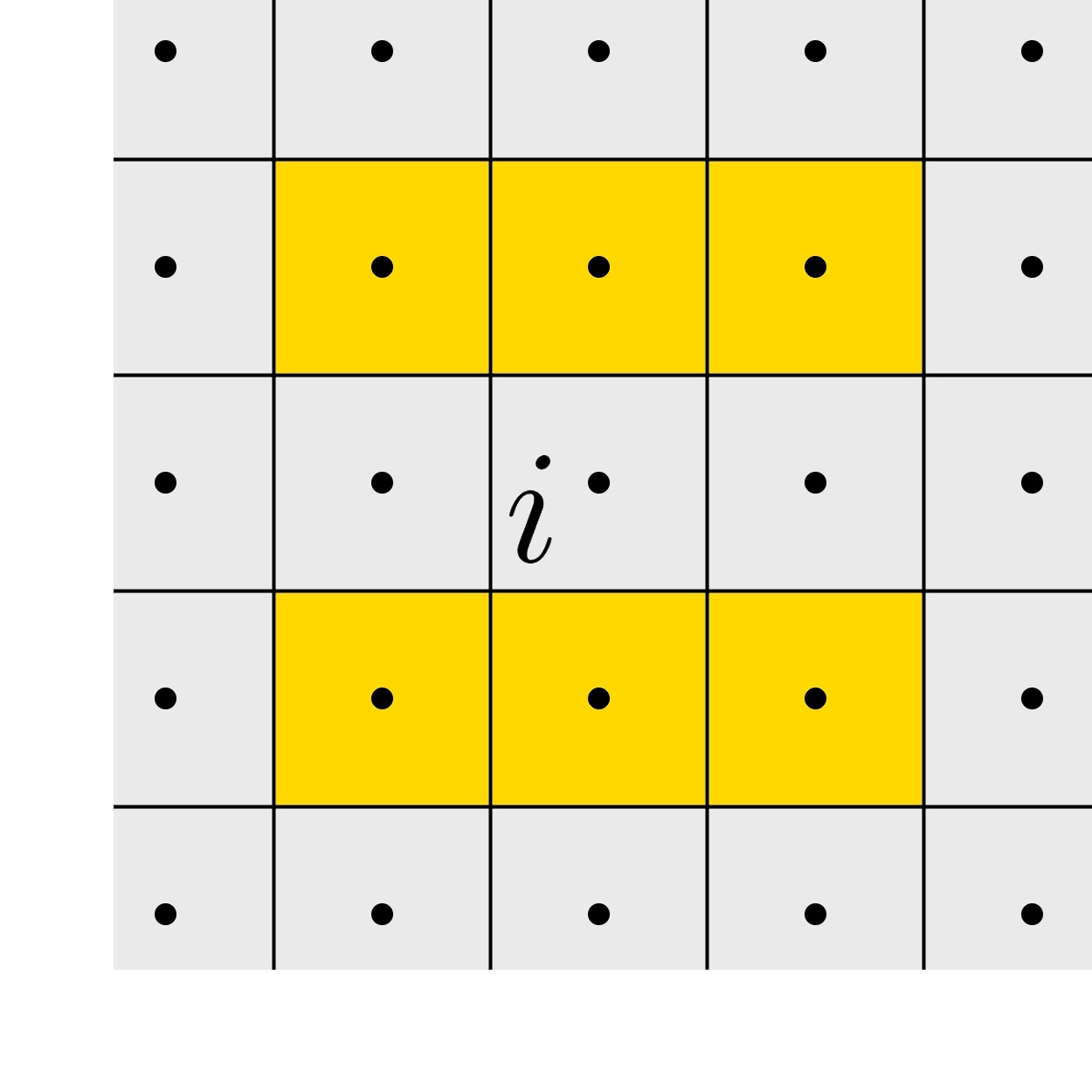}
			\caption{}			
			\label{FIG:Gradient_Computation_b}	
	\end{subfigure}	
	\caption{Local region for gradient computation. (a) $\delta_{\mathrm{x}i}$, and (b) $\delta_{\mathrm{y}i}$.}	
	\label{FIG:Gradient_Computation}			
\end{figure}

	We use the structural gradient approach to provide limitation on the overhang angle, wherein edge of the structure is detected to limit its slope \citep{Qian2017}. In this work, the Prewitt operator, a discrete differentiation operator, is used to compute the approximate gradient of the image intensity function \citep{Yang2019}. It uses a 3$\times$3 elements mask to approximate the gradient $\bm{\delta}_i$ of element $i$, as shown in Fig. \ref{FIG:Gradient_Computation}. The 2D vector $\bm{\delta}_i=(\delta^{\mathrm{x}}_i \: , \: \delta^{\mathrm{y}}_i)$ is computed as:
	\begin{equation}
	\delta^{\mathrm{x}}_i = \displaystyle \sum_{j=1}^N \bar{\rho}_j \: \mathrm{D_x}_{(i,j)}
	\quad\;\; , \quad\;\; 
	\delta^{\mathrm{y}}_i = \displaystyle \sum_{j=1}^N \bar{\rho}_j \: \mathrm{D_y}_{(i,j)}
	\;\; ,
	\end{equation}
	
	\noindent where $\mathrm{D_x}_{(i,j)}$ and $\mathrm{D_y}_{(i,j)}$, the unit weights of the Prewitt operator, are defined as:
	\begin{equation}
	\mathrm{D}_{\mathrm{x}(i,j)} = 
	\left\lbrace
	\begin{matrix} 
	\displaystyle\frac{(x_j-x_i)}{|x_j-x_i|} \;, & \text{if} \;\;\;\;\; 0<|x_j-x_i| \leq t_e   \\
	& \text{and} \;\; 0 \leq |y_j-y_i| \leq t_e
	\vspace{2mm}\\
	0  \;,				& \text{otherwise}
	\end{matrix}
	\right. \; ,
	\end{equation}
	\begin{equation}
	\mathrm{D}_{\mathrm{y}(i,j)} = 
	\left\lbrace
	\begin{matrix} 
	\displaystyle\frac{(y_j-y_i)}{|y_j-y_i|} \;, & \text{if} \;\;\;\;\; 0<|y_j-y_i| \leq t_e   \\
	& \text{and} \;\; 0 \leq |x_j-x_i| \leq t_e
	\vspace{2mm}\\
	0  \;,				& \text{otherwise}
	\end{matrix}
	\right. 
	\: ,
	\end{equation}

	In the matrix form, one writes 
	\begin{equation}
	\bm{\delta}^\mathbf{x} = \mathbf{D_x} \bm{\bar{\rho}} \;,\;\;\; \bm{\delta}^\mathbf{y} = \mathbf{D_y} \bm{\bar{\rho}} \;.
	\end{equation}
	Thus, sensitivities are evaluated as: 
	\begin{equation} \label{EQ:SensitivitiesPrewitt}
	\begin{matrix}
	\displaystyle \deriv{\bm{\delta}^\mathrm{x}}{\bm{\bar{\rho}}} = \mathbf{D_x}^\intercal \; , & \displaystyle \deriv{\bm{\delta}^\mathrm{y}}{\bm{\bar{\rho}}} = \mathbf{D_y}^\intercal \; .
	\end{matrix}
	\end{equation}
	
	The overhang restriction formulation compares directions of the vectors. In this view, the gradient vector $\bm{\delta}_i$ is normalized to obtain the gradient direction vector $\bm{\varrho}_i$  as:
	\begin{equation}
	\begin{matrix}
	\varrho^{\mathrm{x}}_i =
	\delta^{\mathrm{x}}_i \; {\bar{\delta}}_i \; , &
	\varrho^{\mathrm{y}}_i =
	\delta^{\mathrm{y}}_i \; {\bar{\delta}}_i \; ,
	\end{matrix}
	\end{equation} 
	\noindent where ${\bar{\delta}}_i$, the normalization factor, is calculated as:
	\begin{equation}\label{EQ:Normalization_Factor}
	{\bar{\delta}}_i =
	\left\lbrace
	\begin{matrix}
	\displaystyle \frac{1}{\| \bm{\delta}_i \|}=\frac{1}{\sqrt{	(\delta^{\mathrm{x}}_i)^2 + (\delta^{\mathrm{y}}_i)^2 }} \; ,& \text{if} \;\;\;  \| \bm{\delta}_i \| \geq \varepsilon_\mathrm{n}
	\vspace{3mm}\\
	0 \;\;,& \text{otherwise.}
	\end{matrix}
	\right.
	\end{equation}
	
	\noindent The parameter $\varepsilon_\mathrm{n}$, a positive number, is used to avoid singularities when $\| \bm{\delta}_i \|$ is close to zero. We provide a MATLAB code, \texttt{GBOHC}, that implements the Prewitt operation (see Appendix~\ref{APP:2}). The matrices $\mathbf{D_x}$ and $\mathbf{D_y}$ are created on lines 19-21 and the $N$ gradients $\bm{\delta}$ are computed on line 32.

	\subsubsection{Maximum Overhang Angle constraint}
\begin{figure}
	\centering
	\begin{subfigure}{0.45\linewidth}
		    \includegraphics[scale = 0.1]{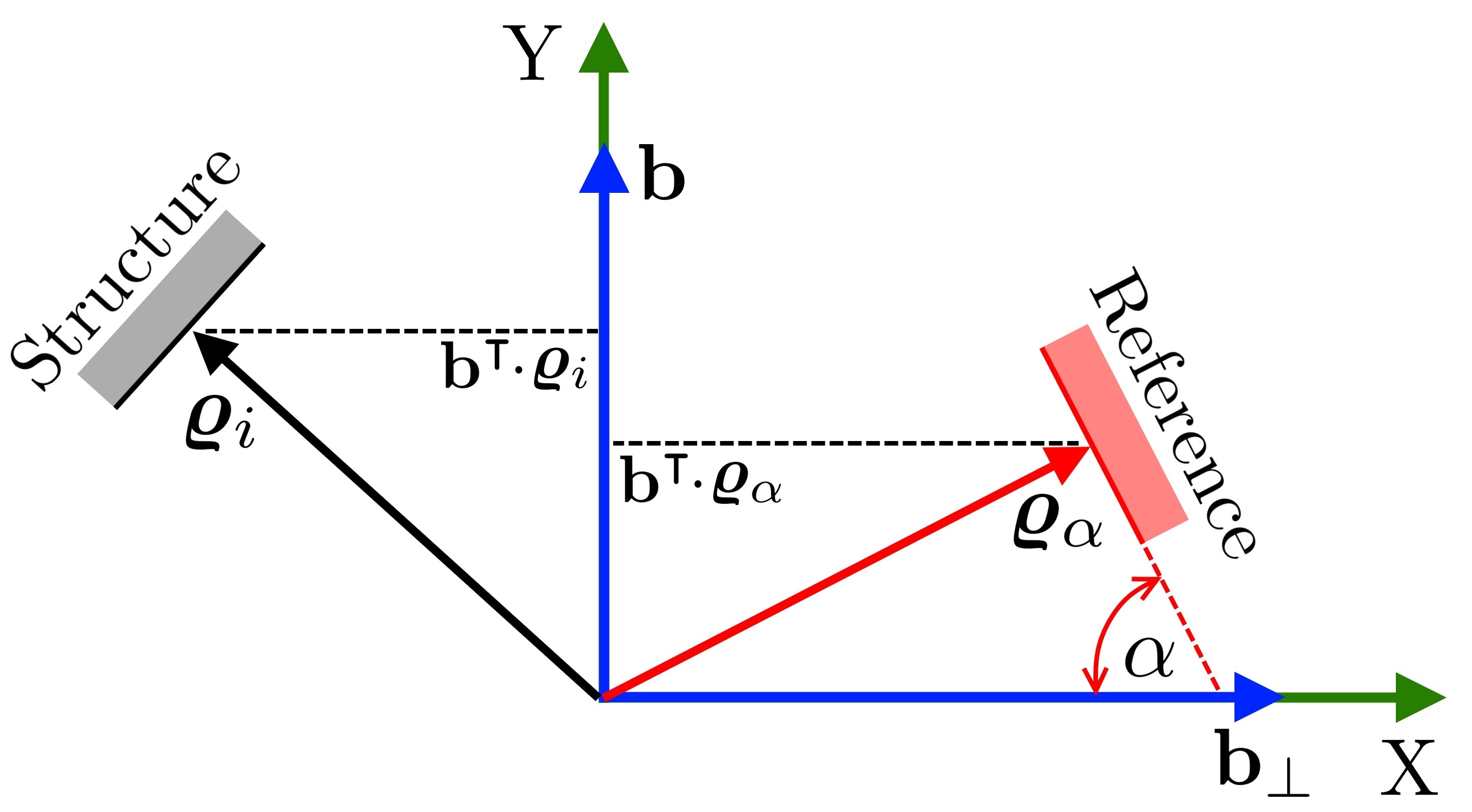}
			\caption{}			
			\label{FIG:Local_Constraint_a}	
	\end{subfigure}
    \begin{subfigure}{0.45\linewidth}
		    \includegraphics[scale = 0.1]{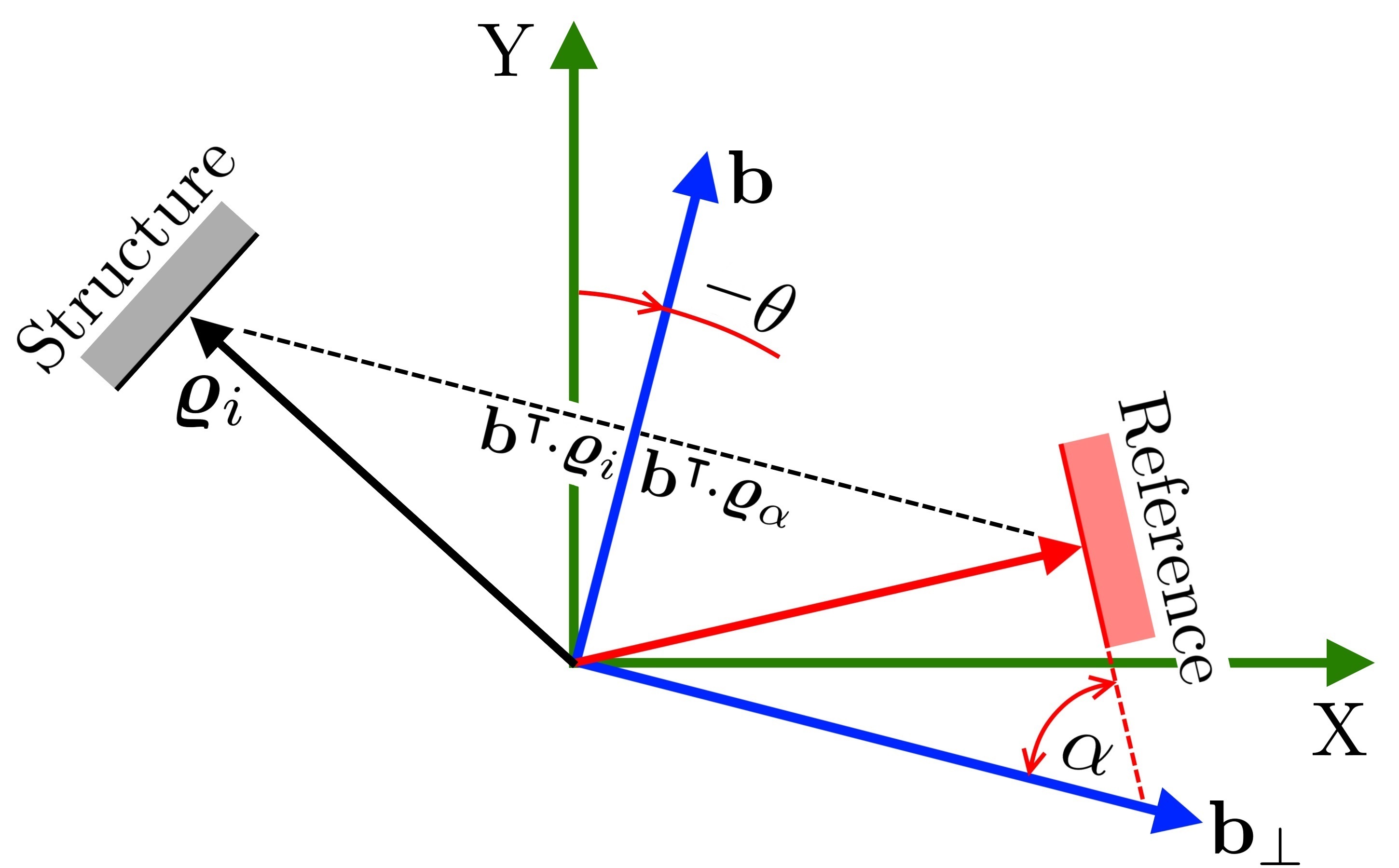}
			\caption{}			
			\label{FIG:Local_Constraint_b}	
			\end{subfigure}
	
	\caption{Gradient-based overhang constraint. (a) For a vertical building direction $\mathbf{b}$, where the condition (Eq.~\ref{EQ:Local_Constraint}) is violated, and (b) $\mathbf{b}$ is rotated and the condition (Eq.~\ref{EQ:Local_Constraint}) is satisfied.}	
	\label{FIG:Local_Constraint}			
\end{figure}

	The MaxOA constraint formulation is illustrated in Fig.~\ref{FIG:Local_Constraint_a}. The lowest possible inclination, $\alpha$, exists at the reference surface. The projection of the gradient $\bm{\varrho}_i$ in the building direction $\mathbf{b}$  is $\mathbf{b}^\intercal \bm{\varrho}_i$, and that for reference gradient $\bm{\varrho_\alpha}$ is $\mathbf{b}^\intercal \bm{\varrho_\alpha}$. To restrict the structure from inclining below $\alpha$, $\mathbf{b}^\intercal \bm{\varrho}_i$ should be less than $\mathbf{b}^\intercal \bm{\varrho_\alpha}$. Mathematically, we write
	
	\begin{equation}\label{EQ:Local_Constraint}
	g_i = \mathbf{b}^\intercal \bm{\varrho}_i - \mathbf{b}^\intercal \bm{\varrho_\alpha} \leq 0 \; ,
	\end{equation}
	\noindent where $g_i$ is the overhang restriction for element $i$. We aggregate the $N$ constraints $\bm{g}$ using a \textit{p}-mean function \citep{fernandez2020imposing}. For that, the constraint $g_i$ is shifted such that it remains in [0,1] as follows:
	\begin{equation}\label{EQ:Local_Constraint_Shifted}
	s_i = \frac{1}{2}\:(g_i + \mathbf{b}^\intercal \bm{\varrho_\alpha} + 1) \; .
	\end{equation}
	
	Now, the global overhang constraint $G$ is obtained by aggregating $\bm{s}$ and shifting back as:
	\begin{equation} \label{EQ:Global_Constraint}
	\mathrm{G} = 2 \left( \frac{1}{N} \sum_{i=1}^{N} s_i^p \right)^{\frac{1}{p}} - \mathbf{b}^\intercal \bm{\varrho_\alpha} - 1 \leq 0 \; .
	\end{equation}
	
	\noindent The chain rule is used to determine the derivative of $\mathrm{G}$. Considering column arrays and denominator-layout notation \citep{Xing2019}, the derivative of $\mathrm{G}$ with respect to the design variables $\bm{\rho}$ is:
	\begin{equation}
	\begin{aligned}
	\deriv{\mathrm{G}}{\bm{\rho}} = 
	& 
	\deriv{\bm{\tilde{\rho}}}{\bm{\rho}}
	\deriv{\bm{\bar{\rho}}}{\bm{\tilde{\rho}}}
	\deriv{\bm{g}}{\bm{\bar{\rho}}}
	\deriv{\mathrm{G}}{\bm{g}} \; ,
	\end{aligned}
	\end{equation}
	\noindent where
	\begin{equation}
	\deriv{\bar{\rho}_i}{\tilde{\rho}_j} =
	\left\lbrace
	\begin{matrix}
	\bar{\rho}_i' = \frac{\beta \; \mathrm{sech}^2(\beta(\tilde{\rho}_i-\eta))}{\mathrm{tanh}(\beta\eta)+\mathrm{tanh}(\beta(1-\eta))}\; , & \text{if} \;\;\; i=j \; , \vspace{2mm}\\
	0 \;\;\;,& \text{otherwise,} 
	\end{matrix}
	\right.
	\end{equation}
	\noindent and
	\begin{equation}
	\deriv{\mathrm{G}}{g_i} = \mathrm{G}_i' = \frac{s_i^{p-1}}{N} \left( \frac{1}{N} \sum_{j=1}^{N} s_j^p \right)^{\frac{1}{p}-1}  \; .
	\end{equation}
	
	\noindent In matrix notation, the sensitivities become:
	\begin{equation} \label{EQ:SensitivitiesG}
	\deriv{\mathrm{G}}{\bm{\rho}} = 
	\mathbf{H}^\intercal \;
	\mathrm{diag}(\bm{\bar{\rho}}') \;
	\deriv{\bm{g}}{\bm{\bar{\rho}}} \;
	\mathbf{G}' \;.
	\end{equation}

	\noindent The derivative of the local constraints $\bm{g}$ are also obtained by the chain rule as:
	\begin{equation} \label{EQ:Sensitivitiesg}
	\begin{aligned}
	\deriv{\mathrm{g}}{\bm{\bar{\rho}}} = 
	&  
	\left(  \deriv{\bm{\delta^{\mathrm{x}}}}{\bm{\bar{\rho}}} 
	\derivp{\bm{\varrho^{\mathrm{x}}}}{\bm{\delta^{\mathrm{x}}}} 
	+ 
	\deriv{\|\bm{\delta}\|}{\bm{\bar{\rho}}}
	\deriv{\bm{\bar{\delta}}}{\|\bm{\delta}\|} 
	\derivp{\bm{\varrho^{\mathrm{x}}}}{\bm{\bar{\delta}}}
	\right) 
	\derivp{\bm{g}}{\bm{\varrho^{\mathrm{x}}}}
	\:+ \\ 
	&
	\hspace{0mm}
	\left(  \deriv{\bm{\delta^{\mathrm{y}}}}{\bm{\bar{\rho}}} 
	\deriv{\bm{\varrho^{\mathrm{y}}}}{\bm{\delta^{\mathrm{y}}}} 
	+ 
	\deriv{\|\bm{\delta}\|}{\bm{\bar{\rho}}}
	\deriv{\bm{\bar{\delta}}}{\|\bm{\delta}\|} 
	\deriv{\bm{\varrho^{\mathrm{y}}}}{\bm{\bar{\delta}}}
	\right) 
	\derivp{\bm{g}}{\bm{\varrho^{\mathrm{y}}}} \; ,
	\end{aligned}
	\end{equation}
	\noindent where
	\begin{subequations}
		\begin{align}
		\derivp{\bm{\varrho^{\mathrm{x}}}}{\bm{\delta^{\mathrm{x}}}}
		& =
		\derivp{\bm{\varrho^{\mathrm{y}}}}{\bm{\delta^{\mathrm{y}}}}
		=
		\mathrm{diag \left( \bm{\bar{\delta}} \right)} \; ,
		\\
		\deriv{\|\bm{\delta}\|}{\bm{\bar{\rho}}} 
		& =
		\deriv{\bm{\delta}^\mathrm{x}}{\bm{\bar{\rho}}}
		\:\mathrm{diag}\left(\bm{\varrho^\mathrm{x}}\right) +
		\deriv{\bm{\delta}^\mathrm{y}}{\bm{\bar{\rho}}}
		\:\mathrm{diag}\left(\bm{\varrho^\mathrm{y}}\right) \; , 
		\\
		\deriv{\bm{\bar{\delta}}}{\|\bm{\delta}\|}
		&=
		-\mathrm{diag}\left(\bm{\bar{\delta}}\right) \mathrm{diag}\left(\bm{\bar{\delta}}\right) \; , 
		\\
		\derivp{\bm{\varrho^{\mathrm{x}}}}{\bm{\bar{\delta}}}
		&=
		\mathrm{diag}\left(\bm{\delta^{\mathrm{x}}}\right) \; ,
		\\
		\derivp{\bm{\varrho^{\mathrm{y}}}}{\bm{\bar{\delta}}}
		&=
		\mathrm{diag}\left(\bm{\delta^{\mathrm{y}}}\right) \; ,
		\\
		\derivp{\bm{g}}{\bm{\varrho^{\mathrm{x}}}}
		&=
		\mathrm{b_x} \mathbf{I} \; ,
		\\
		\derivp{\bm{g}}{\bm{\varrho^{\mathrm{y}}}}
		&=
		\mathrm{b_y} \mathbf{I} \; .
		\end{align}
	\end{subequations}
	
	Note that evaluation of the constraint and sensitivities of a gradient-based formulation is computationally cheaper than that of the layer-by-layer formulations. With gradient formulation, the use of several building directions can lead to high computational cost since it needs many matrix operations to be carried out. As the matrix products originate from the derivative of the Prewitt operator, to reduce the computational cost, Eq.~\ref{EQ:Sensitivitiesg} is treated algebraically with respect to $\deriv{\bm{\delta^{\mathrm{x}}}}{\bm{\bar{\rho}}}$ and $\deriv{\bm{\delta^{\mathrm{y}}}}{\bm{\bar{\rho}}}$  as: 
	\begin{gather} \label{EQ:Sensitivitiesg_Simplified}
	\begin{aligned}
	\hspace{-6mm}
	\displaystyle \deriv{\bm{g}}{\bm{\bar{\rho}}} = 
	&
	\left(
	\mathbf{D_x}^\intercal 
	\mathrm{diag}\left(
	\mathrm{b_x} \bm{1} \minus \mathrm{b_x} \bm{\varrho^\mathrm{x}} \circ \bm{\varrho^\mathrm{x}} \minus \mathrm{b_y} \bm{\varrho^\mathrm{y}} \circ \bm{\varrho^\mathrm{x}} 
	\right)
	\right.
	+ 
	\vspace{2mm} \\
	&
	\displaystyle
	\hspace{1mm}
	\left.
	\mathbf{D_y}^\intercal  
	\mathrm{diag}\left(
	\mathrm{b_y} \bm{1} \minus \mathrm{b_y} \bm{\varrho^\mathrm{y}} \circ \bm{\varrho^\mathrm{y}} \minus   \mathrm{b_x} \bm{\varrho^\mathrm{y}} \circ \bm{\varrho^\mathrm{x}} 
	\right)
	\right)
	\mathrm{diag}\left( \bm{\bar{\delta}}\right) \;,
	\end{aligned}
	\end{gather}
	\noindent where $\circ$ represents the Hadamard product and $\bm{1}$ is a unit array. In view of Eqs.~\ref{EQ:SensitivitiesG} and  \ref{EQ:Sensitivitiesg_Simplified}, the sensitivities of G can be determined as:
	
	\begin{equation}
	\deriv{\mathrm{G}}{\bm{\rho}} = 
	\mathbf{H}^{\intercal}
	\left(
	{h}^\prime{(\bm{\tilde{\rho}})} \circ 
	\left( 
	\mathbf{D_x}^\intercal \left( \mathbf{a_x} \circ \mathbf{B} \right)
	+
	\mathbf{D_y}^\intercal \left( \mathbf{a_y} \circ \mathbf{B} \right)
	\right)
	\right),
	\end{equation}
	where
	\begin{equation}\label{EQ:AxAyB}
	\begin{matrix}
	\mathbf{a_x}
	=
	\mathrm{b_x}\bm{1}- \mathrm{b_x}\bm{\varrho^\mathrm{x}} \circ \bm{\varrho^\mathrm{x}} - \mathrm{b_y}\bm{\varrho^\mathrm{y}} \circ \bm{\varrho^\mathrm{x}},
	\vspace{2mm} 
	\\
	\mathbf{a_y}
	=
	\mathrm{b_y}\bm{1} - \mathrm{b_y} \bm{\varrho^\mathrm{y}} \circ \bm{\varrho^\mathrm{y}}  - \mathrm{b_x} \bm{\varrho^\mathrm{x}} \circ \bm{\varrho^\mathrm{y}},
	\vspace{2mm} \\
	\mathbf{B}
	=
	0.5 \: \bm{\bar{\delta}} \circ \mathrm{G}^{\prime}.
	\end{matrix}
	\end{equation}

	\noindent In the code \texttt{GBOHC}, the local constraints $\bm{g}$ are computed on line 37, the global constraint $\mathrm{G}$ on line 40 and its sensitivities on lines 41-45.
	
	The following overhang-constrained optimization problem is formulation by applying restriction on the eroded, intermediate and dilated designs:
	\begin{align} \label{EQ:OPTI_Constrained}
	\begin{split}
	{\min_{\bm{\rho}}} & \quad c_\mathrm{ero}=\mathbf{f}^{\intercal} \mathbf{u}_\mathrm{ero} \\
	\mathrm{s.t.:} 
	&\quad  \mathbf{v}^{\intercal} \bm{\bar{\rho}}_\mathrm{dil} \leq V^*_\mathrm{dil} \left( V^*_\mathrm{int} \right) 	\\
	&\quad \mathrm{G}(\bm{\bar{\rho}}_\mathrm{ero}) \leq 0 \\
	&\quad \mathrm{G}(\bm{\bar{\rho}}_\mathrm{int}) \leq 0 \\
	&\quad \mathrm{G}(\bm{\bar{\rho}}_\mathrm{dil}) \leq 0 \\
	&\quad 0 \leq {\rho_i} \leq1  \;\;,\;\; i=1,...\:,N \;\;, 
	\end{split}
	\end{align}
	
	The default printing direction is considered vertical, i.e. $\theta=0$ as shown in Fig. \ref{FIG:Local_Constraint_a}. If this is rotated by $\theta$, so does the reference surface (Fig.~\ref{FIG:Local_Constraint_b}). The printing direction is defined in the rotated sense as: 
	\begin{equation}
	\mathbf{b}=(\mathrm{cos}(\theta+\pi/2) \:,\: \mathrm{sin}(\theta+\pi/2)),
	\end{equation}
	\noindent and the gradient vector of the reference surface as:
	\begin{equation}
	\varrho_\alpha^\mathrm{x}
	=
	\mathrm{cos} \left( \theta + \pi/2 -\alpha \right) 
	\; , \;
	\varrho_\alpha^\mathrm{y}
	=
	\mathrm{sin} \left( \theta + \pi/2 -\alpha \right).
	\end{equation}
	
\begin{figure}
	\centering
	\includegraphics[width=0.5\linewidth]{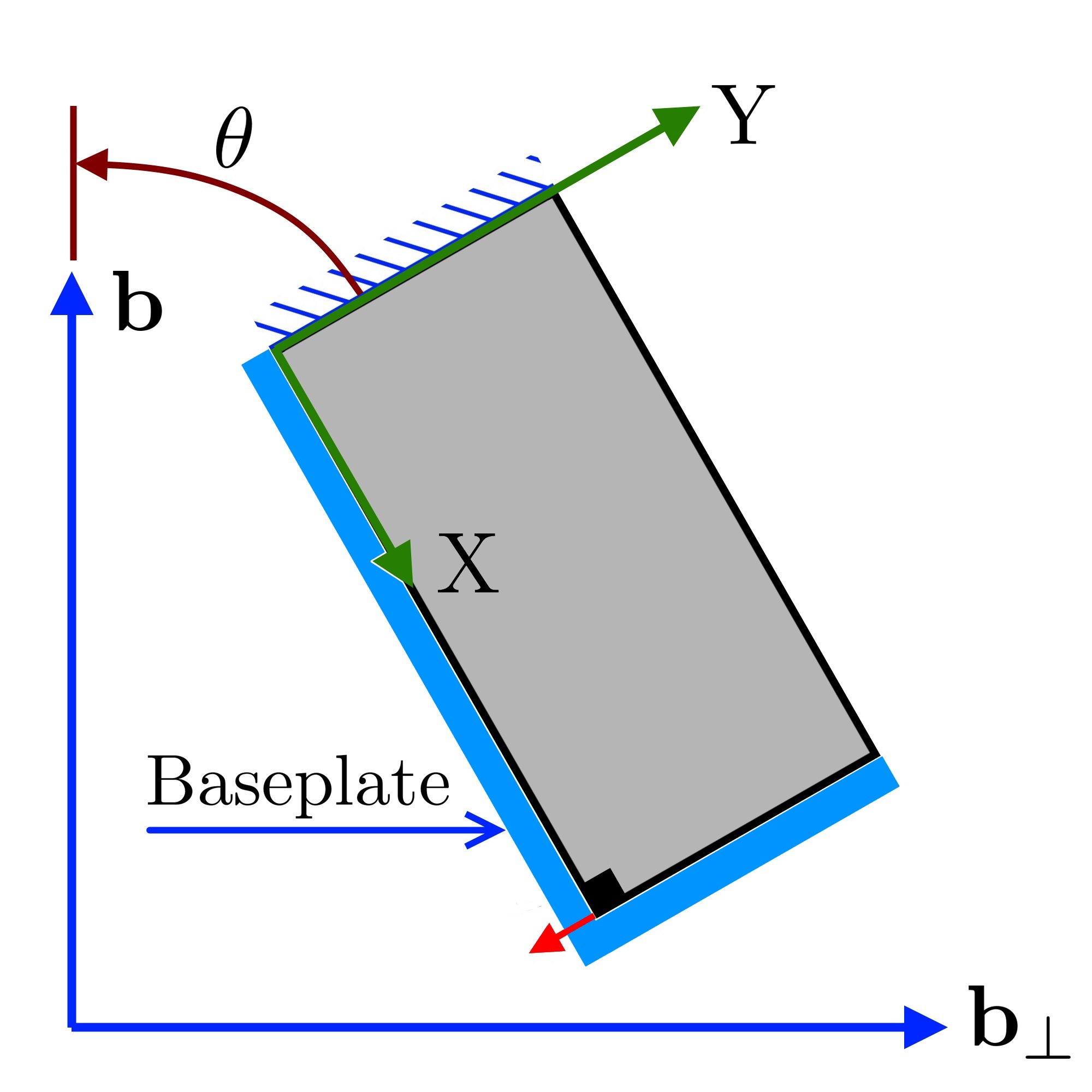}
		\caption{The boundaries of the design domain serve as baseplate when inclined building orientations are considered.}			
	\label{FIG:Rotation_Domain}	
\end{figure}

	\noindent In the code \texttt{GBOHC}, the angle $\alpha$ is defined on line 8, and the building direction $\theta$ is defined on line 9. Support structures outside the design domain are not considered herein. Thus, for printing directions that are inclined with respect to the design domain, the lower edge of the design domain is treated as the base plate, as shown in Fig.~\ref{FIG:Rotation_Domain}. In order to keep the code compact and general for any printing direction, the base plate in the code \texttt{GBOHC} is considered in all (lines 68-72).
	
	\begin{table*}
\caption{Strategy to avoid non-self-supporting features. First row shows the intermediate material field, second row the compliance field and third row the overhanging angle restriction evaluated in the intermediate field.}
\centering
\begin{tabular}{p{0.02\textwidth}p{0.24\textwidth}p{0.2\textwidth}p{0.2\textwidth}p{0.2\textwidth}}
	\toprule	    
     & \makecell{Constrained ($\mathrm{G}$) \\ It: $189$} & \makecell{Free Evolution (\cancel{$\mathrm{G}$}) \\ It: $200$} & \makecell{Constrained ($\mathrm{G}$) \\ It: $273$} & \makecell{Constrained after PP \\  It: $339$}  	
	\\	
	\cmidrule(r){2-5}	      
    \vspace{-10mm}{\large{$\;\;\bm{\bar{\rho}}_\mathrm{int}$}}
    &   \includegraphics[width=1\linewidth]{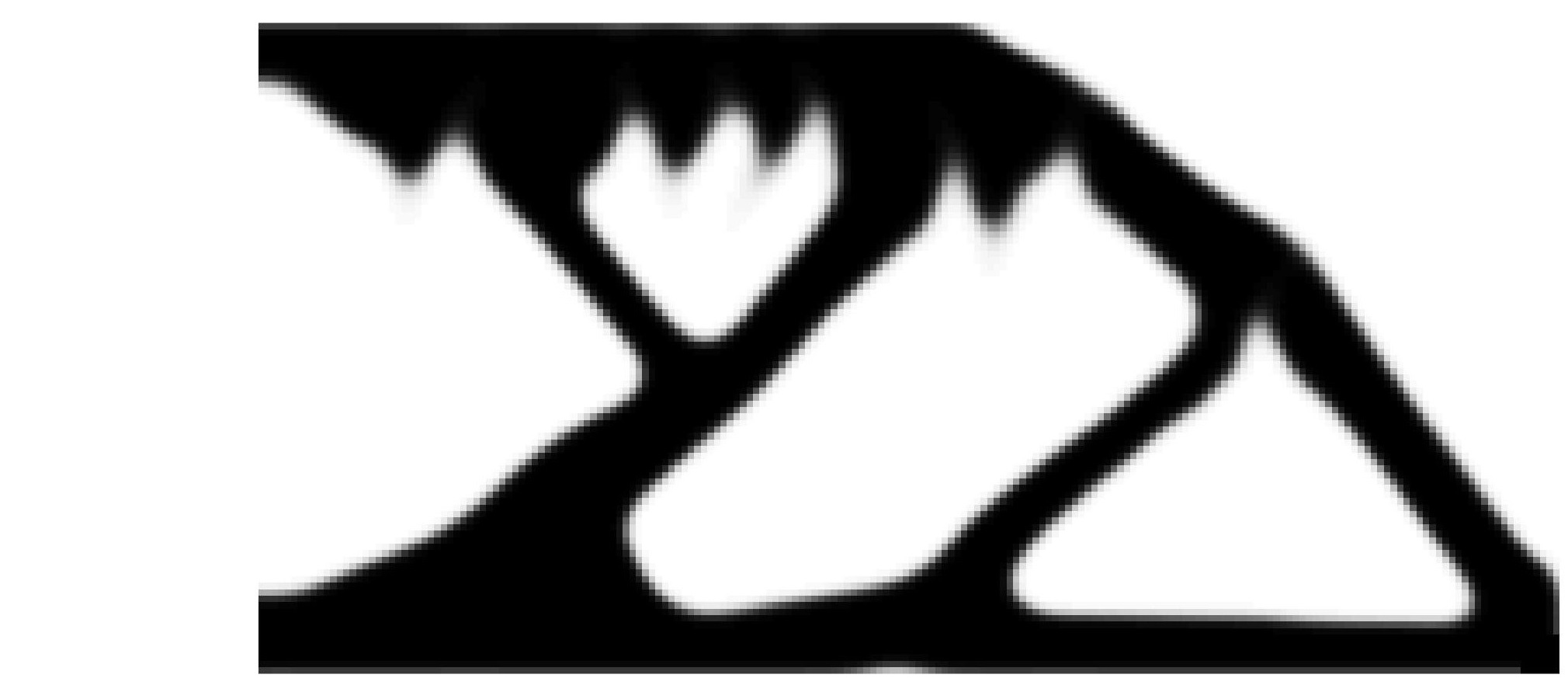}		
	&   \includegraphics[width=1\linewidth]{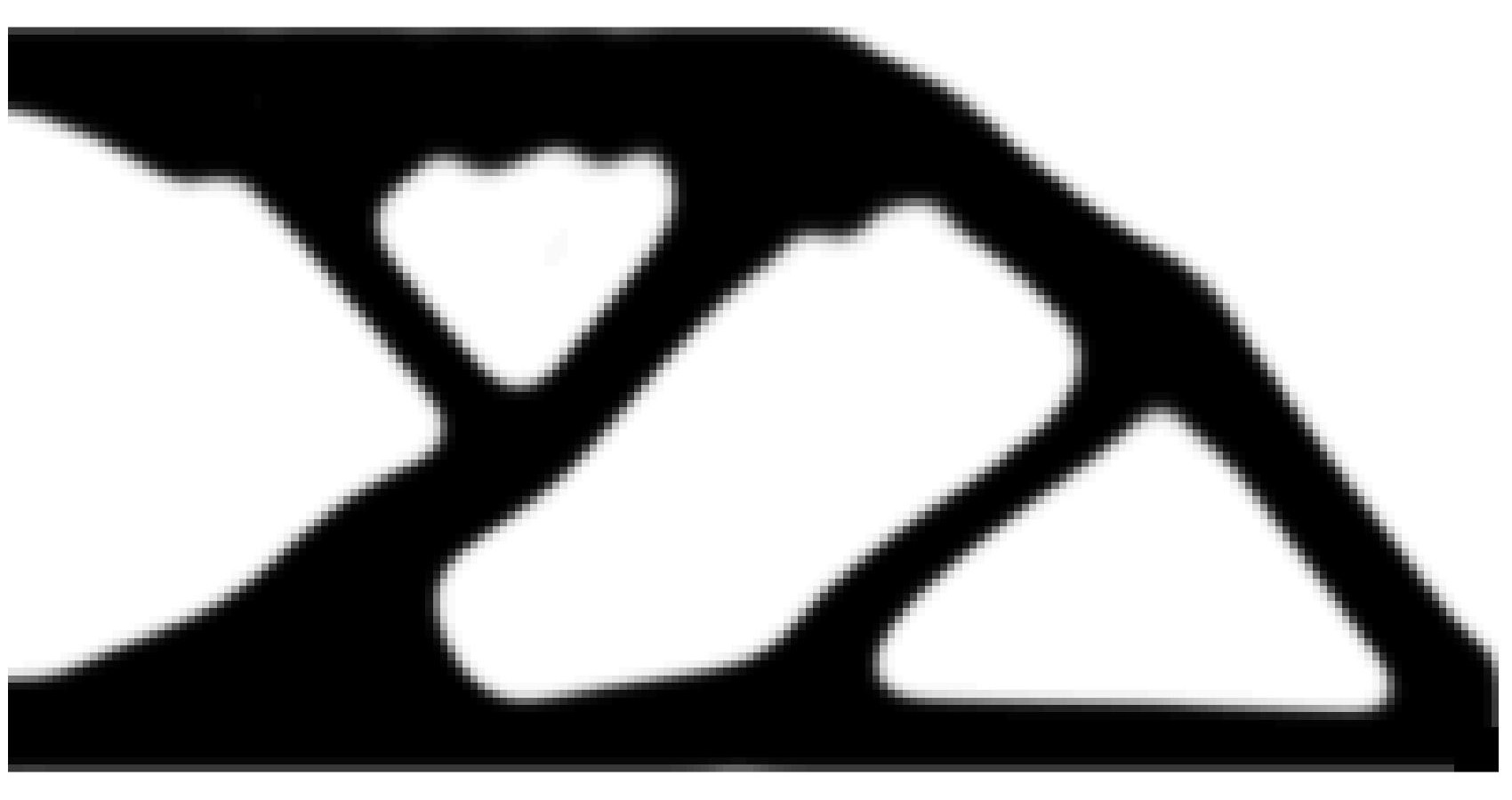}	
	&   \includegraphics[width=1\linewidth]{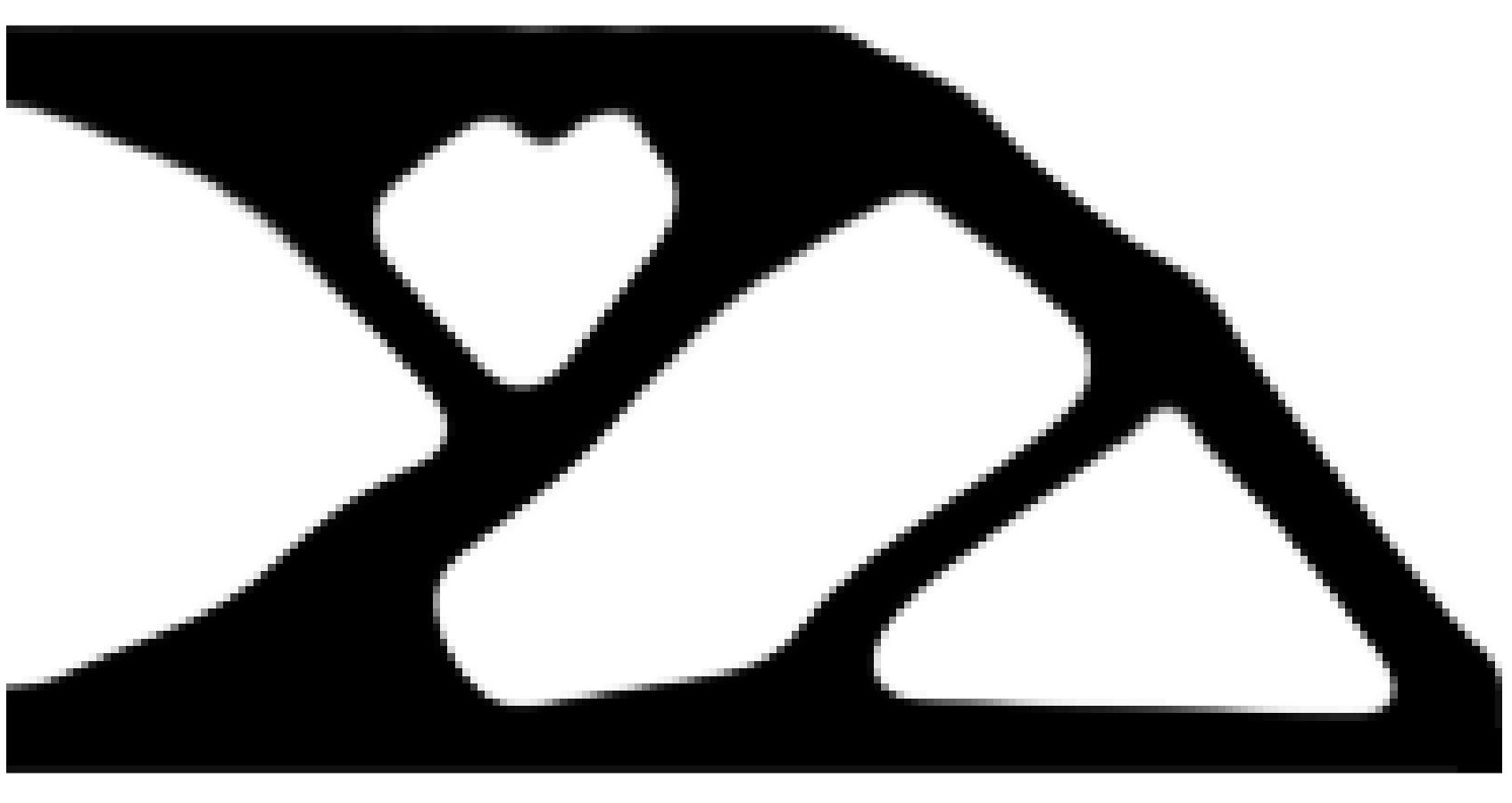}
	&   \includegraphics[width=1\linewidth]{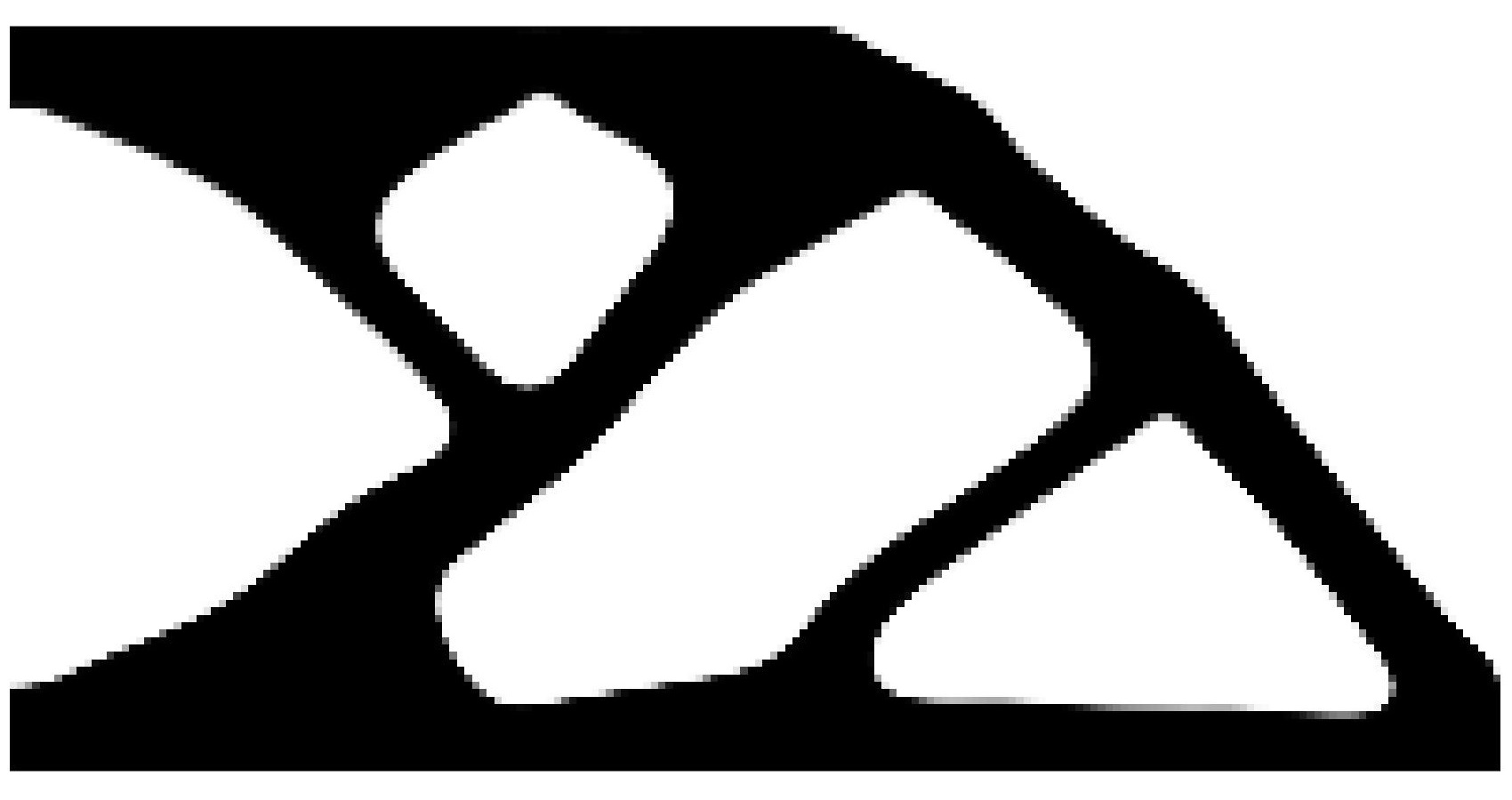}
	\\ 
	\vspace{-10mm}\large{$\;\;\bm{c}$}
	&	\includegraphics[width=1\linewidth]{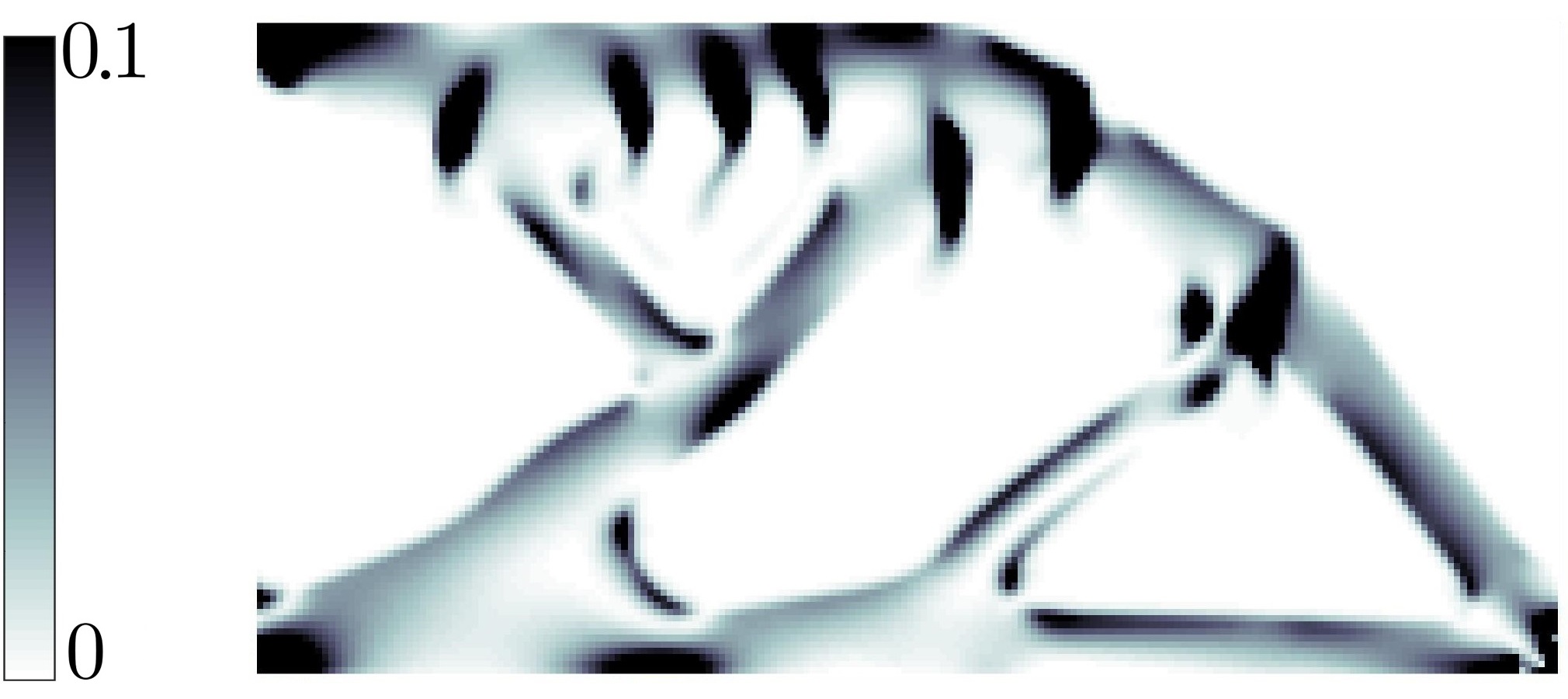}		
	&   \includegraphics[width=1\linewidth]{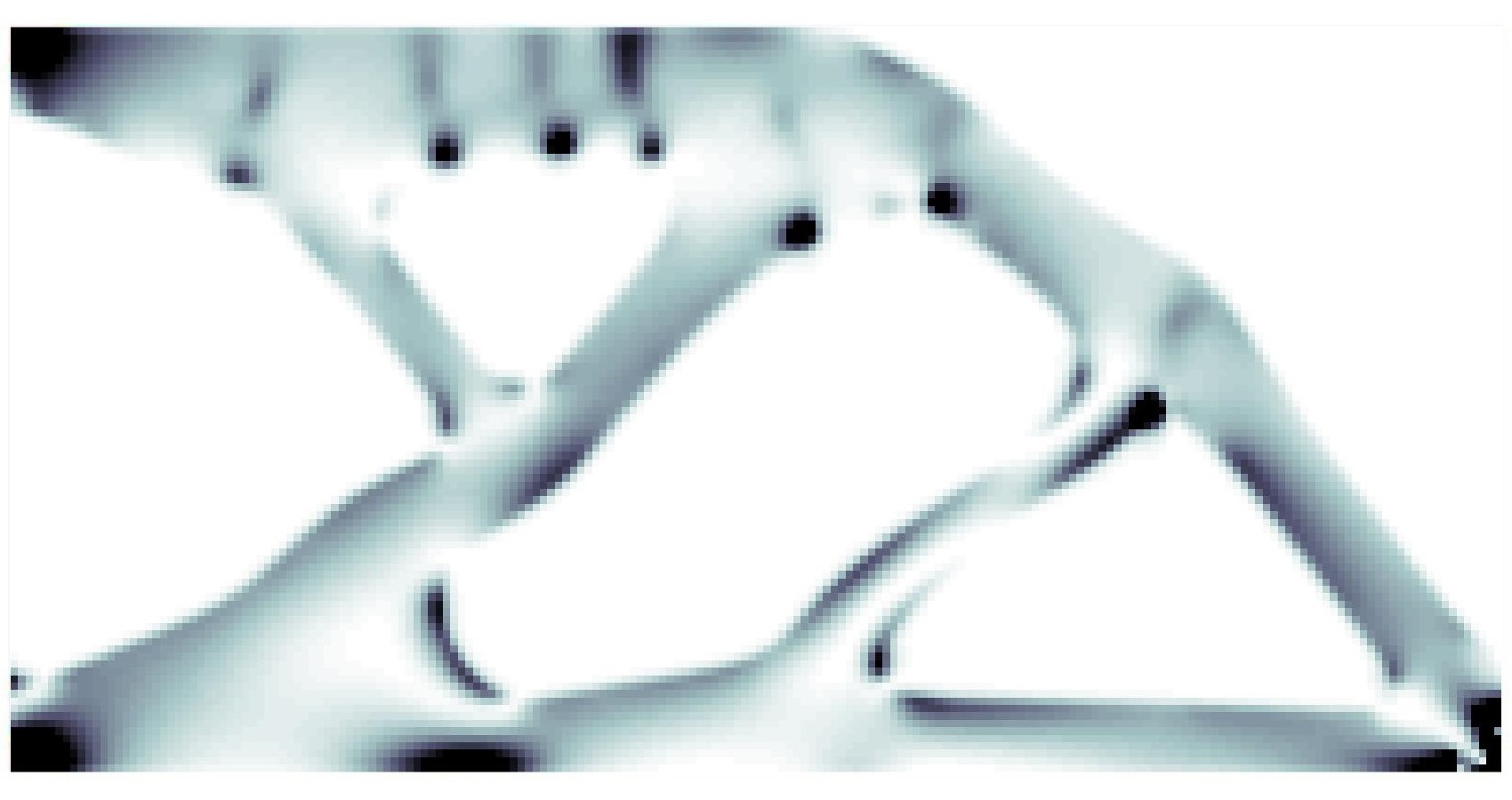}	
	&   \includegraphics[width=1\linewidth]{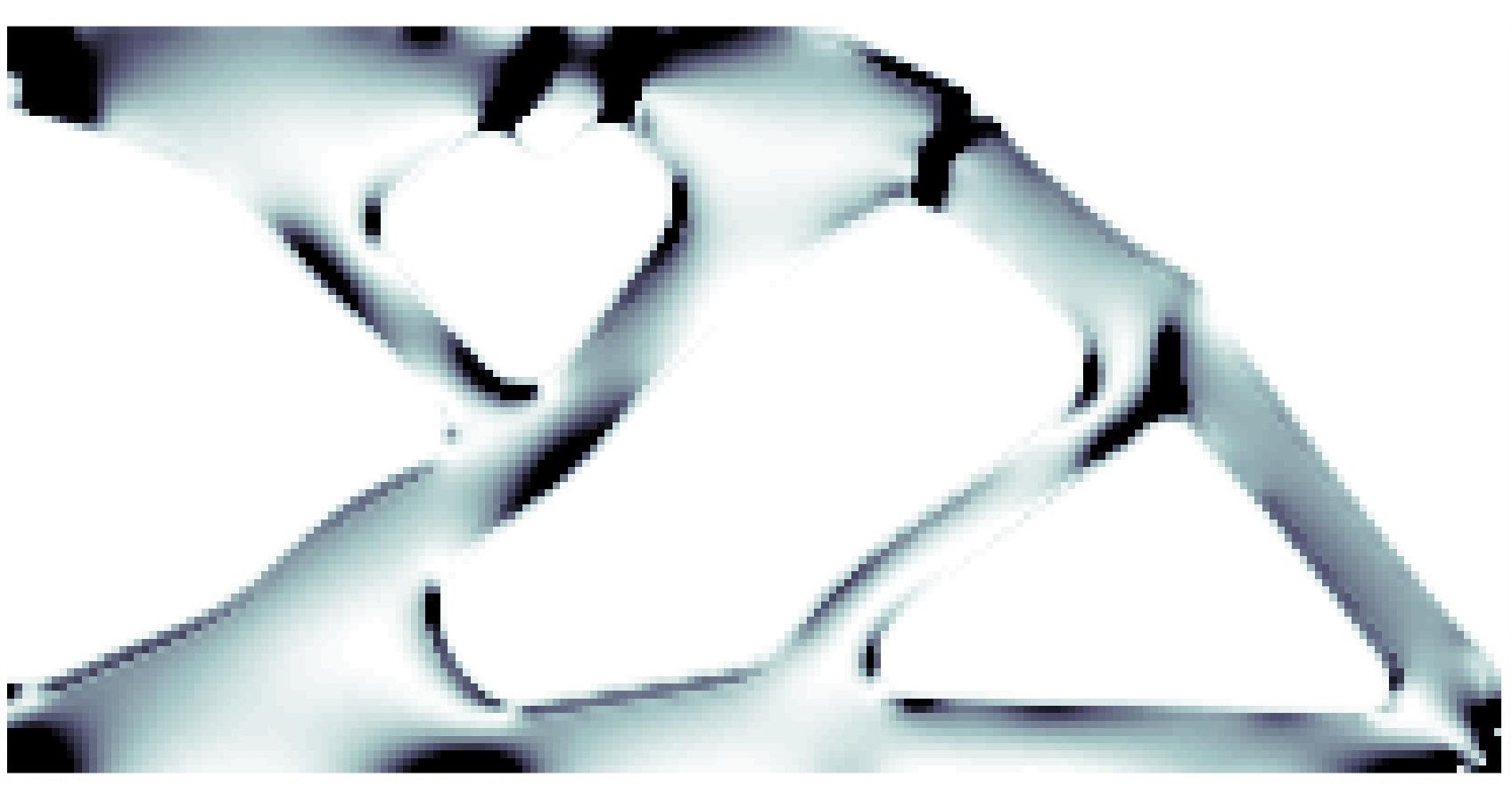}
	&   \includegraphics[width=1\linewidth]{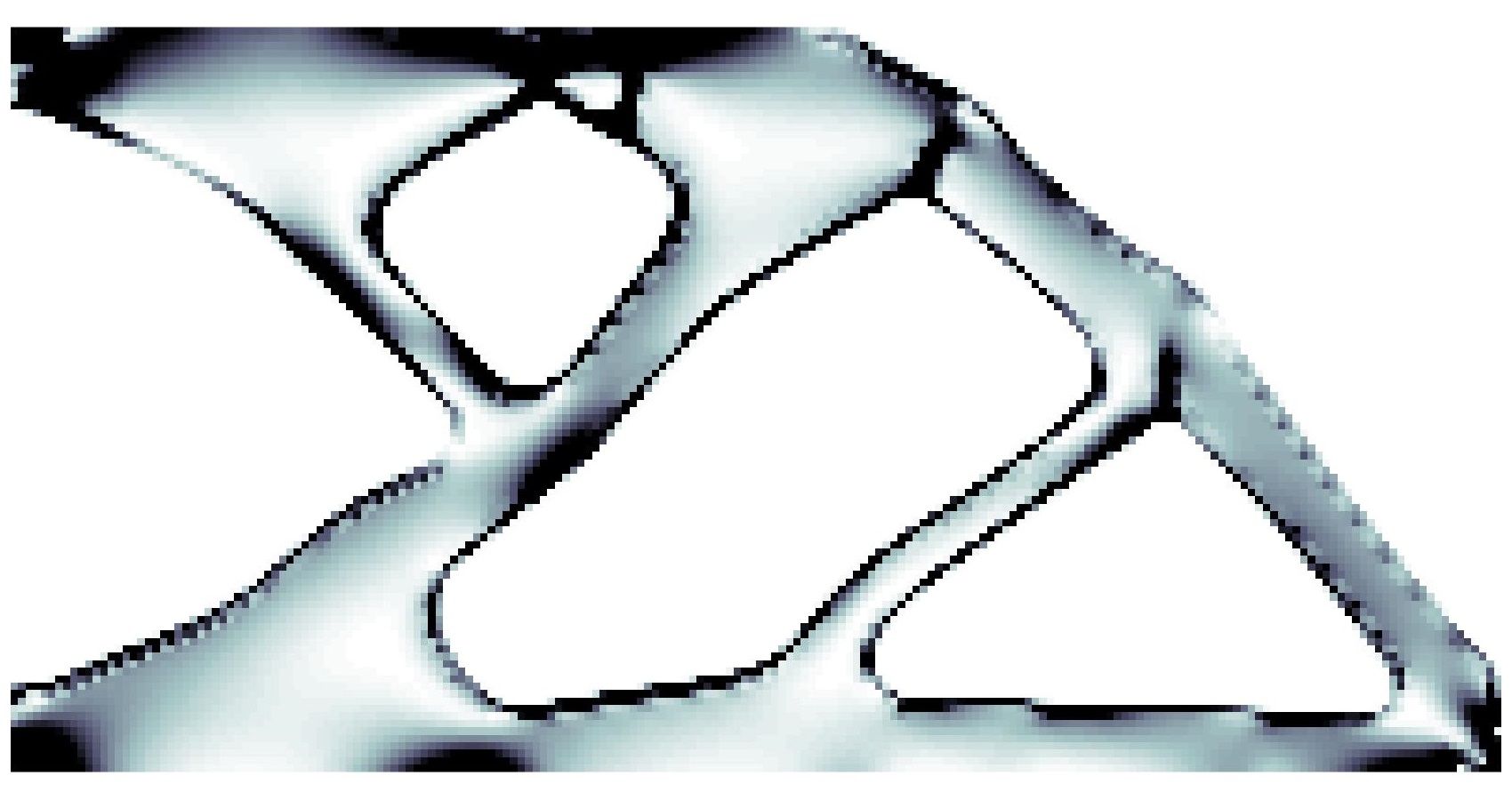}
	\\
	\vspace{-10mm}\large{$\;\;\bm{g}$}
	&	\includegraphics[width=1\linewidth]{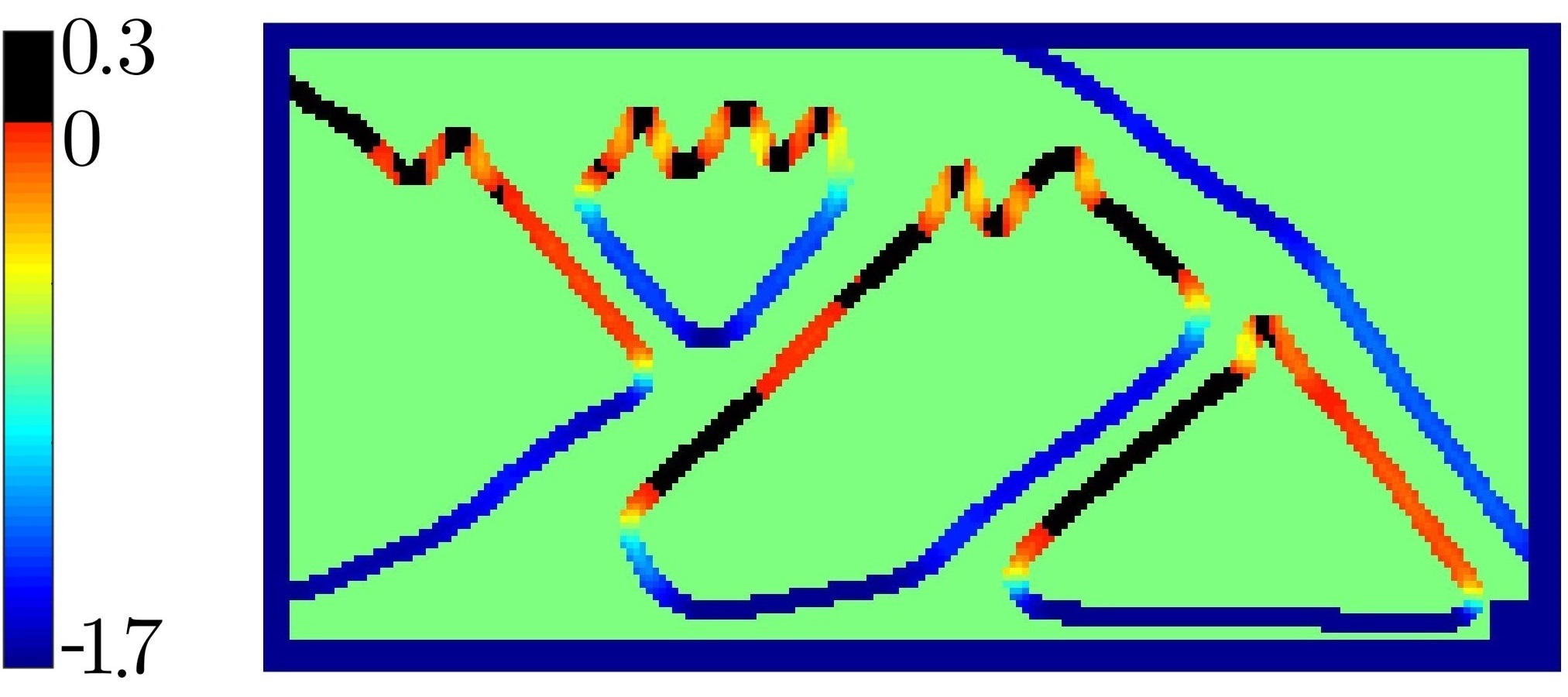}		
	&   \includegraphics[width=1\linewidth]{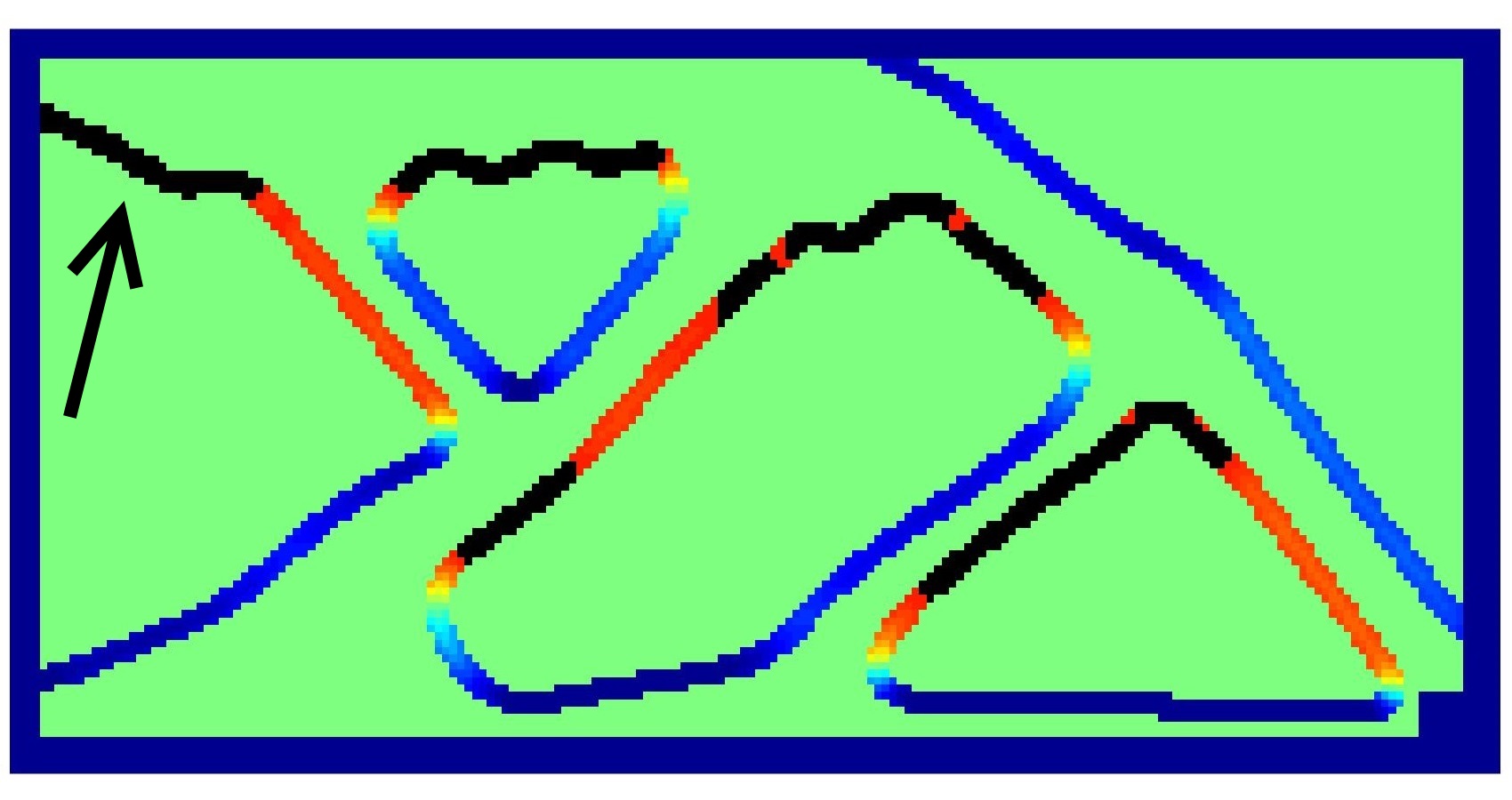}	
	&   \includegraphics[width=1\linewidth]{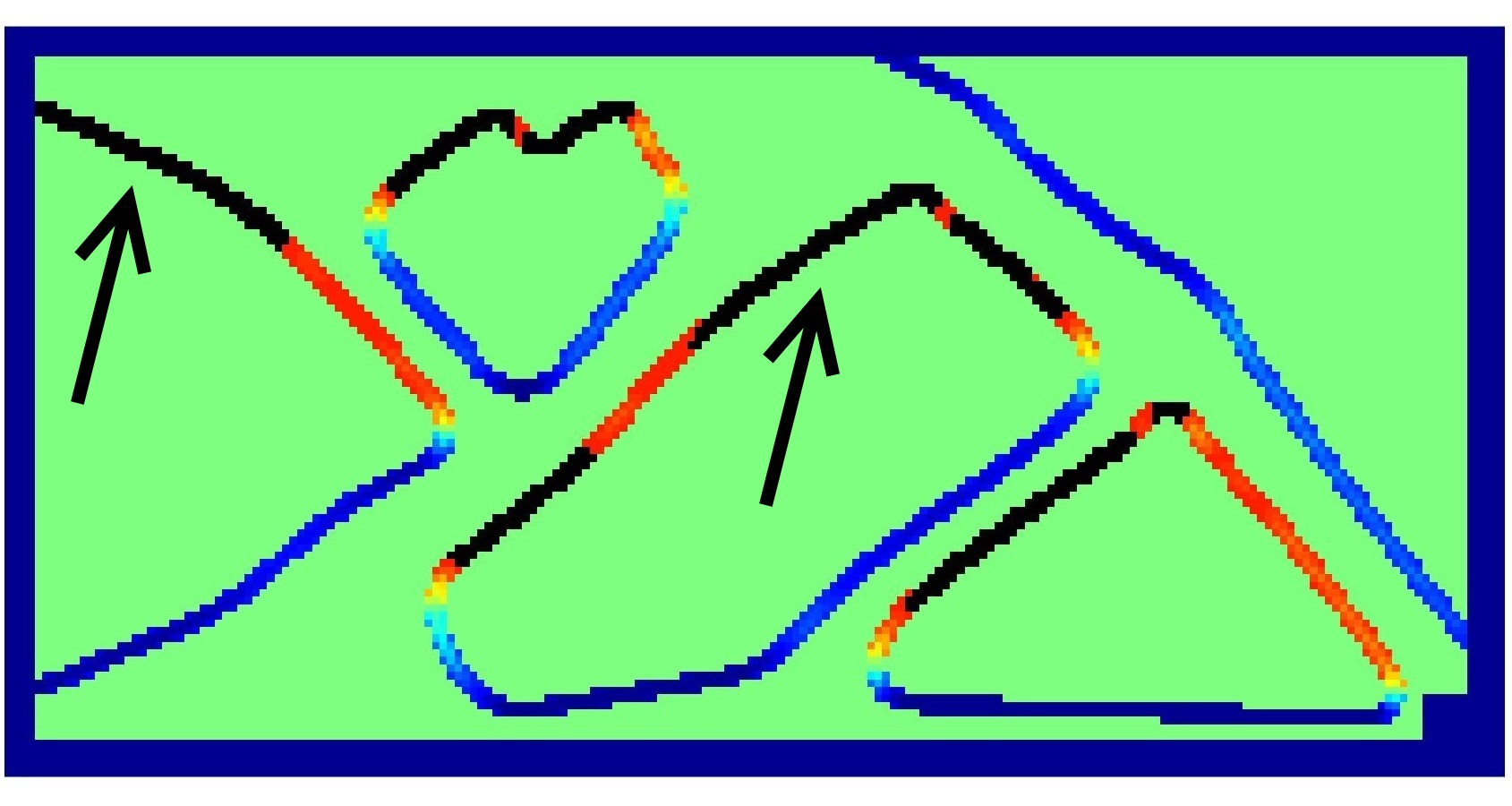} 	
	&   \includegraphics[width=1\linewidth]{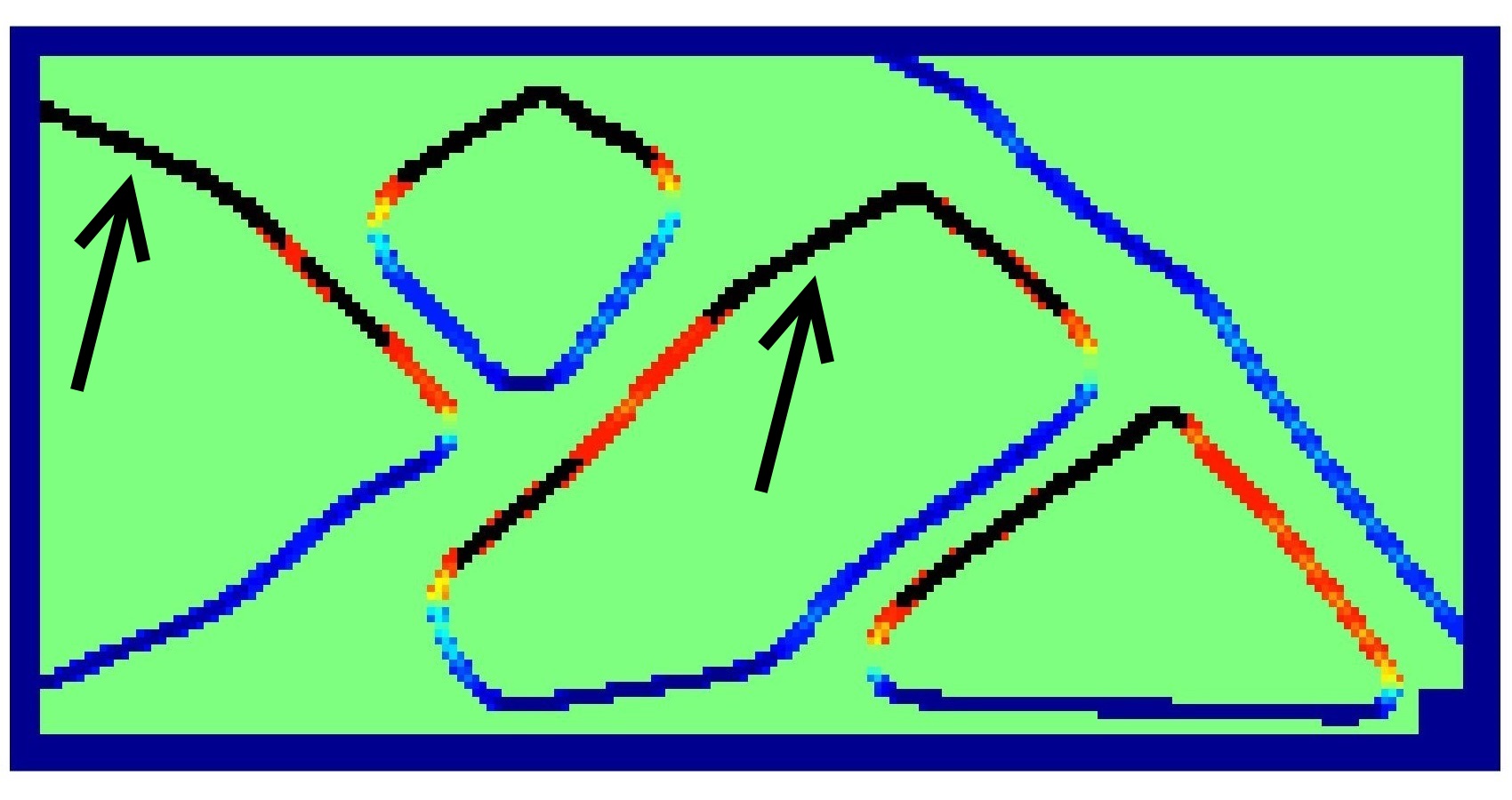}
	\\
	\bottomrule	
\end{tabular}
\label{TAB:Strategy}	
\end{table*}
	
	\subsubsection{Strategy to avoid non-self-supporting parts}\label{Sec:Strategy_to-avoid-nonSelfSupporting}
	
	The optimized solution of the cantilever beam  (Fig.~\ref{FIG:Cone_Example_a}) with the overhang-constrained formulation (Eq.~\ref{EQ:OPTI_Constrained}) after 189 iteration is shown at position~(1,1)  of Table~\ref{TAB:Strategy} at iteration 189. One can note that the optimized solution contains non-self-supporting triangular parts. At the position~(2,1) of Table~\ref{TAB:Strategy}, we plot corresponding elemental compliance field. It can be noted that these parts have no compliance. In other words, these hanging parts although use material, do not contribute in stiffening the optimized structure. These are also not self-supported. We also notice during optimization process that once these  triangular parts get created, they do not disappear as TO progresses, i.e., the optimizer gets stuck in a local optimum \citep{Allaire2017}. These parts arise when $\left\{\eta,\beta\right\}$ is less, and  allow the placement of gray elements at the tip of the triangular structures to satisfy the local constraint. When the penalization on intermediate densities increases, the gray elements disappear leaving the triangular part without support. The sensitivities are local, it is no longer possible to remove these parts.  
	
	We propose a strategy to remove such undesirable non-self-supporting triangular parts in  two steps:

	\paragraph{\textbf{Free evolution}}    
	
	As noted from the plots displayed at positions~(1,1) and (2,1) of Table~\ref{TAB:Strategy}, the hanging triangular parts use material but do not carry any force. Thus, an optimization problem that maximizes stiffness under a volume restriction i.e. the reference problem (Eq.~\ref{EQ:OPTI}), would eliminate such parts. This ideal is exploited herein such that topology of the design remain intact. The optimization problem without overhang constraints is run for a small number of iterations, $it_\mathrm{free}$, wherein design variables associated only to the structural surface are allowed to change. The surface is detected using the norm of the gradient, and the evolution of the design variables is controlled with the following move limit: 
	\begin{equation}
	\mathrm{m}_{\mathrm{L}i} = 
	\left\lbrace
	\begin{matrix}
	1.0 & \text{if} \; \| \bm{\delta}_{i} \| \leq \varepsilon_\mathrm{m}
	\\
	0.001 & \text{otherwise}
	\end{matrix}
	\right.
	\end{equation}
	\noindent where $\mathrm{m}_{\mathrm{L}i}$ denotes the move limit of variable $\rho_i$. The free evolution is included in the code \texttt{GBOHC} on lines 23-27. The overhang constrained-free problem  (Eq.~\ref{EQ:OPTI}) is applied for 10 iterations (line 12), before increasing parameters $\left\{\eta,\beta\right\}$. After 10 iterations the constrained problem (Eq.~\ref{EQ:OPTI_Constrained}) is continued. The material layout using this strategy is displayed at position~(1,2) of Table~\ref{TAB:Strategy} after 200 iteration. We note that size of the triangular appendages has got reduced.
	
	\paragraph{\textbf{Post-Processing (\PP) for triangular appendages removal}}\label{Sec:PP-triangules_removal}
	We note that switching on and off to the free evolution process helps reduce size of the undesirable triangular appendages (see positions~(1,2) and (1,3) of Table~\ref{TAB:Strategy}) but does not circumvent the issue completely. Therefore, we propose the second steps and call it Post-Processing (PP). In this step, we first identify such parts through a detection scheme and then, we remove them by a Boolean operation. The code that implements the presented detection scheme is also provided, it is called \texttt{PPTri} (see Appendix~\ref{APP:3}).
	
\begin{figure}
\captionsetup[subfigure]{labelformat=empty}
\centering
\begin{tabular}{p{0.195\textwidth}p{0.23\textwidth}p{0.23\textwidth}p{0.23\textwidth}}
		\begin{subfigure}{1.00\linewidth}   	
    		\includegraphics[width=1\linewidth]{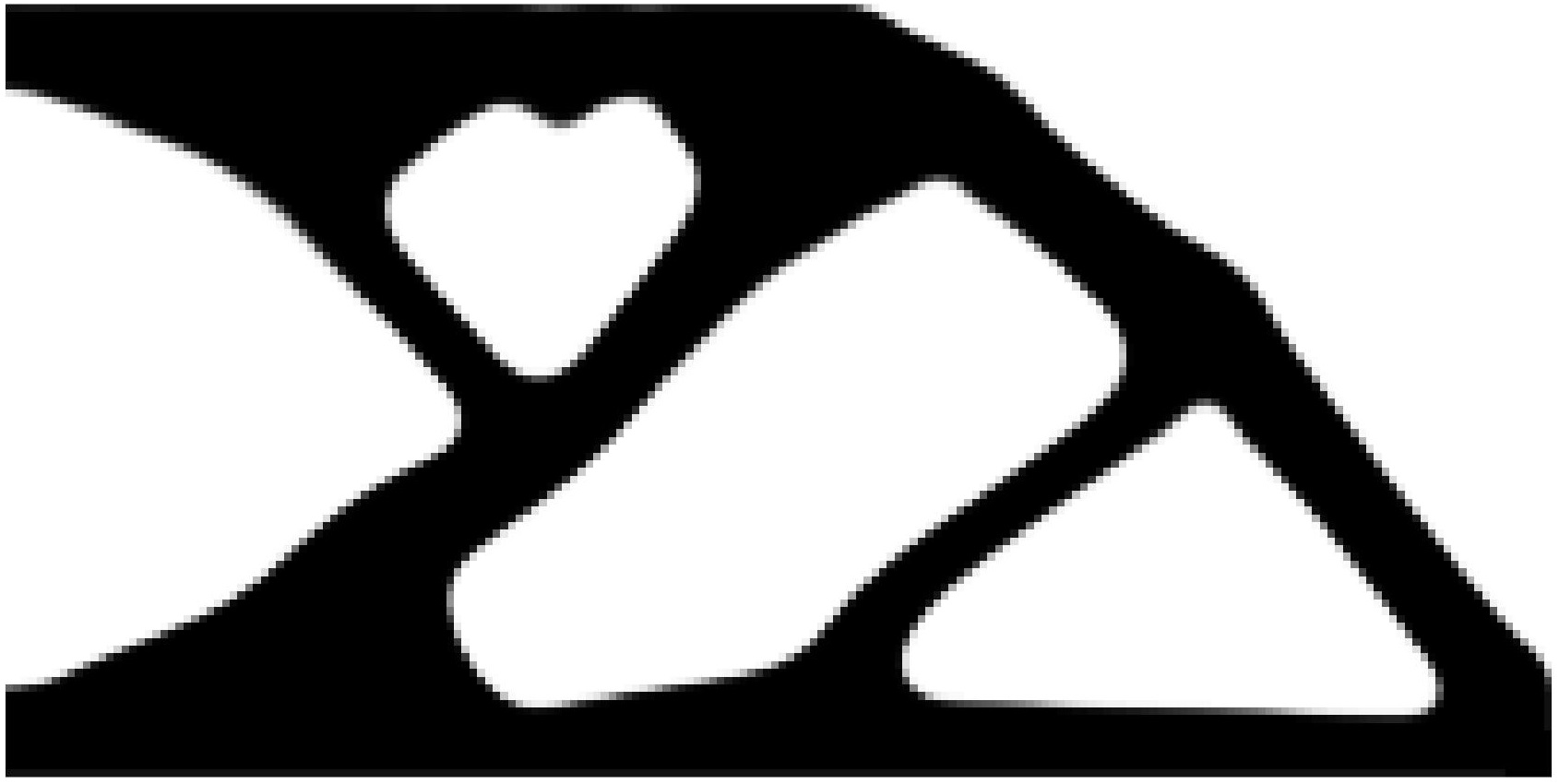} \caption{}
    		\label{FIG:PP_a}
    	\end{subfigure}			
	&   \begin{subfigure}{1.00\linewidth} 
			\includegraphics[width=1\linewidth]{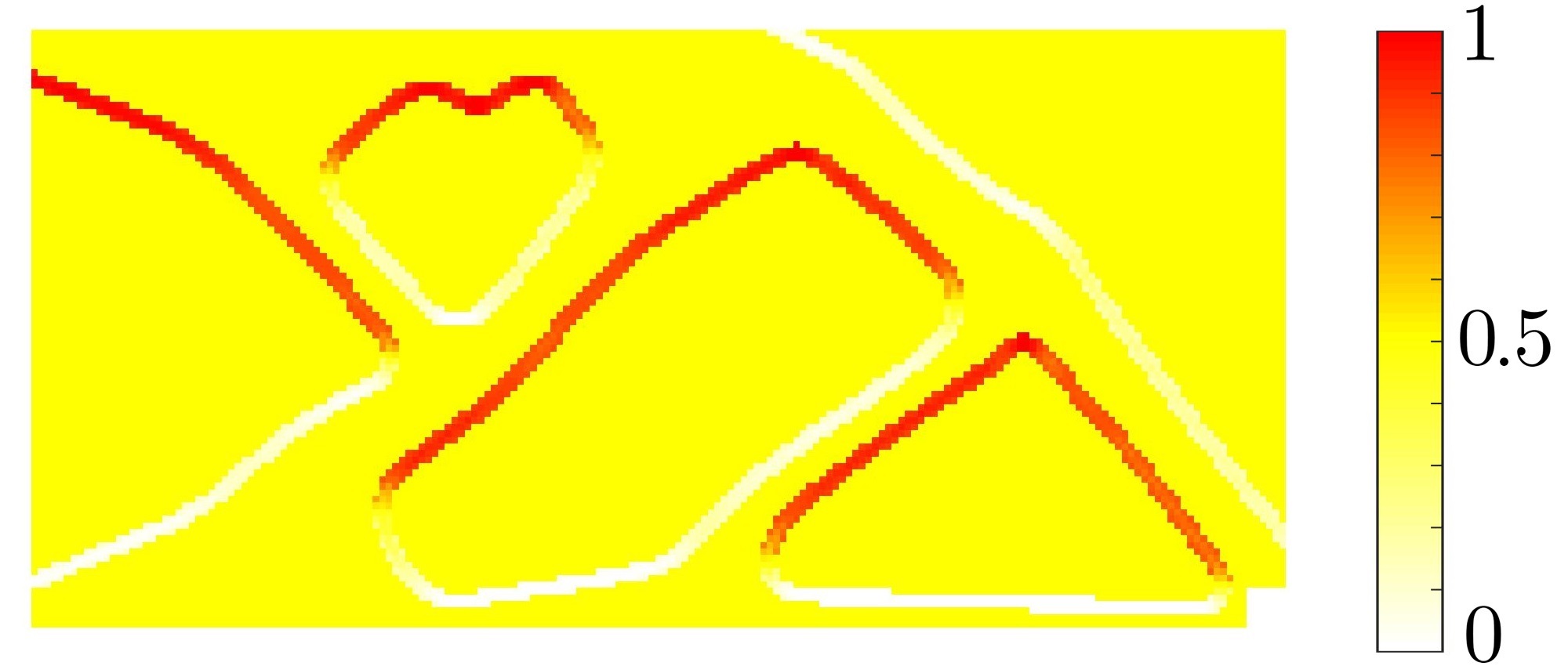}	\caption{}
			\label{FIG:PP_b}
		\end{subfigure}
	&   \begin{subfigure}{1.00\linewidth}
			\includegraphics[width=1\linewidth]{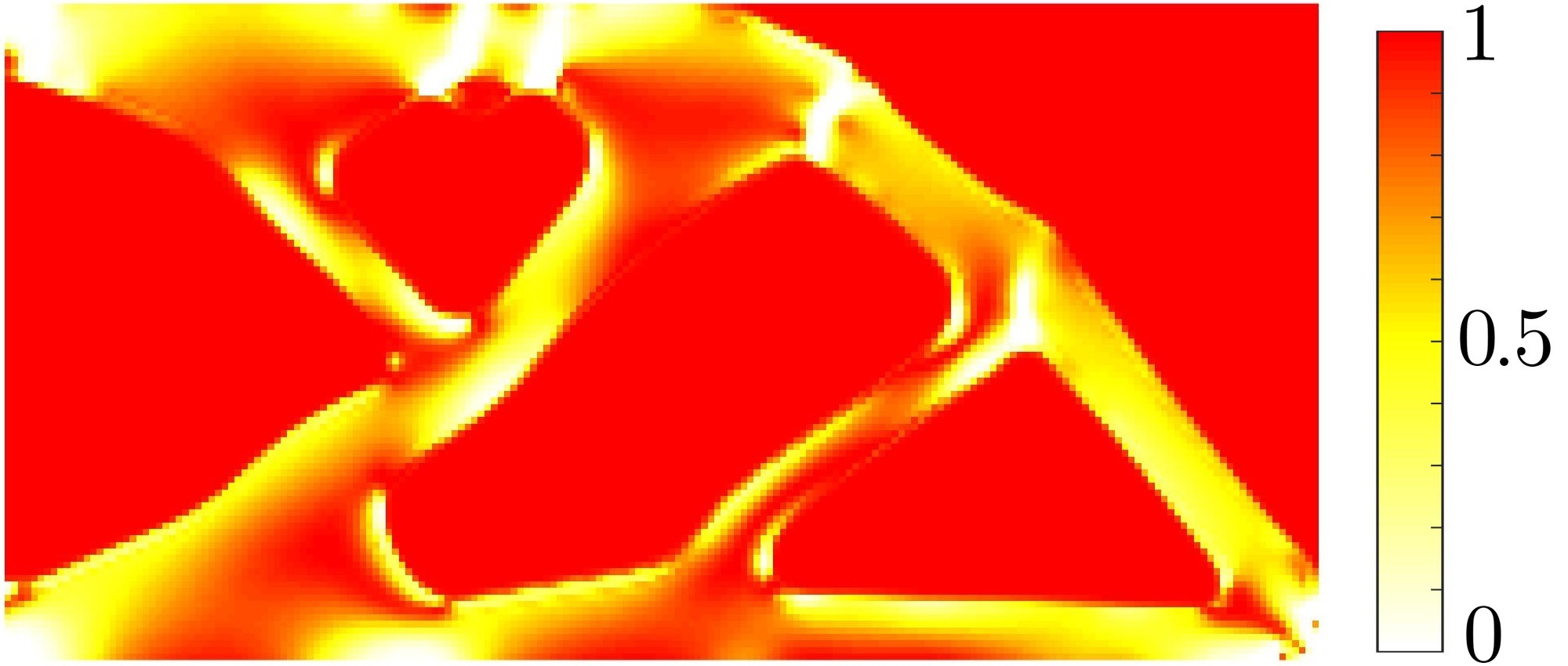} \caption{}  
			\label{FIG:PP_c}
		\end{subfigure}	
	&   \begin{subfigure}{1.00\linewidth}
			\includegraphics[width=1\linewidth]{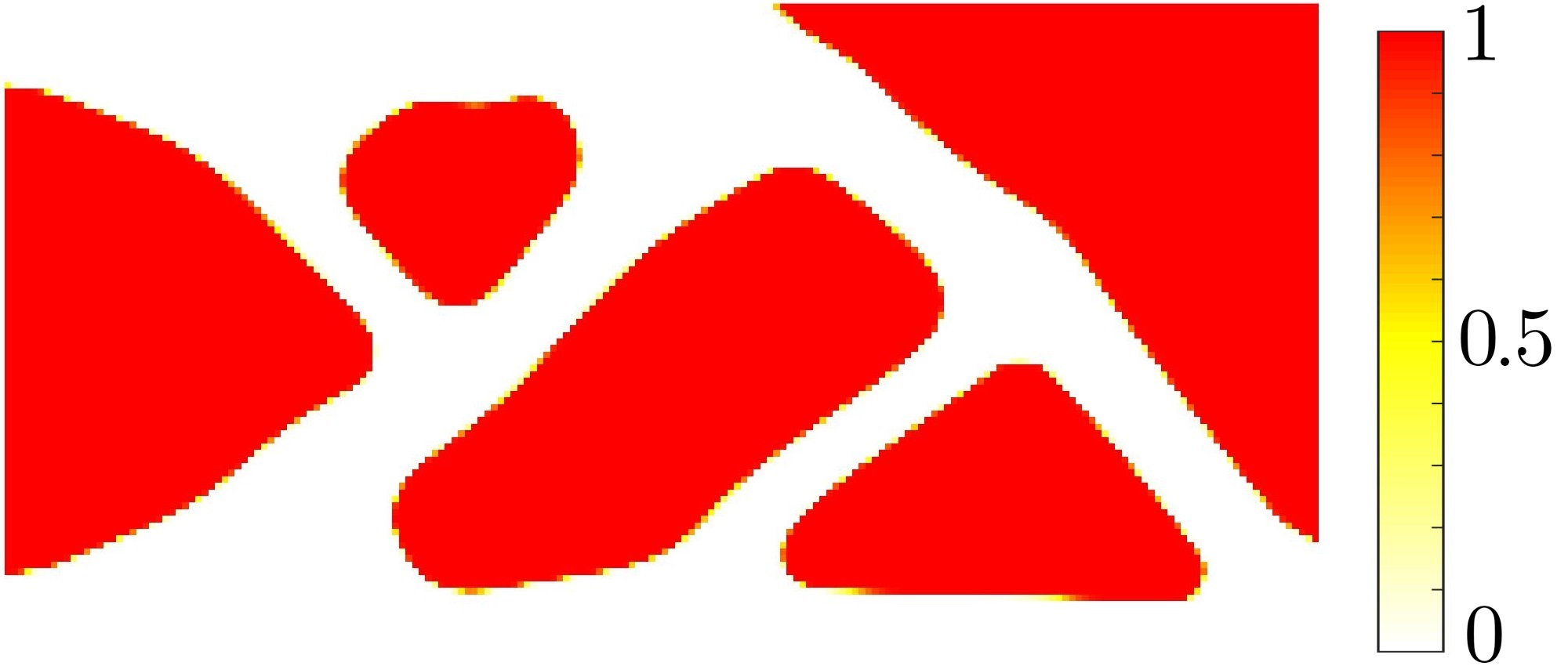} \caption{}  
			\label{FIG:PP_d}
		\end{subfigure}	
	\\
	\vspace{-6mm}\hspace{10mm}(a) $\bm{\bar{\rho}}_\mathrm{int}$ & \vspace{-6mm}\hspace{10mm} (b) $\bm{s}$ & \vspace{-6mm}\hspace{10mm} (c) $\bm{c_\mathrm{p}}$ & \vspace{-6mm}\hspace{10mm} (d) $\mathbf{I_v}$
	\\
	 & \vspace{-12mm}\hspace{-13mm}$\xrightarrow{\makebox[1.5cm]{ }}$  &  & \hspace{12mm}$\Bigg\downarrow$ $\mathrm{I}_{\mathrm{c}i}=s_i \cdot c_{\mathrm{p}i} \: \mathrm{I}_{\mathrm{v}i}$ 
	\vspace{2mm}
	\\  \begin{subfigure}{1.00\linewidth}   
			\includegraphics[width=1\linewidth]{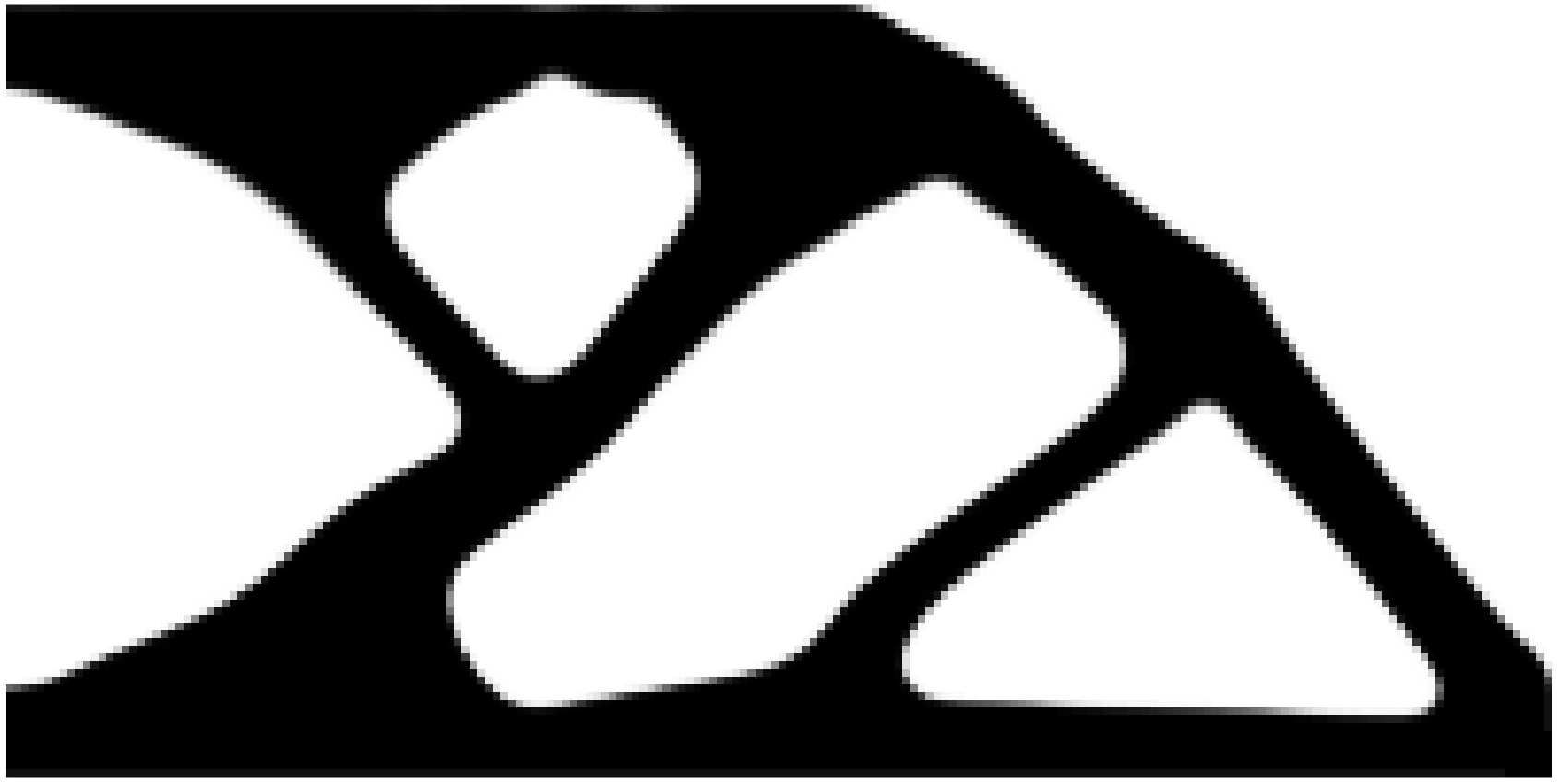} \caption{} 
			\label{FIG:PP_e}	
		\end{subfigure}			
	&   \begin{subfigure}{1.00\linewidth}
			\includegraphics[width=1\linewidth]{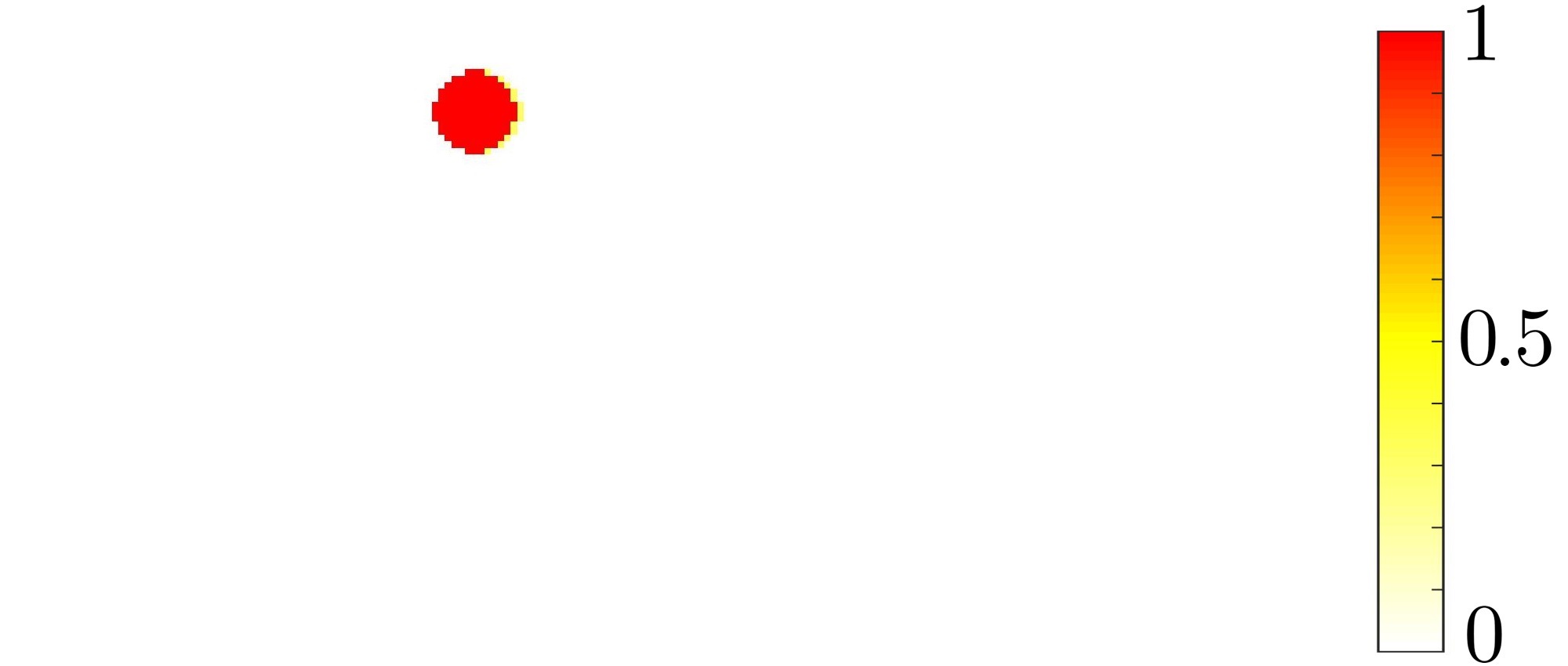}	\caption{} 
			\label{FIG:PP_f}
		\end{subfigure}	
	&   \begin{subfigure}{1.00\linewidth}
			\includegraphics[width=1\linewidth]{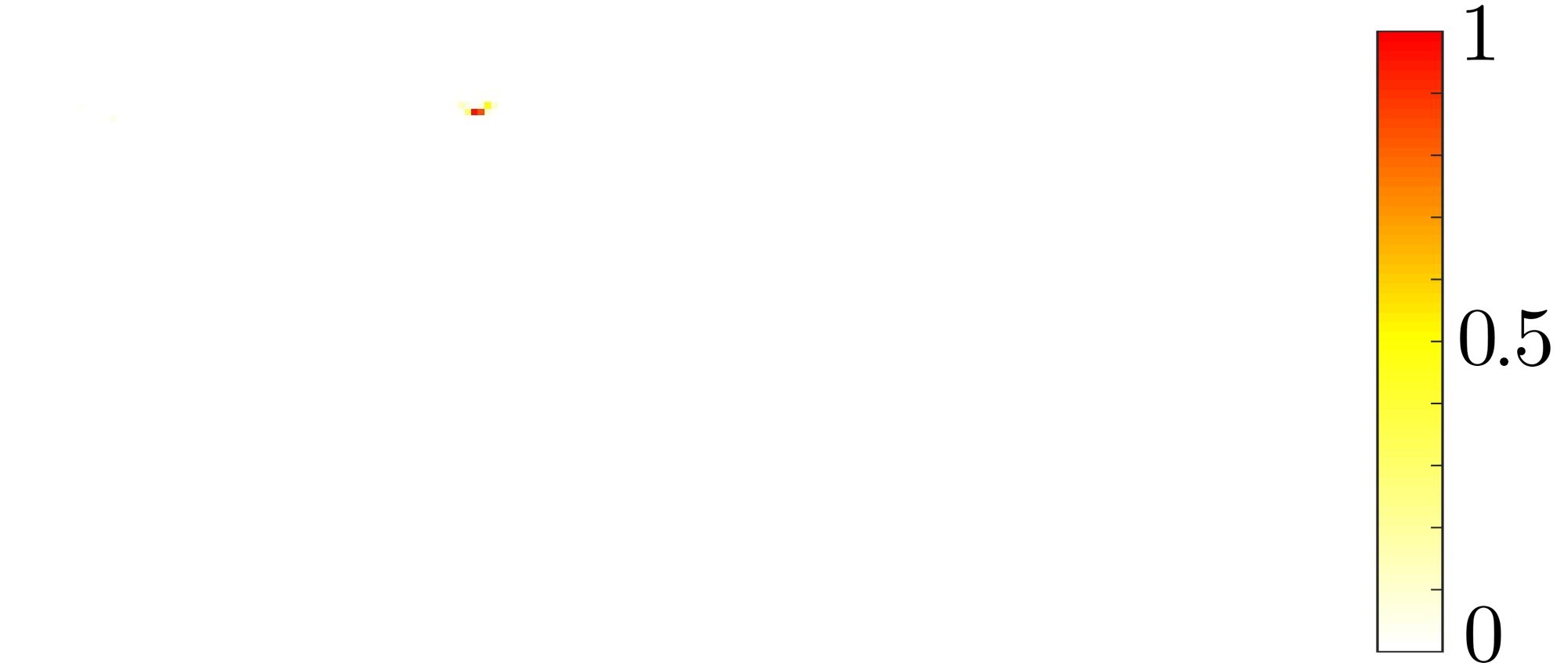} \caption{} 
	 		\label{FIG:PP_g}
	 	\end{subfigure}	
	&   \begin{subfigure}{1.00\linewidth}
			\includegraphics[width=1\linewidth]{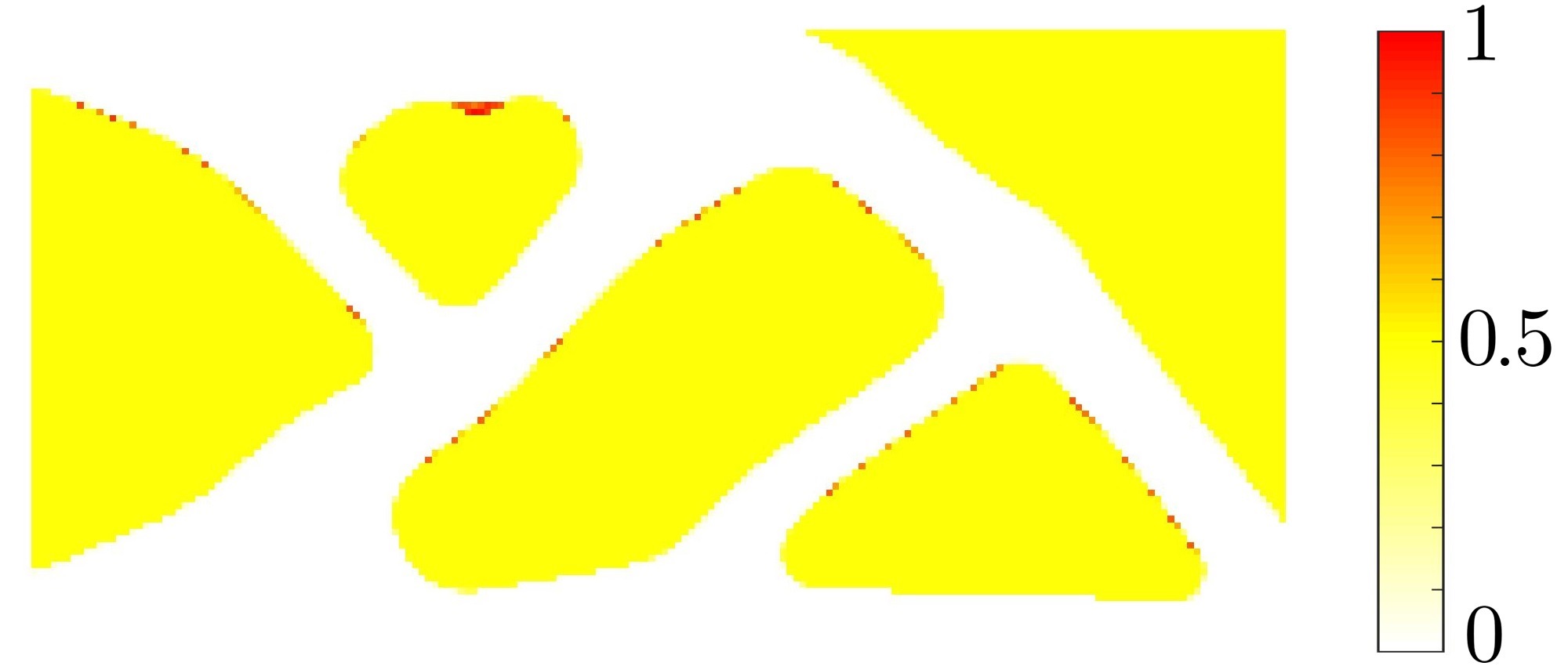} \caption{}
			\label{FIG:PP_h}
		\end{subfigure}	
	\\
	\vspace{-6mm}\hspace{10mm} (e) $\bm{{\bar{\hat{\rho}}}}_\mathrm{int}$ & \vspace{-6mm}\hspace{10mm} (f) $\mathbf{I_r}$ & \vspace{-6mm}\hspace{8mm} (g) $h \left(\mathbf{I_t},\beta,\mu \right)$ & \vspace{-6mm}\hspace{10mm} (h) $\mathbf{I_t}$ 
	\\
	& \vspace{-10mm}\hspace{-13mm} $\xleftarrow{\makebox[1.5cm]{ }}{}$  & \vspace{-10mm}\hspace{-13mm} $\xleftarrow{\makebox[1.5cm]{ }}{}$  & \vspace{-10mm}\hspace{-13mm} $\xleftarrow{\makebox[1.5cm]{ }}{}$ 
\end{tabular}
\label{FIG:PP}
\vspace{-6mm}	
\caption{Procedure for removing the triangular parts that are not self-supporting. (b) to (d) the factors involved in the detection of non-self-supporting parts. (f) to (h) the indicators used for parts removal. (e) the intermediate field after Post-Processing (PP). Building direction is upwards.}	
\end{figure}

\begin{figure}
	\centering
	\includegraphics[width=0.70\linewidth]{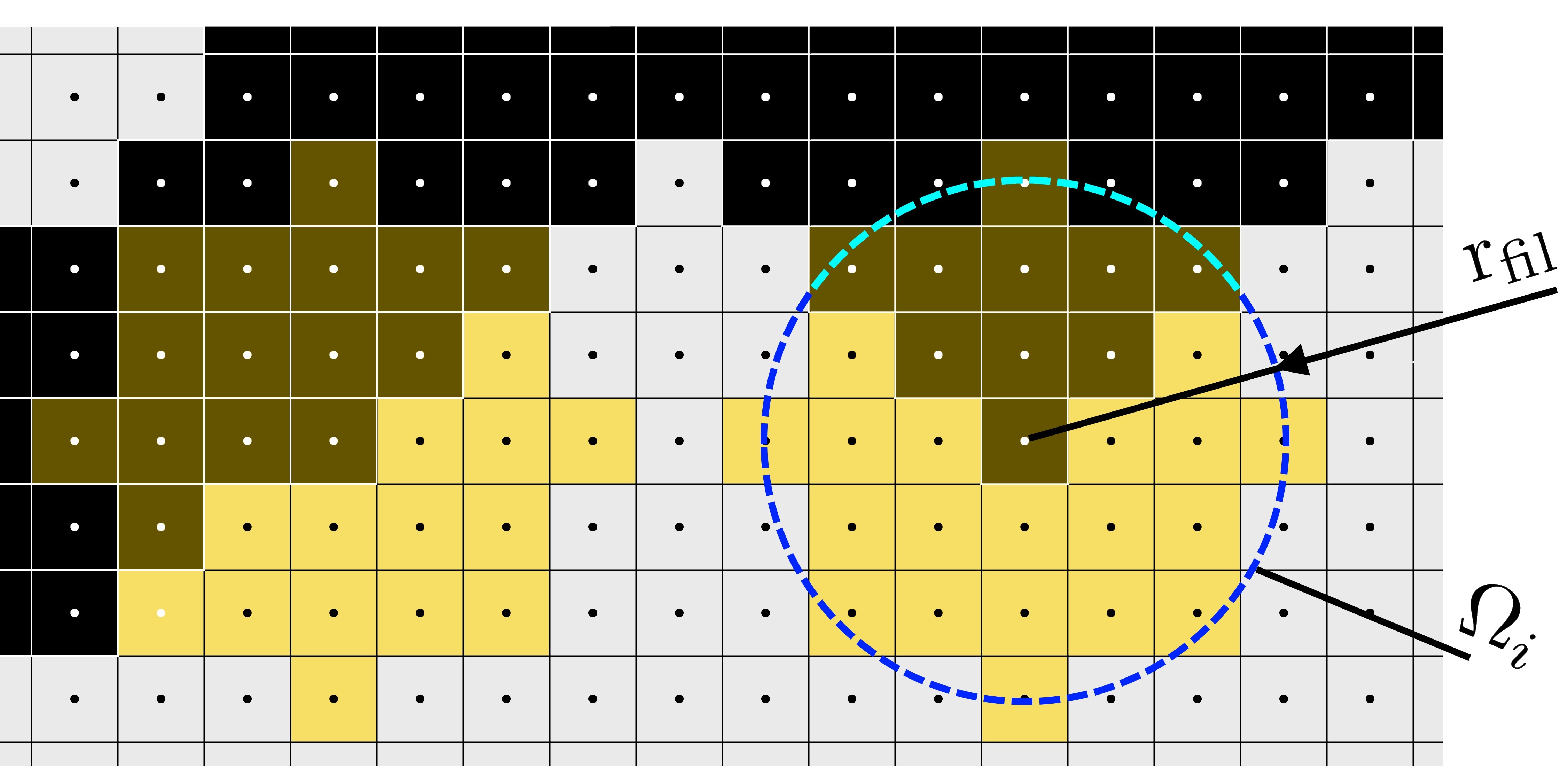} 
	\caption{Region $\Omega$ where the fraction of voids is measured to get the factor $\mathbf{I_v}$. On the left, the factor $\mathrm{I}_{\mathrm{v}i}=0$ and on the right $\mathrm{I}_{\mathrm{v}i}=1$. } 
	\label{FIG:TestRegion_Voids}	
\end{figure}

	The detection scheme uses 3 parameters $s_i$, $c_{\mathrm{p}i}$ and $\mathrm{I}_{\mathrm{v}i}$ for an element $i$, which are respectively associated to the overhang constraint (shifted constraint, cf. Eq.~\ref{EQ:Local_Constraint_Shifted}), the compliance and a measure of surrounding material. Note, $s_i$, $c_{\mathrm{p}i}$ and $\mathrm{I}_{\mathrm{v}i} \in$ [0,\,1]. These parameters with value 1 indicate that the element is a part of the undesirable triangular regions (Fig.~\ref{FIG:PP_h}) as described below. Note that if the gradient of element $i$ is orthogonal to the printing direction $\mathbf{b}$, the $s_i$ takes value 1, which is expected at the tip of the hanging triangular part (see Fig.~\ref{FIG:PP_b}). $c_{\mathrm{p}i}$ highlights zones of low compliance that is defined as:
	\begin{equation}
	{c}_{pi} = {e}^{-c_{i}\varepsilon_\mathrm{c}},
	\end{equation} 
	\noindent where $\varepsilon_\mathrm{c}$ controls the decay of the exponential function. Thus, $c_{\mathrm{p}i}$ takes value 1 if element $i$ does not have deformation energy (see Fig.~\ref{FIG:PP_c}). $\mathrm{I}_{\mathrm{v}i}$ takes value 1 if there is at least $\varepsilon_v$ fraction of voids within a circular region $\Omega$ (see Figs.~\ref{FIG:PP_d} and \ref{FIG:TestRegion_Voids}), which is defined as:
	\begin{equation}\label{EQ:Iv}
	\mathrm{I}_{\mathrm{v}i} = 
	\left\lbrace	
	\begin{matrix}
	1 \; ,& \text{if} \;\;
	\displaystyle\sum_{j \in \Omega} \mathrm{v}_j(1-\bar{\rho}_j) \: \leq \varepsilon_\mathrm{v} \sum_{j \in \Omega} \mathrm{v}_j \;,
	\vspace{2mm}\\
	0 \; ,& \text{otherwise.}
	\end{matrix}
	\right.
	\end{equation}
	
	\noindent One computes the amount of voids, $\mathbf{A_v}$, inside each $\Omega_i$ as follows. The regions $\Omega_i$ are stored in a matrix $\mathbf{D_v}$ such that $\mathbf{A_v}=\mathbf{D_v}(\bm{1}-\bm{\bar{\rho}})$.  $\mathrm{I}_{\mathrm{v}i}=h(\mathrm{A}_{\mathrm{v}i},\beta,\varepsilon_\mathrm{v})$ and thus, $\mathbf{I}_{\mathbf{v}}$ is evaluated. The threshold parameter $\varepsilon_\mathrm{v} \in (0.5  (\pi + 2\alpha) / (2\pi)]$ is taken so that $\mathrm{I}_{\mathrm{v}i}$ takes value 1 only in non-self-supporting triangle appendages (Fig.~ \ref{FIG:TestRegion_Voids}).

	The three parameters are multiplied to get the triangles indicator $\mathbf{I_t}$ (Fig.~\ref{FIG:PP_h}). Once the non-self-supporting part is detected, it is removed using filtering techniques. First the $\mathbf{I_t}$ indicator is passed through a Heaviside function to remove the intermediate values, as shown in Fig. \ref{FIG:PP_g}. The obtained value is then filtered to get the removal indicator $\mathbf{I_r}$, as shown in Fig.~\ref{FIG:PP_f}. This last filtering process determines  $\mathrm{I}_{\mathrm{r}i}$ which is equal to the maximum value $h(\mathrm{I}_{\mathrm{t}j})$ within a circle of radius $\mathrm{r_{fil}}$. Finally, the PP strategy eliminates the triangular parts from the design variables $\bm{\rho}$ using the Boolean operation as:
	\begin{equation}
	\hat{\rho}_i = \mathrm{min}( \:\mathrm{max}({\rho}_i - \mathrm{I}_{\mathrm{r}i} \; , \; 0) \; , \; 1 ) \; ,
	\end{equation}  
	\noindent where $\hat{\rho}_i$ is the post processed design variable used to continue the optimization problem (Eq.~\ref{EQ:OPTI_Constrained}). Fig. \ref{FIG:PP_e} shows the intermediate design obtained from $\bm{\hat{\rho}}$ i.e. $\bm{{\bar{\hat{\rho}}}}_{int}(\bm{{\tilde{\hat{\rho}}}}(\bm{\hat{\rho}}))$. This state is performed when the topology is roughly defined (see column~3 of Table~\ref{TAB:Strategy}). This is because gray elements must be penalized so that the non-self-supporting structure does not transmit load and can be detected by $\bm{c_p}$.

	\begin{figure}[h!]
\centering
\begin{tabular}{c c} 
        \begin{subfigure}{0.45\linewidth}  	
        	\includegraphics[width=1\linewidth]{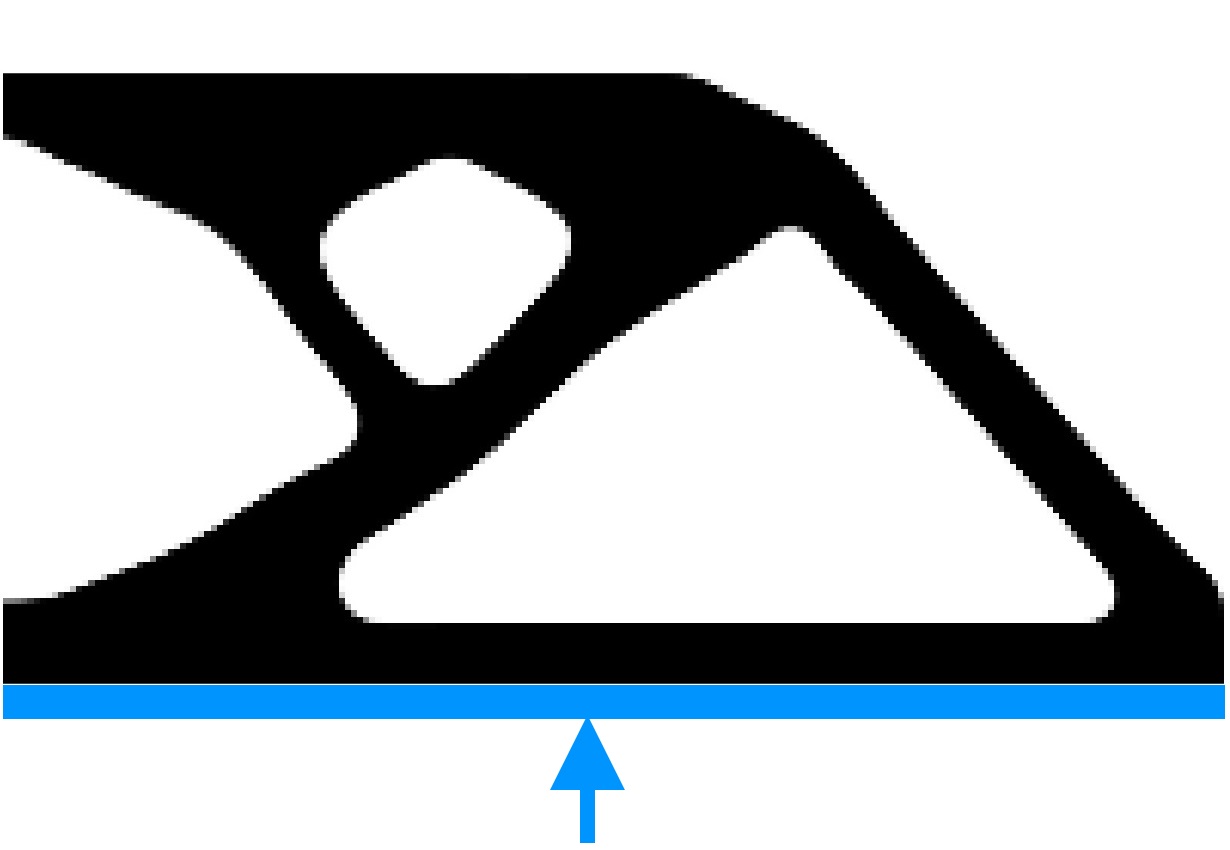}	
	        \caption{$r_\mathrm{min}=3$ , $\;C_\mathrm{int}=92.61$}
	        \label{FIG:Diff_MinSize_a}
	    \end{subfigure}
	&   \begin{subfigure}{0.45\linewidth}
			\includegraphics[width=1\linewidth]{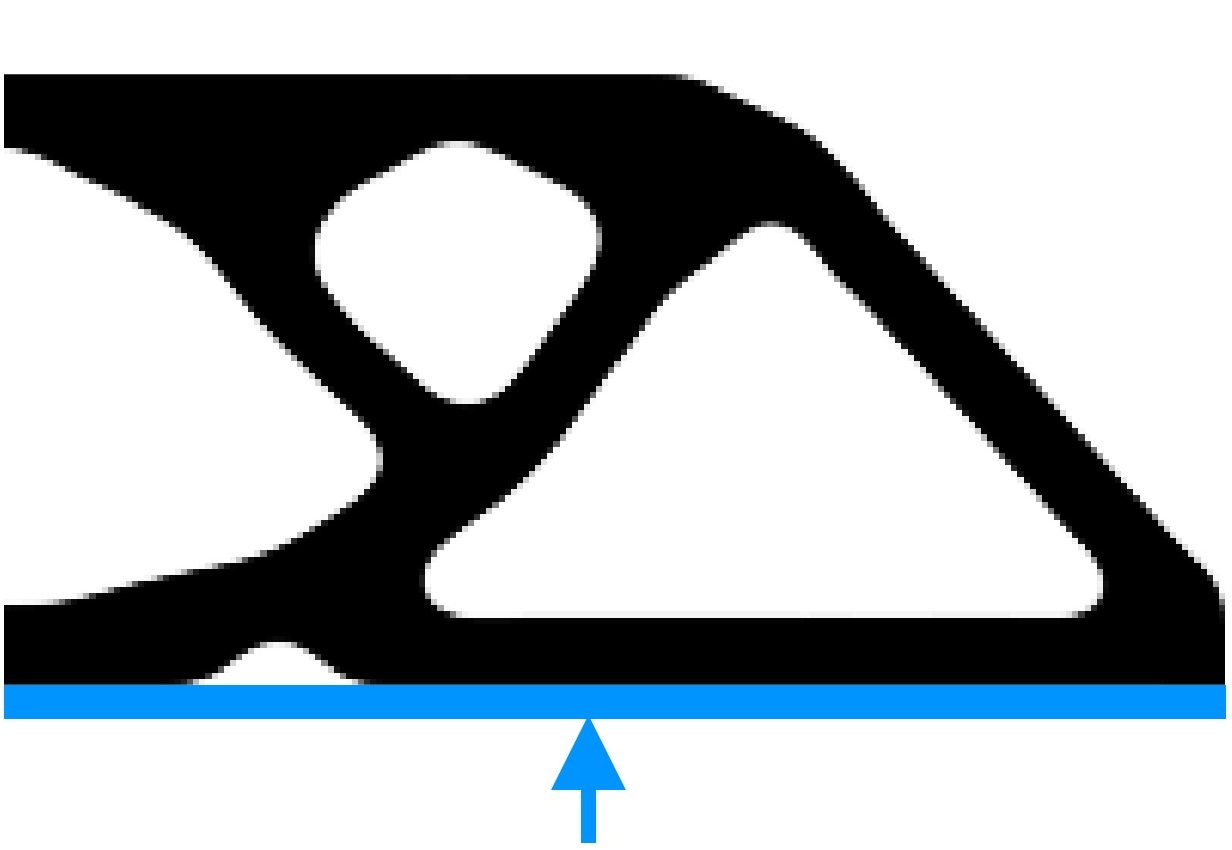}	
			\caption{$r_\mathrm{min}=5$ , $\;C_\mathrm{int}=95.30$}
	        \label{FIG:Diff_MinSize_b}
	    \end{subfigure}
	\vspace{2mm}	
	\\  \begin{subfigure}{0.45\linewidth}	
			\includegraphics[width=1\linewidth]{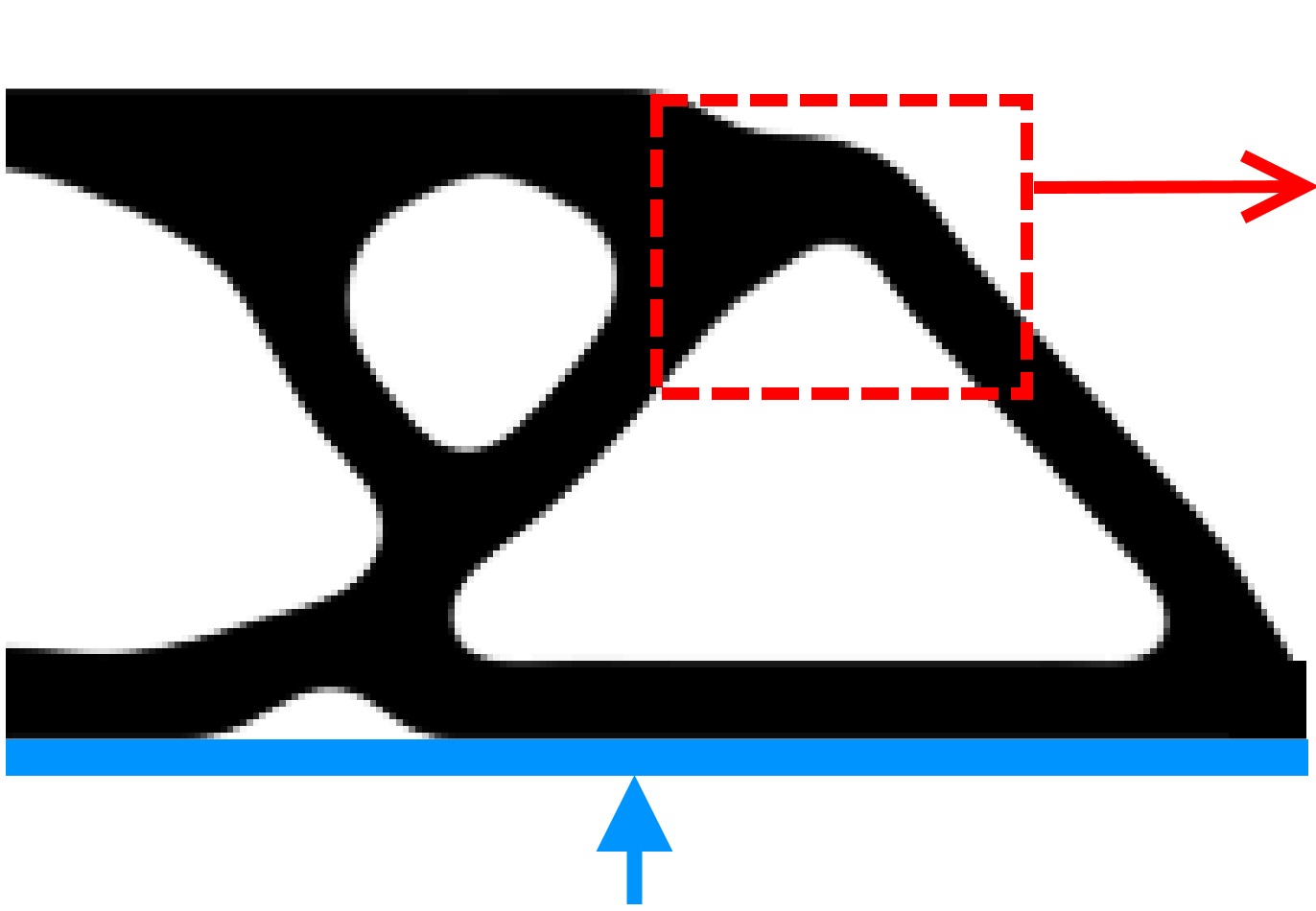}
			\caption{$r_\mathrm{min}=6$ , $\;C_\mathrm{int}=103.90$}
	        \label{FIG:Diff_MinSize_c}
	    \end{subfigure}
	&   \begin{subfigure}{0.35\linewidth}
			\includegraphics[width=1\linewidth]{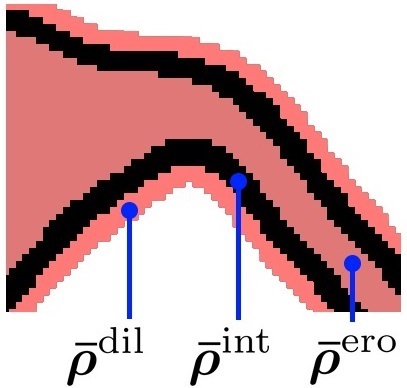}
			\caption{}
	        \label{FIG:Diff_MinSize_d}
	    \end{subfigure}
\end{tabular}	
\caption{ Solution to cantilever beam with overhang angle constraint. (a) $r_\text{min}=3$, (b) $r_\text{min}=5$  and (c) $r_\text{min}=6$, and (d) The eroded, intermediate and dilated designs.} 	
\label{FIG:Diff_MinSize}
\end{figure}  

	\subsubsection{Discussion on different parameters used}
	Like other TO approaches \citep{Sigmund2013}, the presented strategy with overhang constraint, the free evolution, and the Post-Processing include user-defined parameters that influence the optimized designs. This section discusses those parameters. 
	
	A low $p$ can produce a smoother convergence, however, a large amount of local restrictions may be violated due to underestimation of the \textit{p-mean} function \citep{fernandez2020imposing}. Thus, the aggregation parameter $p$ must be large enough to capture the critical constraints $g_i\le 0$ (Eq.~\ref{EQ:Global_Constraint}). In this work, we use $p=60$ in all the presented examples.
	
	Normalization of the gradient vector  (Eq.~\ref{EQ:Normalization_Factor}) is controlled by $\varepsilon_\mathrm{n}$. We notice that this parameter strongly affects the performance of the overhang constraint $\mathrm{G}$ (Eq.~\ref{EQ:Global_Constraint}). Since densities are not exactly identical in a homogeneous zone, such as large gray areas or regions close to being perfectly solid, the gradient vector director $\bm{\varrho}_i$ may not be null even if $\varepsilon_\mathrm{n}$ is too small. That is, in an area where densities are similar, but not equal, the $g_i$ restriction could be violated. In such cases, constraint $\mathrm{G}$ divides the material from homogeneous zones. Due to that a large number of branches is created, and large oscillations in the value of design variables are noticed,  those in turn hinder convergence of the problem. Although gradient normalization plays a significant role in success of the constraint, details for this procedure are found in \cite{Garaigordobil2018,Garaigordobil2019,Qian2017} and related reference therein, which hinders replication of results. In this paper, the factor $\varepsilon_\mathrm{n}$ is defined with a continuation scheme. It is initialized to $\varepsilon_\mathrm{n}^\mathrm{ini}$ and increased up to $\varepsilon_\mathrm{n}^\mathrm{end}$. The code \texttt{GBOHC} (Appendix~\ref{APP:2}) defines $\varepsilon_\mathrm{n}^\mathrm{ini}$ to $10\%$ of the maximum value that $\|\bm{\delta}_i\|$ can achieve i.e. 0.3. As the optimization progresses, $\varepsilon_\mathrm{n}$ is increased to $30\%$. The continuation is provided on line 29 in the code \texttt{GBOHC}. The remaining parameters, e.g., $\varepsilon_\mathrm{m}$, $\varepsilon_\mathrm{c}$, $\varepsilon_\mathrm{v}$, among others, may also influence the proposed strategy. In this paper  we however have fixed their value for all the numerical test presented and omit comments pertaining to their influence on the optimized design to avoid overextending the paper.

	\subsubsection{Minimum size and MaxOA}
	Figures. \ref{FIG:Diff_MinSize_a}-\ref{FIG:Diff_MinSize_c} display results of the cantilever beam with the MinS, the MinV and the MaxOA limitations. We use $40\%$ volume fraction for these optimized results with different $r_\text{min}$, i.e., 3, 5 and 6. The overhang constraint is applied on the dilated design. 
	
	One can note form Figs.~\ref{FIG:Diff_MinSize_c} and \ref{FIG:Diff_MinSize_d} that, overhang constraining for all three designs can become geometrically incompatible. For instance, if minV is too large, then MaxOA restriction may not get satisfied at the re-entrant corners of the intermediate and eroded designs (Figs.~\ref{FIG:Diff_MinSize_c} and \ref{FIG:Diff_MinSize_d}). On the other hand, the dilated design is not in conflict as it does not feature a minimum cavity size \citep{fernandez2020imposing,Wang2011}. Therefore, when the minimum size in the void phase is too large, the MaxOA restriction is only applied in the dilated design.   
	
	The local nature of the gradient-based constraints is influenced by a few elements around element $i$. In this view, applying constraint on the eroded, intermediate and dilated designs has the benefit that the sensitivity analysis gets enriched. This does not guarantee a robust formulation with respect to MaxOA, but improves the formulation when only dilated design is restricted. Therefore, we restrict all three designs when $r_\mathrm{min}\leq 3s$, otherwise only dilated design is restricted.
	
	\begin{figure}
	\centering
\captionsetup[subfigure]{labelformat=empty}
\begin{tabular}{p{0.25\textwidth}p{0.25\textwidth}p{0.25\textwidth}}   	
    	\hspace{4mm}\includegraphics[width=1\linewidth]{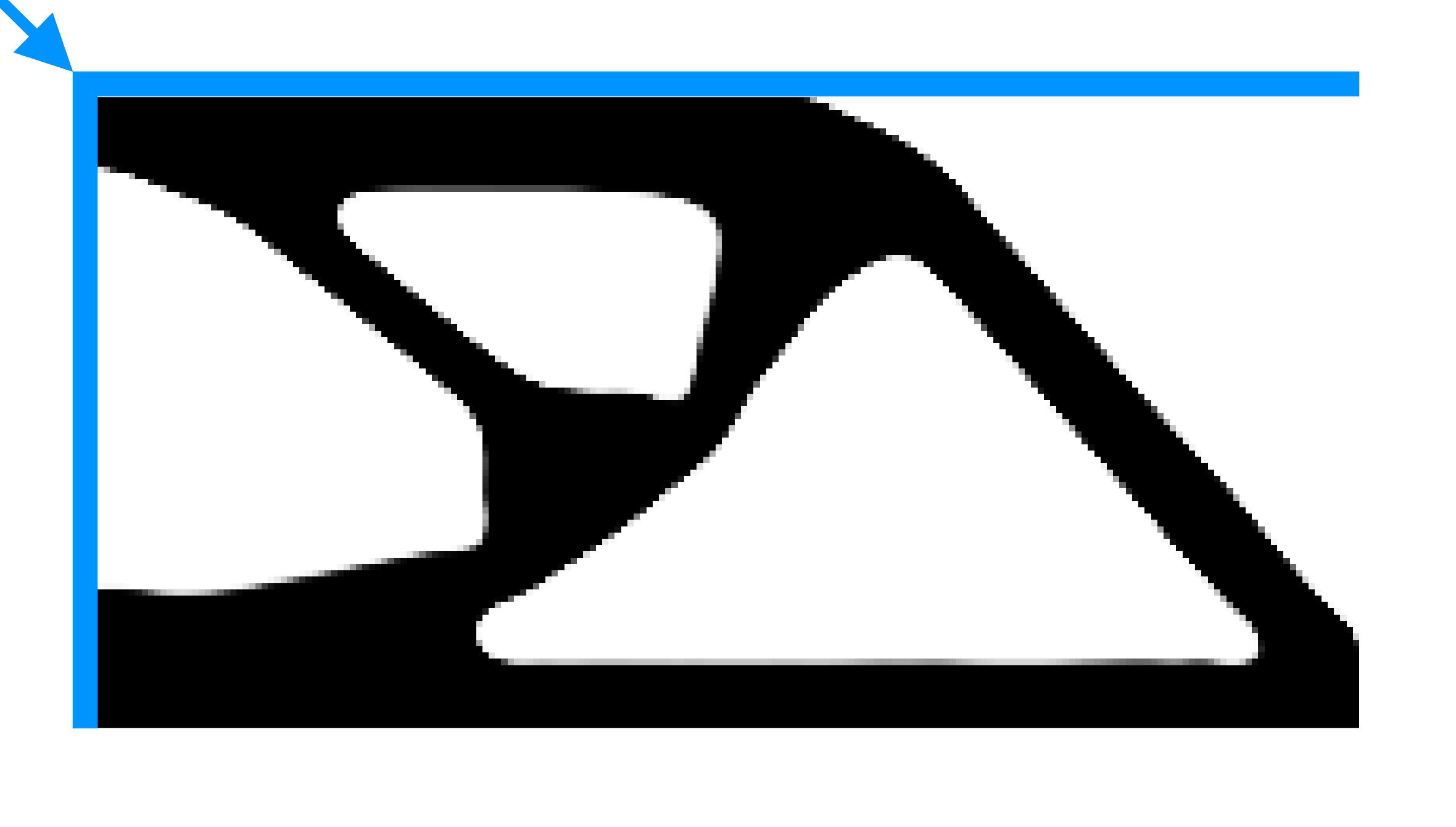}		
	&   \hspace{0mm}\includegraphics[width=1\linewidth]{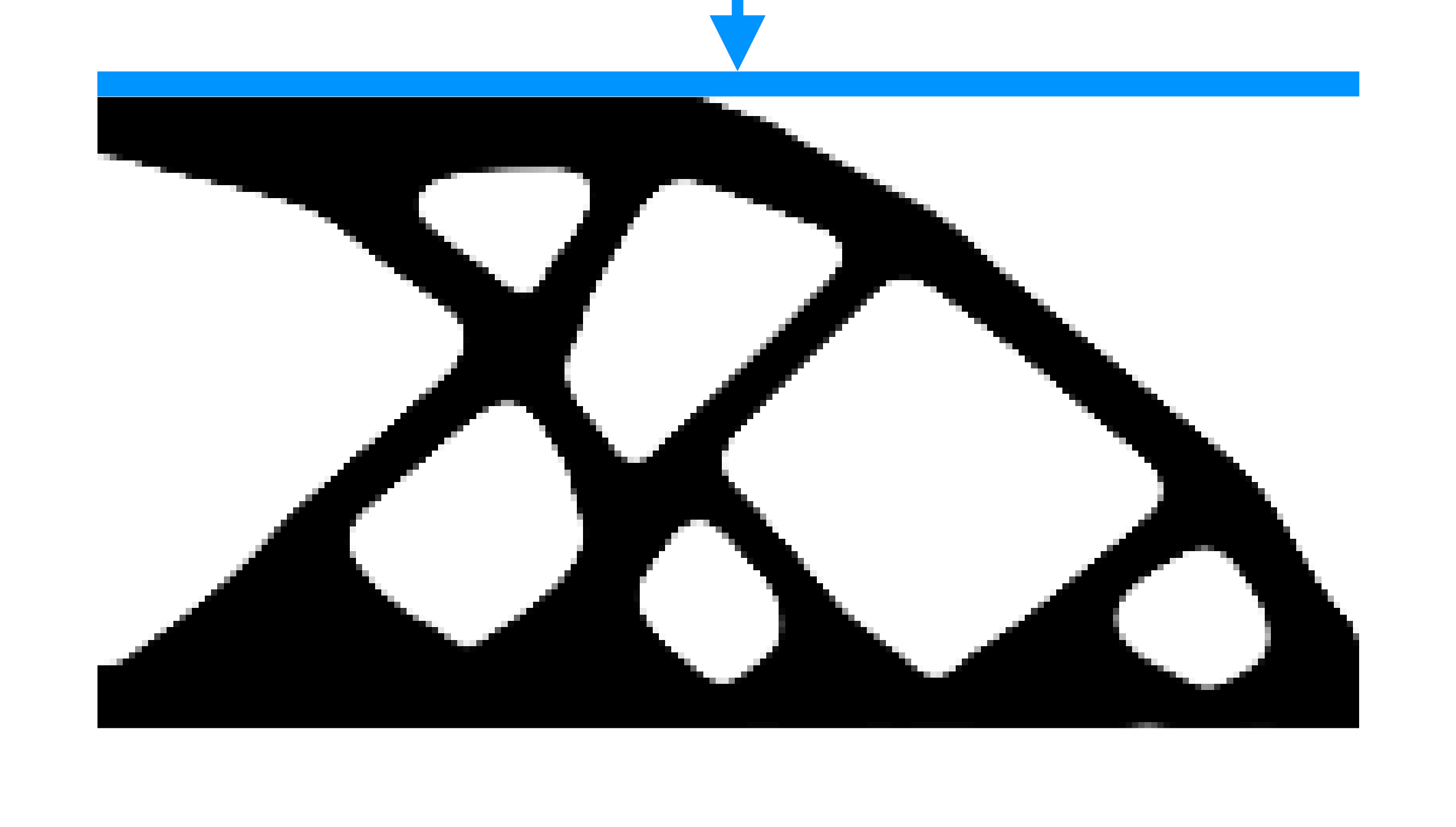}	
	&   \hspace{-4mm}\includegraphics[width=1\linewidth]{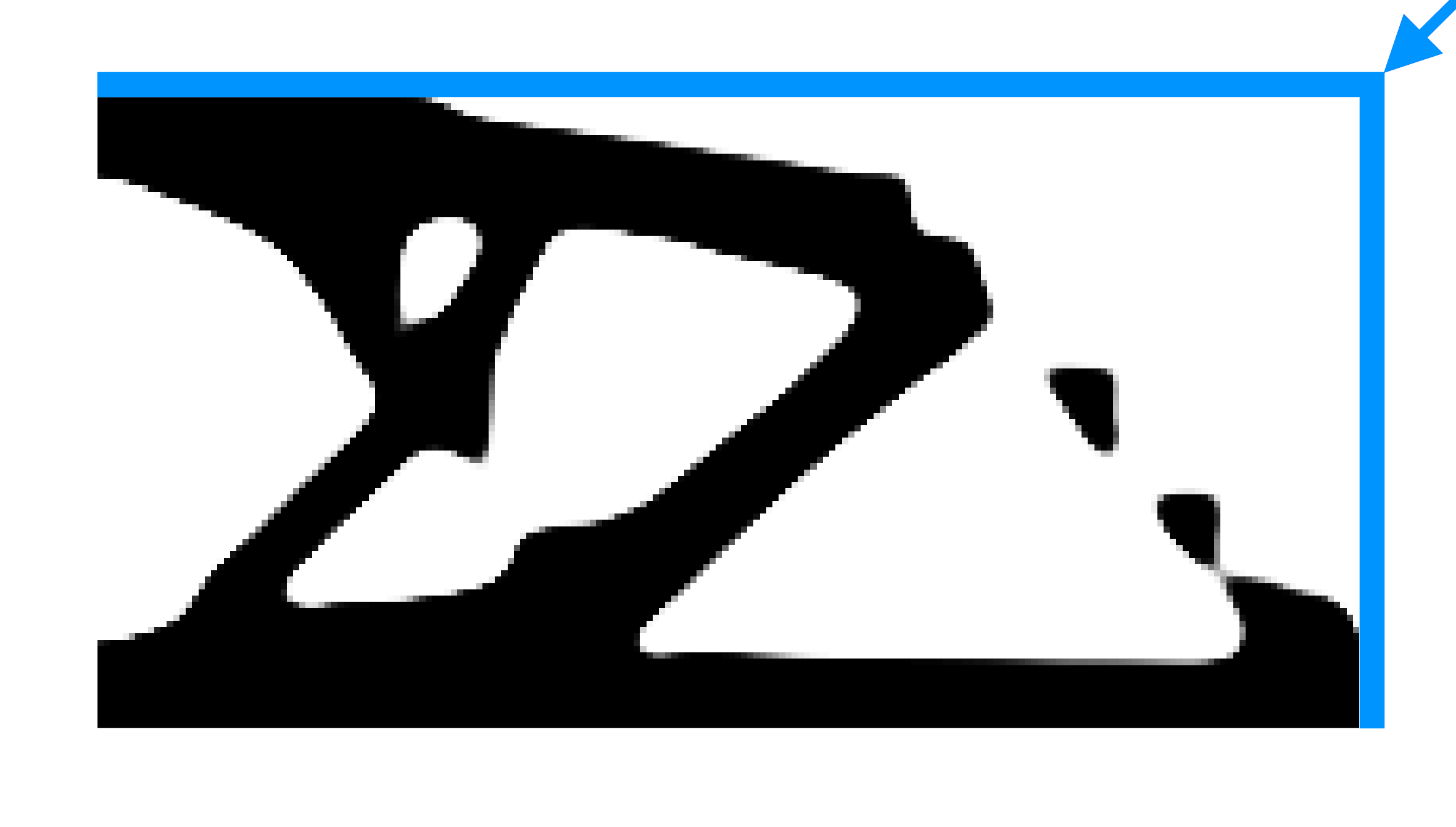}
	\\
		\vspace{-5mm}\hspace{4mm}\includegraphics[width=1\linewidth]{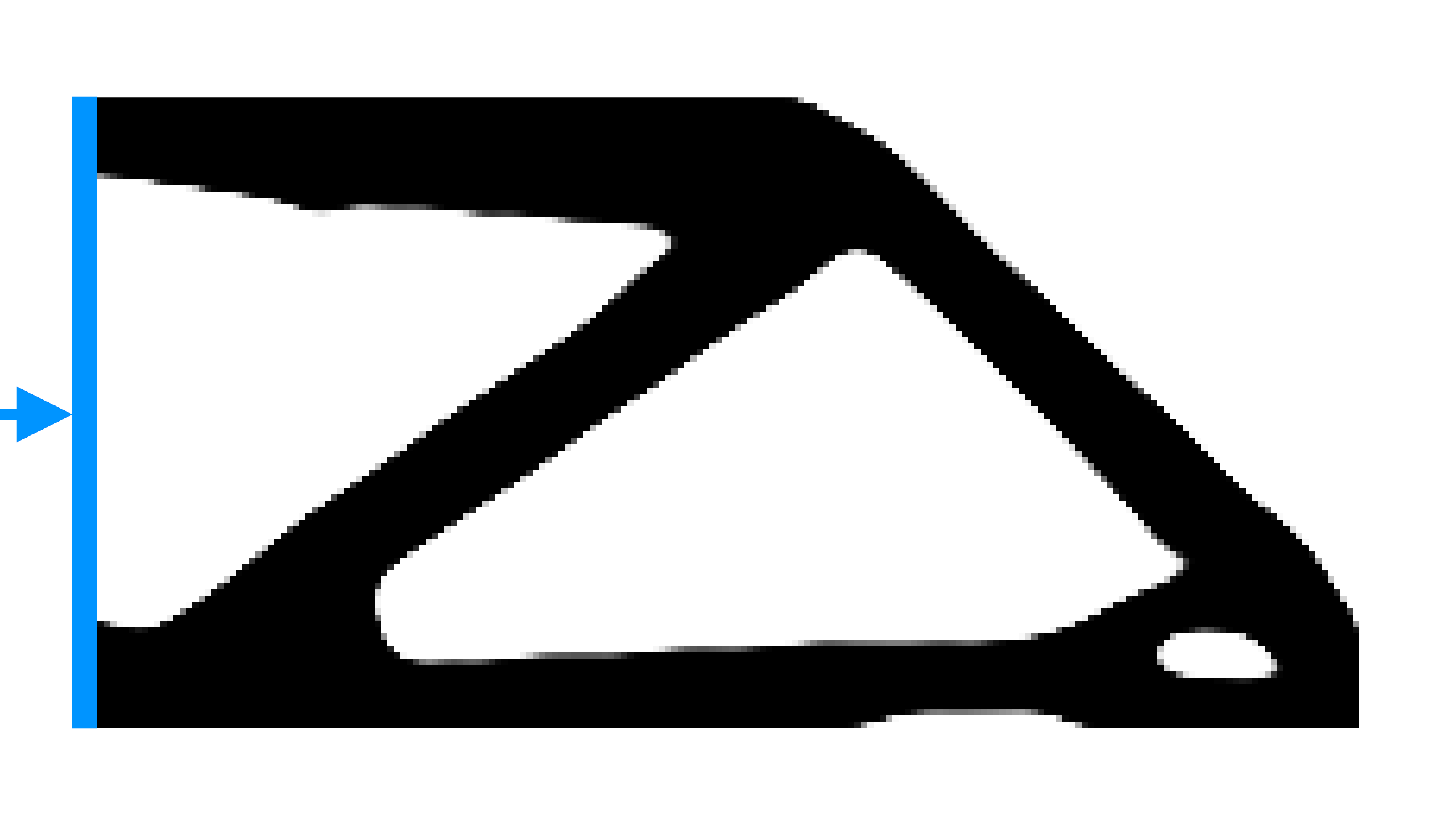}		
	&   \vspace{-5mm}\hspace{0mm}\includegraphics[width=1\linewidth]{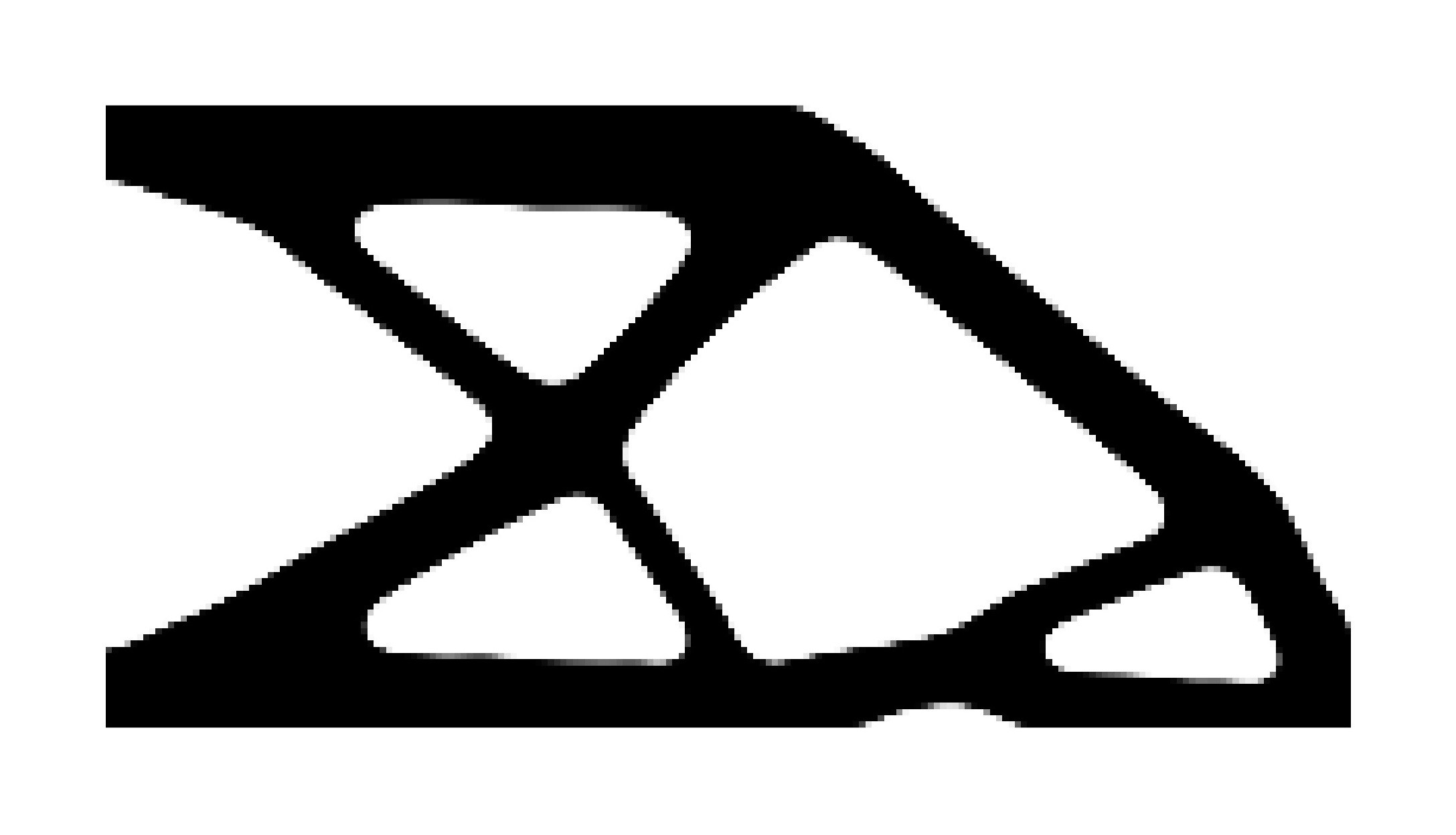}	
	&   \vspace{-5mm}\hspace{-4mm}\includegraphics[width=1\linewidth]{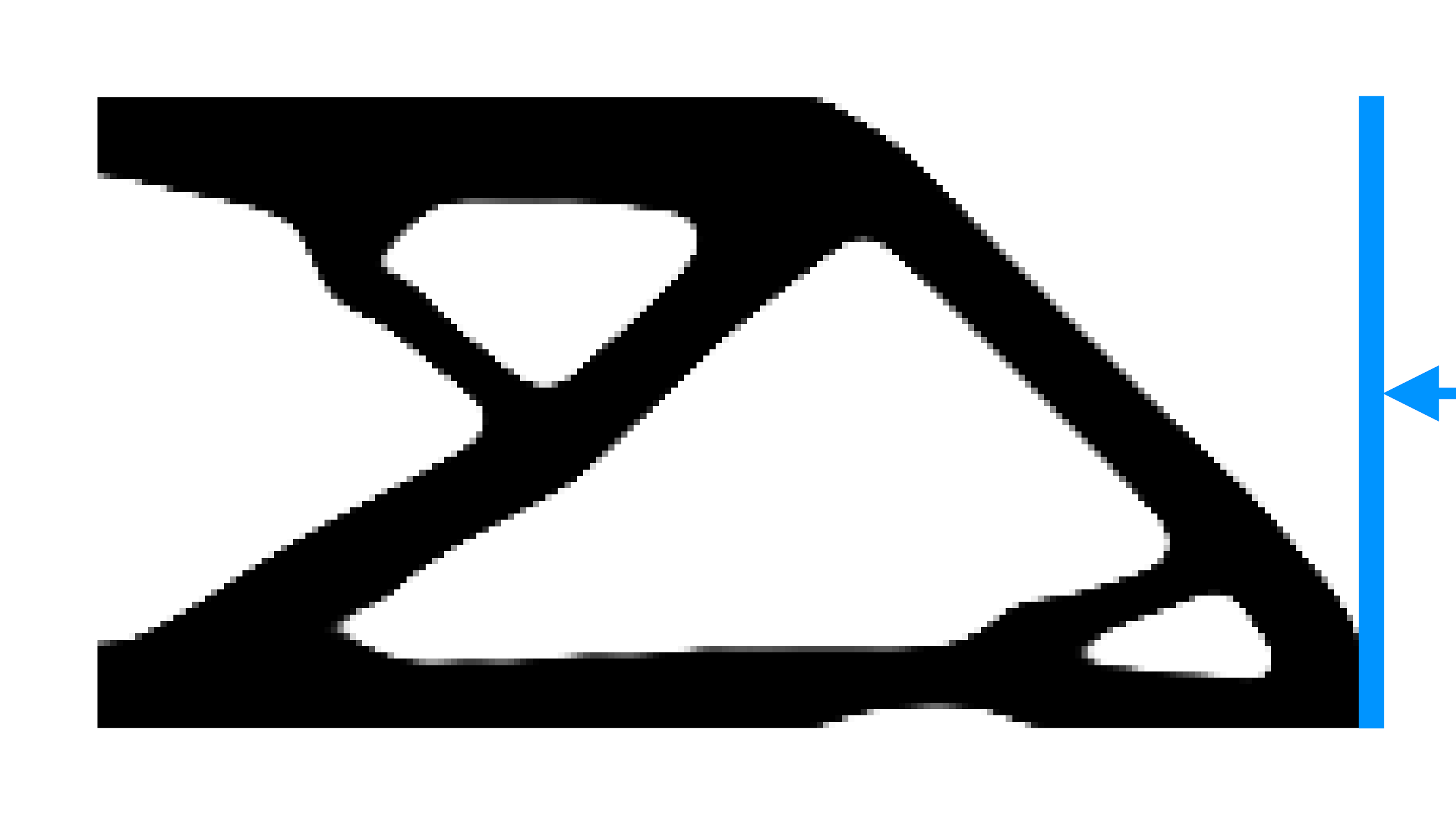}
	\\	
			\vspace{-5mm}\hspace{4mm}\includegraphics[width=1\linewidth]{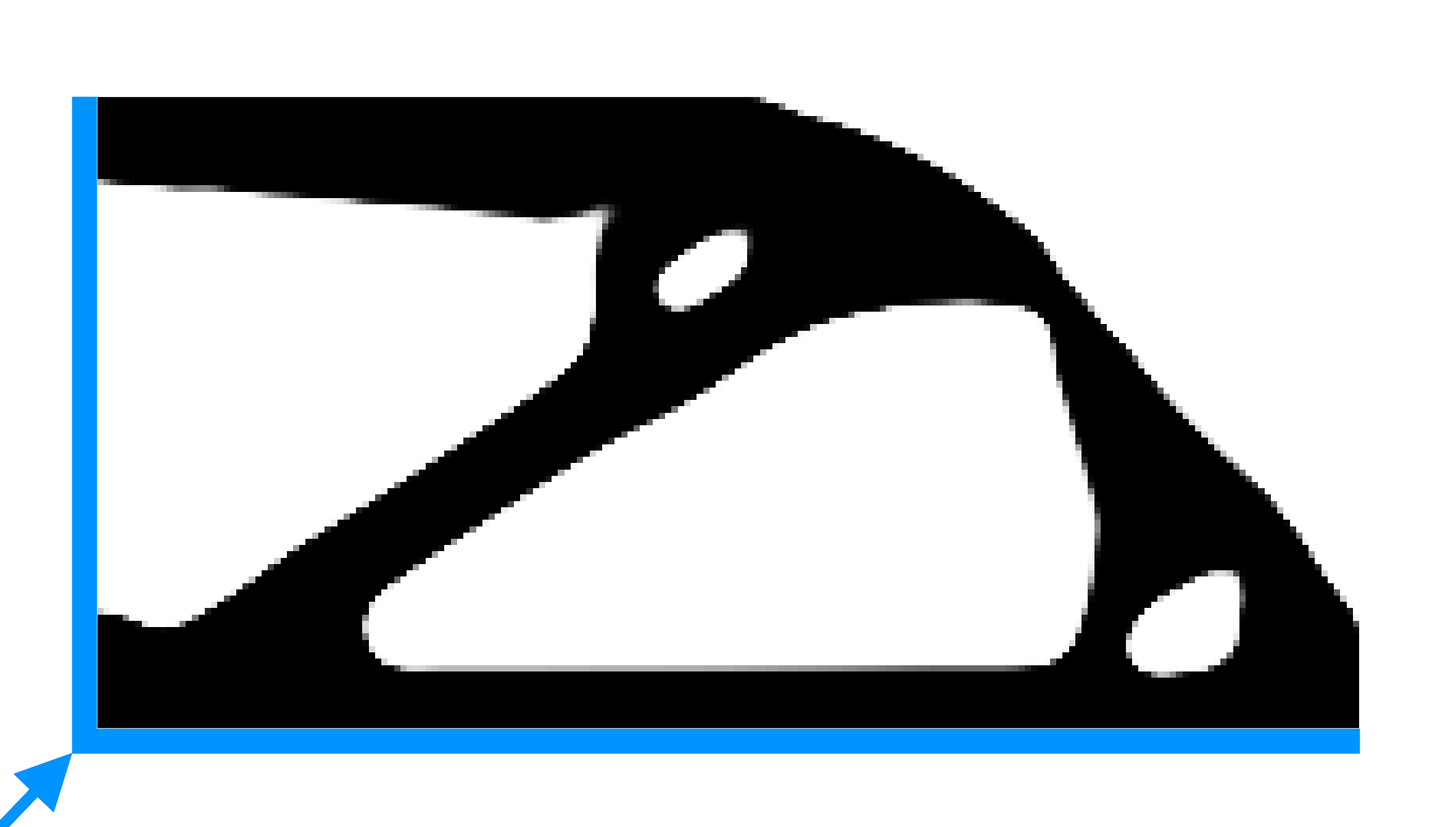}		
	&   \vspace{-5mm}\hspace{0mm}\includegraphics[width=1\linewidth]{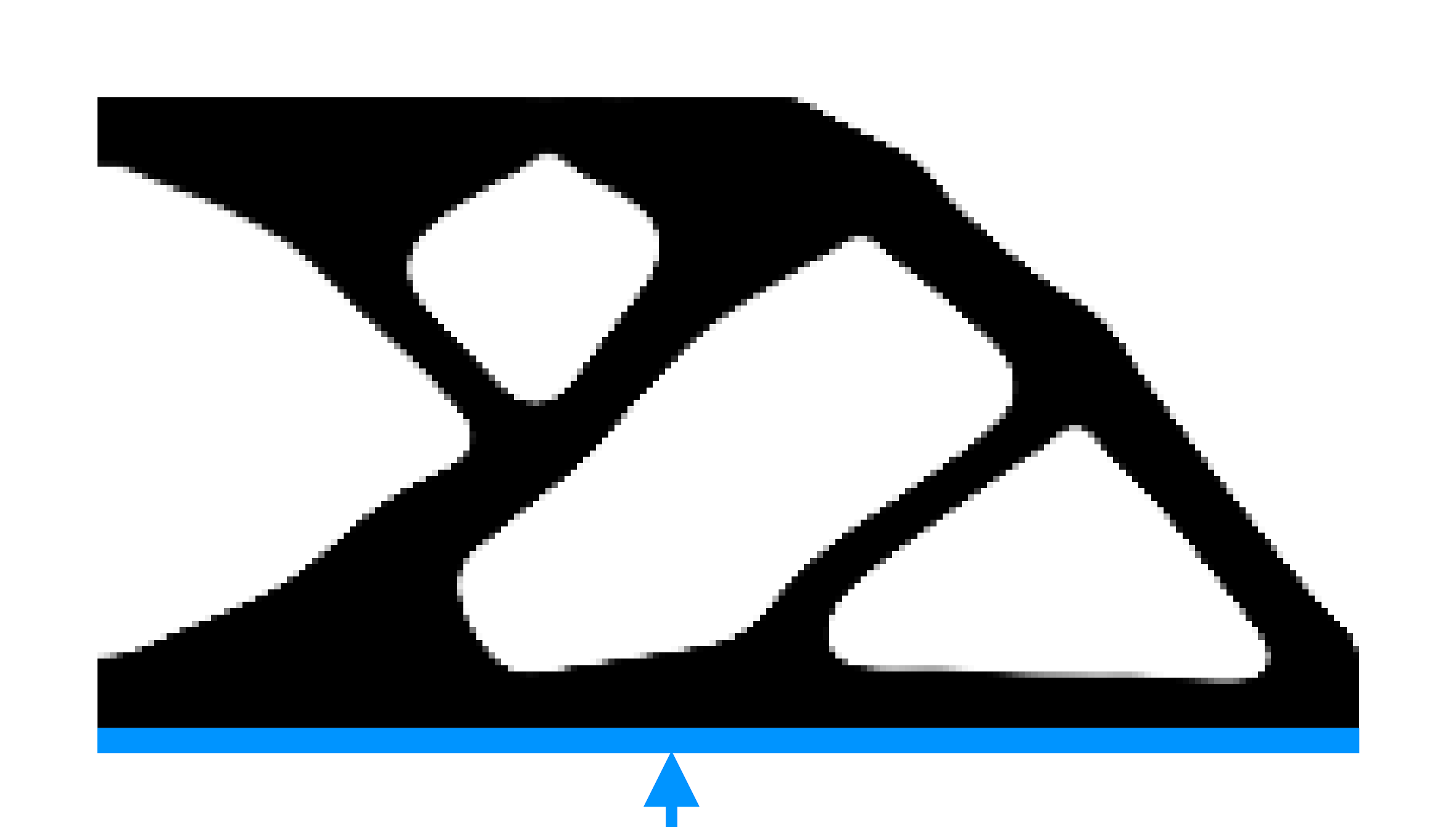}	
	&   \vspace{-5mm}\hspace{-4mm}\includegraphics[width=1\linewidth]{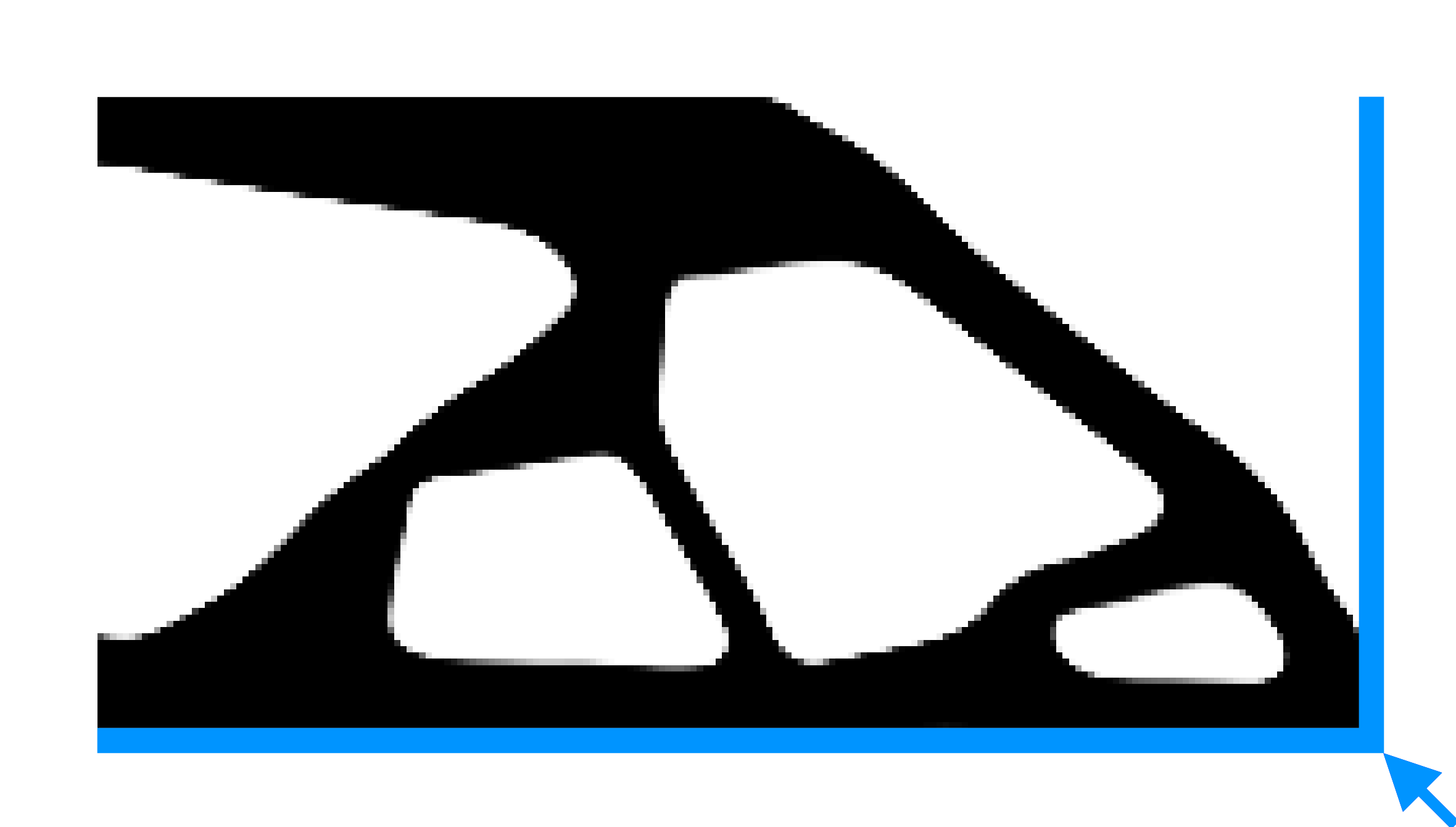}
	\\
	\vspace{-72mm}\hspace{37mm}$93.04$ & \vspace{-72mm}\hspace{33mm}$98.38$ & \vspace{-72mm}\hspace{28mm}$5000$ \\
	\vspace{-54mm}\hspace{37mm}$92.03$ & \vspace{-54mm}\hspace{33mm}$87.28$ & \vspace{-54mm}\hspace{28mm}$90.70$ \\
	\vspace{-32mm}\hspace{37mm}$102.23$ & \vspace{-32mm}\hspace{33mm}$97.82$ & \vspace{-32mm}\hspace{28mm}$99.40$ \\
	\\
	\vspace{-84mm}\hspace{37mm}$(225^\circ)$ & \vspace{-84mm}\hspace{33mm}$(180^\circ)$ & \vspace{-84mm}\hspace{28mm}$(135^\circ)$ \\
	\vspace{-66mm}\hspace{37mm}$(270^\circ)$ & \vspace{-66mm}\hspace{33mm}(Ref.) & \vspace{-66mm}\hspace{29mm}$(90^\circ)$ \\
	\vspace{-44mm}\hspace{38mm}$(315^\circ)$ & \vspace{-44mm}\hspace{35mm}$(0^\circ)$ & \vspace{-44mm}\hspace{29mm}$(45^\circ)$ \\			
\end{tabular}	
\vspace{-26mm}
\centering
\caption{Solution to cantilever beam for compliance minimization with overhanging angle constraints. The reference solution is displayed in the center. Compliance and building direction are given next to each solution. Blue arrows show building directions $\mathbf{b}$.} 	
\label{FIG:Diff_Orientations}
\end{figure}  
	
	\begin{figure}
\centering
\begin{tabular}{p{0.41\linewidth}p{0.48\linewidth}}
	    \multicolumn{2}{c}{
			\begin{subfigure}{0.65\linewidth}   	
    		\includegraphics[width=1\linewidth]{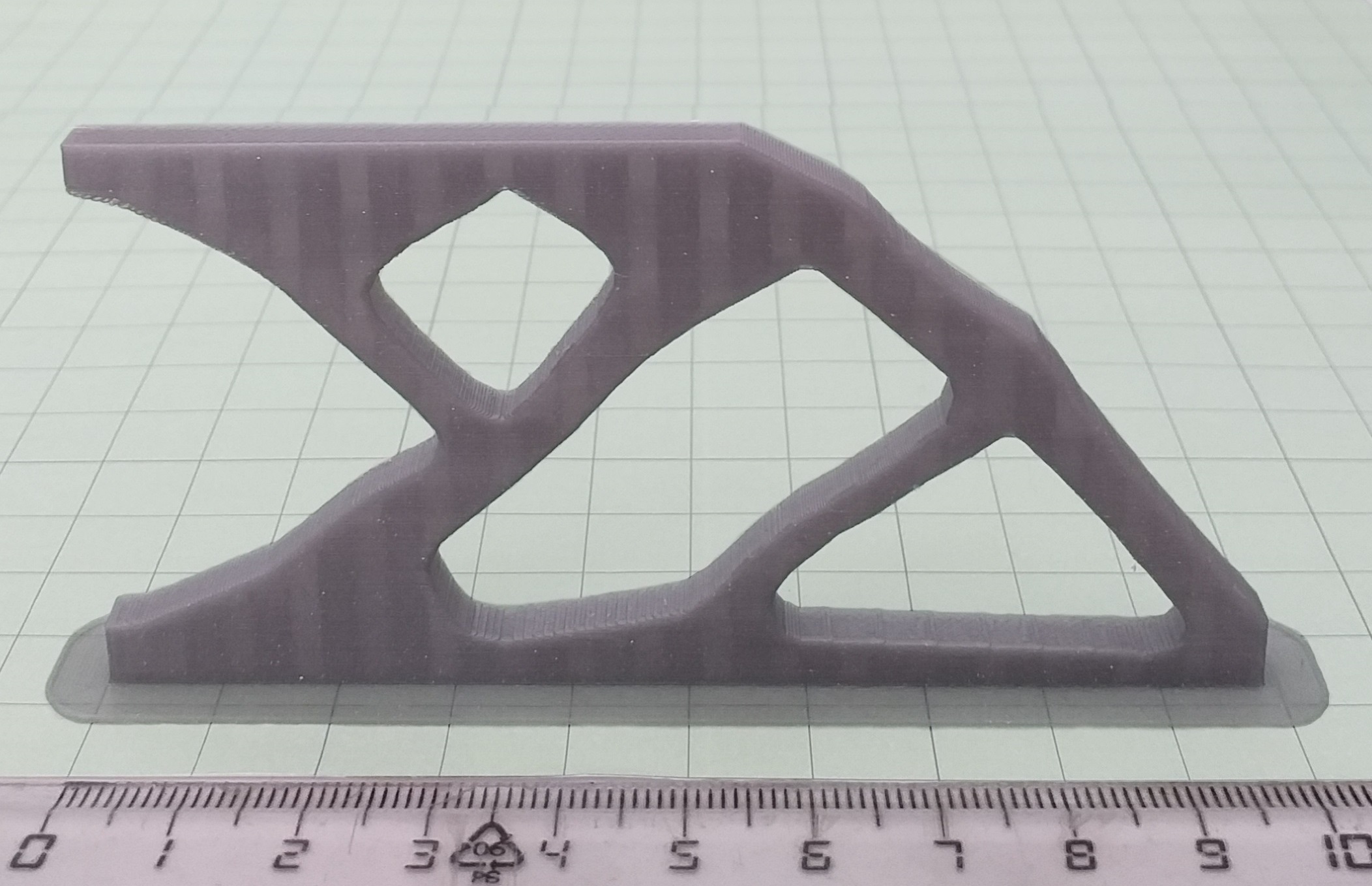} \caption{$\theta=0^\circ$}
    		\label{FIG:Printed_a}
    	\end{subfigure}		 
		}
	\vspace{3mm}\\
		\begin{subfigure}{1.00\linewidth} 
			\includegraphics[width=1\linewidth]{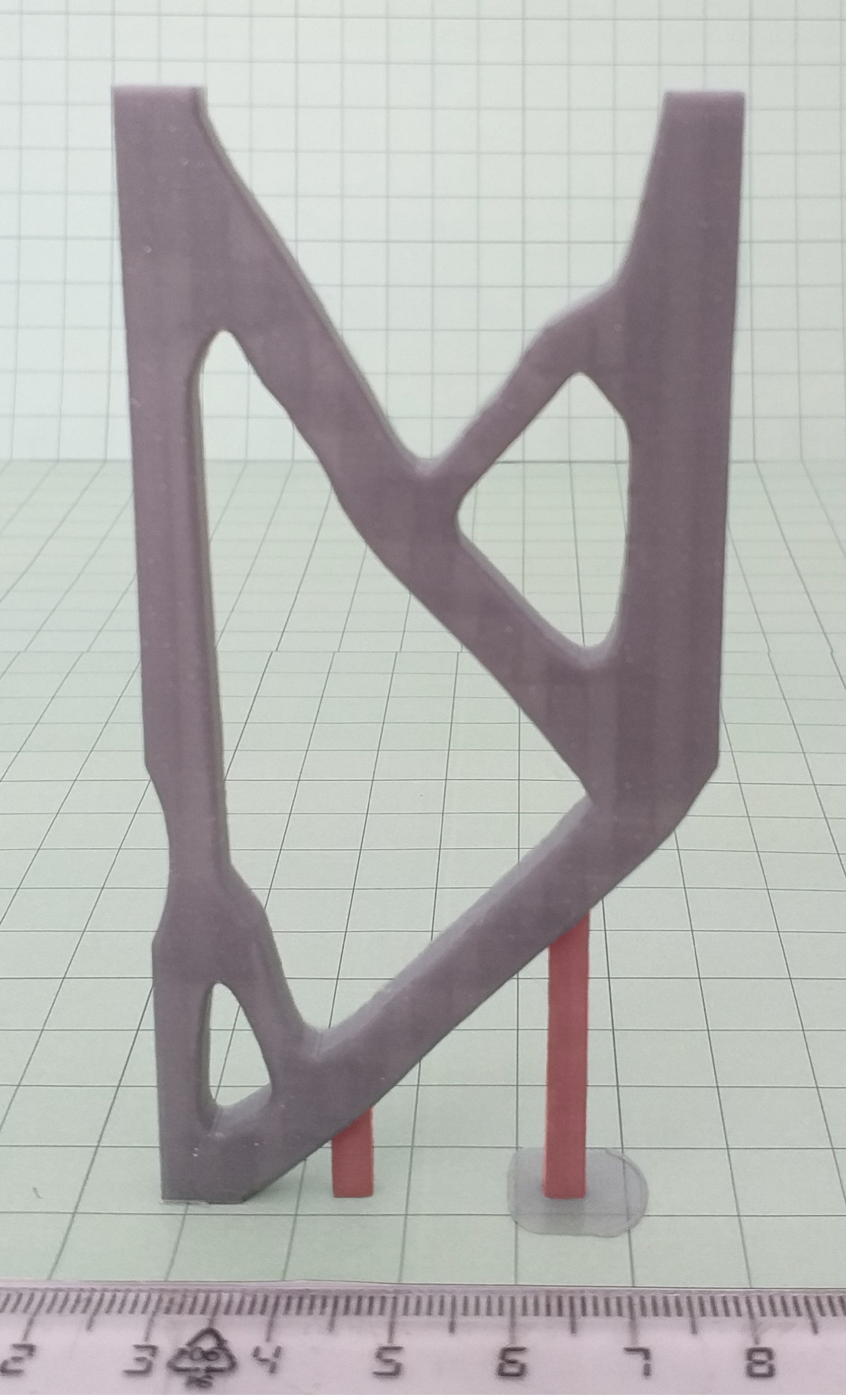} \caption{$\theta=90^\circ$}
			\label{FIG:Printed_b}
		\end{subfigure}
	&   \begin{subfigure}{1.00\linewidth}
			\includegraphics[width=1\linewidth]{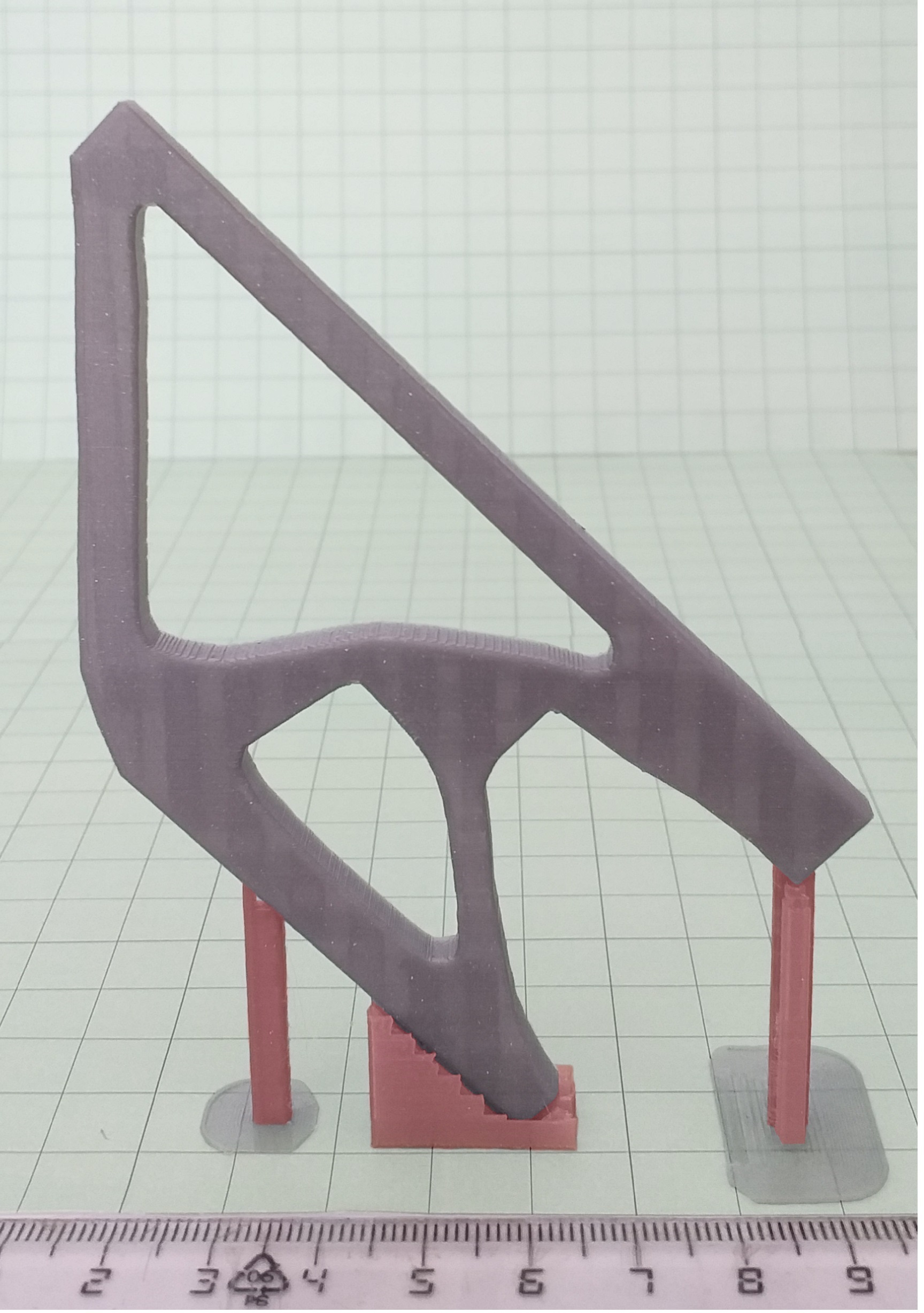} \caption{$\theta=225^\circ$}  
			\label{FIG:Printed_c}
		\end{subfigure}	
\end{tabular}
\caption{Optimized designs from Fig. \ref{FIG:Diff_Orientations} printed by fused deposition modeling with a machine PRUSA I3 MK3S. Dimension of design domain is $100 \times 50$ mm. Supports are in magenta color.}  	
\label{FIG:Printed}	
\end{figure} 
	
	\subsubsection{Results with different building orientations}
	The overhang-constrained TO problem (Eq.~\ref{EQ:OPTI_Constrained}) is solved with the proposed strategy to avoid the hanging triangular parts. The volume fraction for the intermediate design is set to $40\%$, and the minimum size is fixed to $\mathrm{r_{min}}=3s$. The optimization problem is solved for $\alpha=45^\circ$, however one can also get optimized designs for different $\alpha$ using the provided codes in the appendices.  The different building orientations~$\theta$ are taken.
	
	The results are shown in Fig.~\ref{FIG:Diff_Orientations} wherein the reference solution is placed in the center. The final compliance and building orientation are also displayed next to each solution. The parameters used for the optimization problem and the strategy to avoid non-self-supporting parts are described above and are set default in provided codes \texttt{topCbeam}, \texttt{GBOHC} and \texttt{PPTri}. In the code \texttt{GBOHC}, one can alter $\theta$ on line 9. The main code can be called as \texttt{topCbeam(200,100,0.4,1.0,6.1)}.
	
	One can note that solutions exhibit improved printability compared to the reference design and avoid hanging parts (Fig.~\ref{FIG:Diff_Orientations}). The solution at $\theta=135^\circ$ is not self-supporting, this indicates that the proposed approach may not ensure self-supporting designs occasionally. One can also notice that optimized designs may have minimum inclination of downward facing surfaces lower than $\alpha$. As per the result displayed at position (3,1) of Table~\ref{TAB:Strategy}, once in a while some local constraints may be violated even though  global constraints $\mathrm{G}$ get satisfied in the optimization iterations. This is due to \textit{p-mean} aggregation wherein the function may underestimate the constraints resulting in a minimum inclination angle lower than $\alpha$. In addition,  we notice that after the free evolution, the overhang constraints $\mathrm{G}$ may not get satisfied because the redistribution of material significantly reduces inclination of the surfaces pointing downwards, as shown by arrows in the third row of Table~\ref{TAB:Strategy}.

	Despite some of the aforementioned  limitations of the presented approach, with respect to the reference solution the overhang-constrained designs can be easily printed. Fig.~\ref{FIG:Printed} shows designs fabricated by fused deposition modeling (FDM). 2D designs of Fig.~\ref{FIG:Diff_Orientations} are extruded by 0.05L in the out-of-plane direction. For  $\theta=0^\circ$, the design is manufactured without additional supports, proving to have better printability than the reference design. For $\theta=90^\circ$, it is necessary to place supporting columns to avoid collapse due to self-weight during fabrication. With inclined designs as in Fig.~\ref{FIG:Printed_c}, the design is supported only by  thin supports. This indicates that the formulation based on the structural inclination improves printability, however may not ensure self-supporting designs always. On the other hand, self-weight TO approaches are process dependent as they depend on the material and size of the components \citep{Allaire2018,Allaire2017,Amir2018}. For instance, the self-weight is not a limitation for the internal zone of the components. Therefore, for such examples the overhang angle restriction improves printability by avoiding placing support structures on the interior that in turn eases post-machining for the support removal. Readers may refer to \cite{kumar2022topology} for a detailed description of the self-weight modeling in TO. Although, the presented method have some limitations as other TO approaches do \citep{Sigmund2013}, novalty is that we have included length scale and overhang constraints in one method. Next, we present on how the proposed approach also takes care of the building direction.
	
	\subsection{Considering building orientation in \TO} 
	
	As per Fig.~\ref{FIG:Diff_Orientations}, it is clear that printing direction is indeed influences structural performance. However, knowing the best building direction \textit{a priori} is not so straightforward, especially when multiple process constraints have to be considered.  In this work, we propose to consider $m$ printing directions and only to include in the TO problem the least restrictive one, denoted by $\theta^*$, similar to the one proposed by Langelaar~\cite{Langelaar2018}. However, instead of a layer-by-layer, a gradient-based formulation is proposed and implemented.
	
	Let the candidate building directions be $\theta_k|_{k=1,...,m}$ i.e. $m$ restrictions, therefore $\mathrm{G}(\theta_k)$ must be evaluated in each iteration of the TO problem. However, the gradient is computed only once as $\bm{\varrho}_i$ does not change with rotation of the printing orientation (see Figs. \ref{FIG:Local_Constraint_a} and \ref{FIG:Local_Constraint_b}). This significantly reduces the computational cost since only Eqs.~\ref{EQ:Local_Constraint}-\ref{EQ:Global_Constraint} are to be evaluated $m$ times. Once the restrictions $G_k$ are computed, they are aggregated using an \textit{r-norm} function to capture the lowest restriction value. To this end, the quantity of interest to be aggregated is:  
	\begin{equation}\label{EQ:t}
	\mathrm{t}_k = 1-0.5(\mathrm{G}_k + \mathbf{b}^\intercal \bm{\varrho_\alpha} +1),
	\end{equation}
	
	Considering $r$ exponent of the aggregation function, the overhang constraint that follows the least restrictive printing orientation is determined as:
	\begin{equation} \label{EQ:B_g}
	B_\mathrm{G} = 1 - 2 \left( \sum_{k=1}^m (\mathrm{t}_k)^r \right) ^ {1/r} - \mathbf{b}^\intercal \bm{\varrho_\alpha} \leq 0.
	\end{equation}
	
	\noindent The sensitivities are obtained by chain rule as:
	\begin{equation}\label{EQ:DerivB_G}
	\deriv{B_\mathrm{G}}{\bm{\rho}} = \deriv{\mathbf{G}}{\bm{\rho}} \deriv{\mathbf{t}}{\mathbf{G}}\deriv{B_\mathrm{G}}{\mathbf{t}},
	\end{equation}
	\noindent where $\mathbf{G}$ is the array containing the constraints $\mathrm{G}_k$. As the gradient $\bm{\varrho}_i$ is independent of $\theta$, the expression (Eq.~\ref{EQ:DerivB_G}) can be treated algebraically to reduce the number of matrix operations, that leads to the following expression:
	\begin{equation}\label{EQ:DerivB_G_Compact}
	\hspace{-2mm}
	\deriv{B_\mathrm{G}}{\bm{\rho}} = \mathbf{H}^{\intercal}
	\left(
	{h}^\prime{(\bm{\bar{\rho}})} \circ 
	\left( 
	\mathbf{D_x}^\intercal \left( \mathbf{{A}_x}  {\bm{B}_\mathbf{G}^\prime}^\intercal \right)
	+
	\mathbf{D_y}^\intercal \left( \mathbf{{A}_y}  {\bm{B}_\mathbf{G}^\prime}^\intercal \right)
	\right)
	\right),
	\end{equation}
	
	\noindent where matrices $\mathbf{{A}_x}$ and $\mathbf{{A}_y}$, having size $N \times m$, are defined as:
	\begin{equation}\label{EQ:barAx_barAy}
	\begin{matrix}
	\mathrm{{A}}_{\mathrm{x}(i,k)} = & \mathrm{a}_{\mathrm{x}(i)} (\theta_k) \; \mathrm{B}_{(i)}(\theta_k),
	\vspace{2mm}\\
	\mathrm{{A}}_{\mathrm{y}(i,k)} = & \mathrm{a}_{\mathrm{y}(i)} (\theta_k) \; \mathrm{B}_{(i)}(\theta_k),
	\end{matrix}
	\end{equation}
	
	\begin{figure}
\captionsetup[subfigure]{labelformat=empty}
	\centering
	\begin{tabular}{p{0.22\linewidth}p{0.22\linewidth}p{0.22\linewidth}p{0.22\linewidth}}
		\begin{subfigure}{1.00\linewidth}   	
    		\includegraphics[width=1\linewidth]{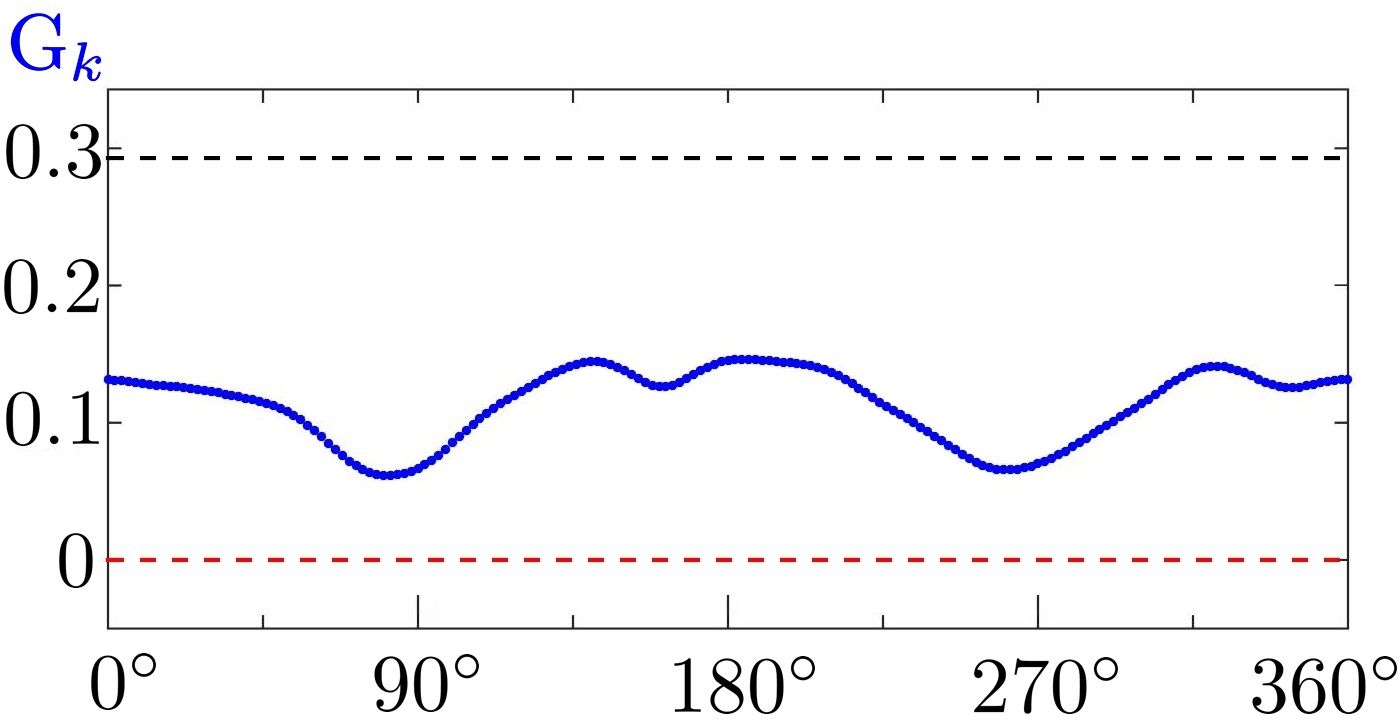} 
    		\caption{}
    		\label{FIG:CB_RESULTS_a}
    	\end{subfigure}			
	&   \begin{subfigure}{1.00\linewidth}   	
    		\includegraphics[width=1\linewidth]{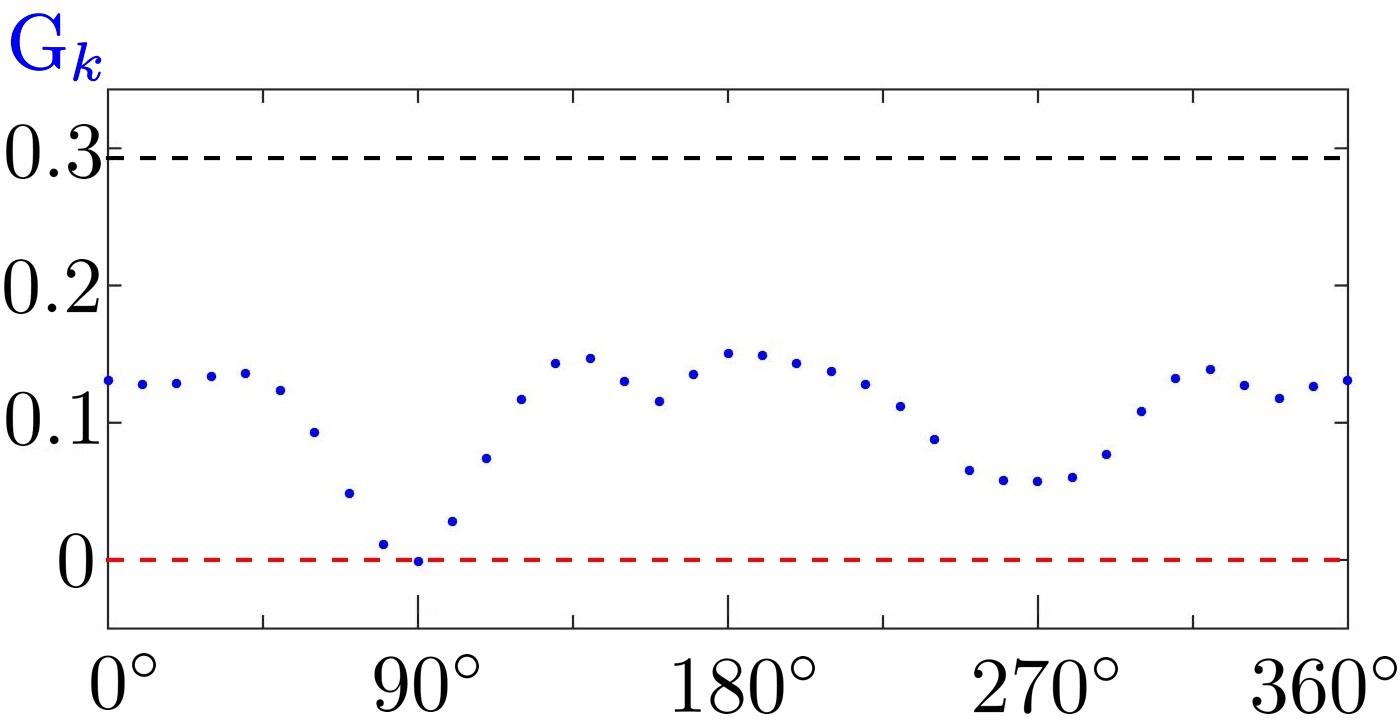} 
    		\caption{}
    		\label{FIG:CB_RESULTS_b}
    	\end{subfigure}	
    &	\begin{subfigure}{1.00\linewidth}   	
    		\includegraphics[width=1\linewidth]{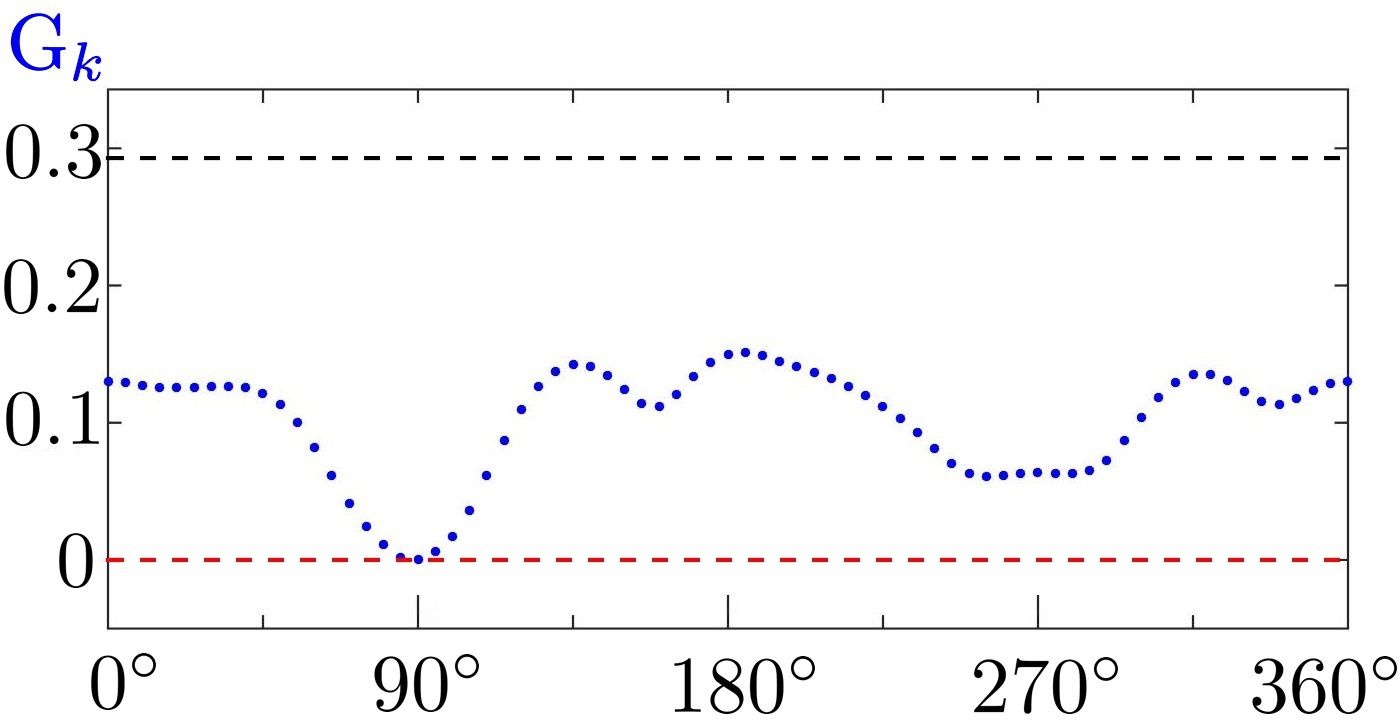} 
    		\caption{}
    		\label{FIG:CB_RESULTS_c}
    	\end{subfigure}	
    &  	\begin{subfigure}{1.00\linewidth}   	
    		\includegraphics[width=1\linewidth]{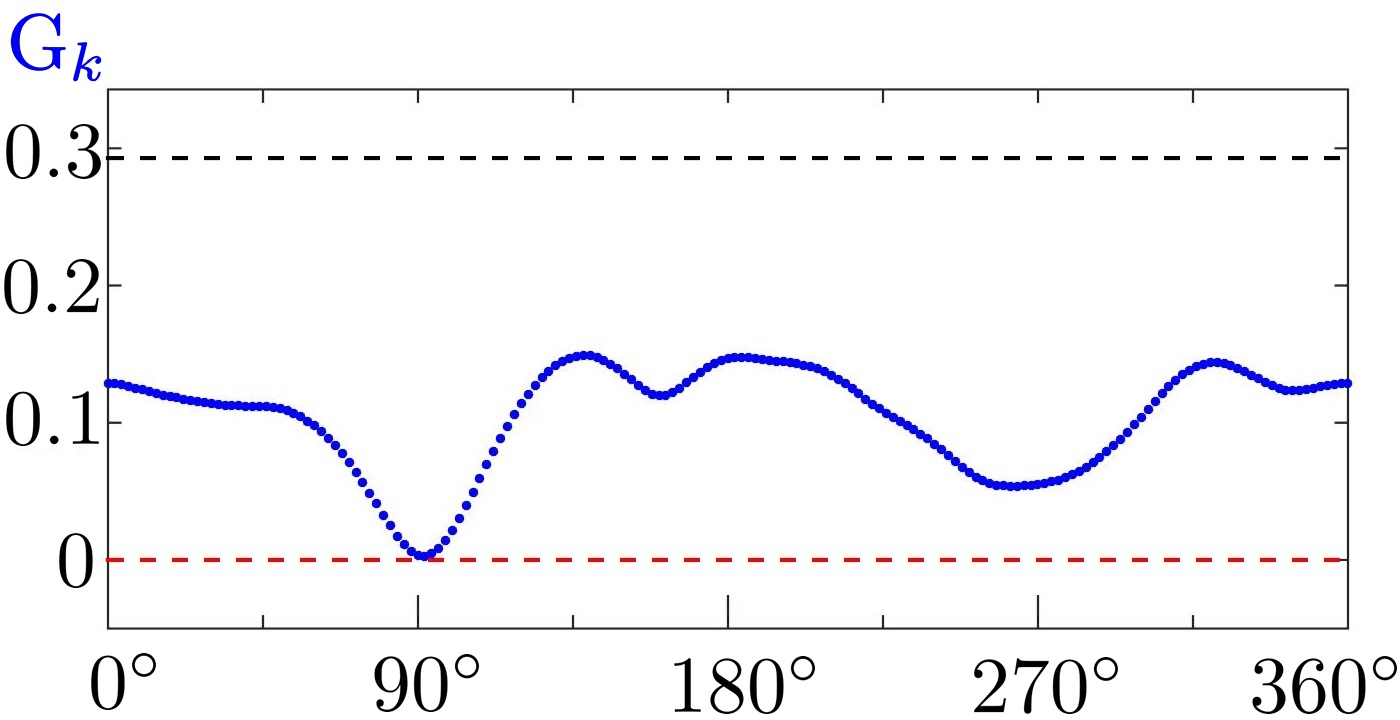}  
    		\caption{}
    		\label{FIG:CB_RESULTS_d}
    	\end{subfigure}	
    \\
    	\begin{subfigure}{1.00\linewidth}   	
    		\vspace{-4mm}	
    		\includegraphics[width=1\linewidth]{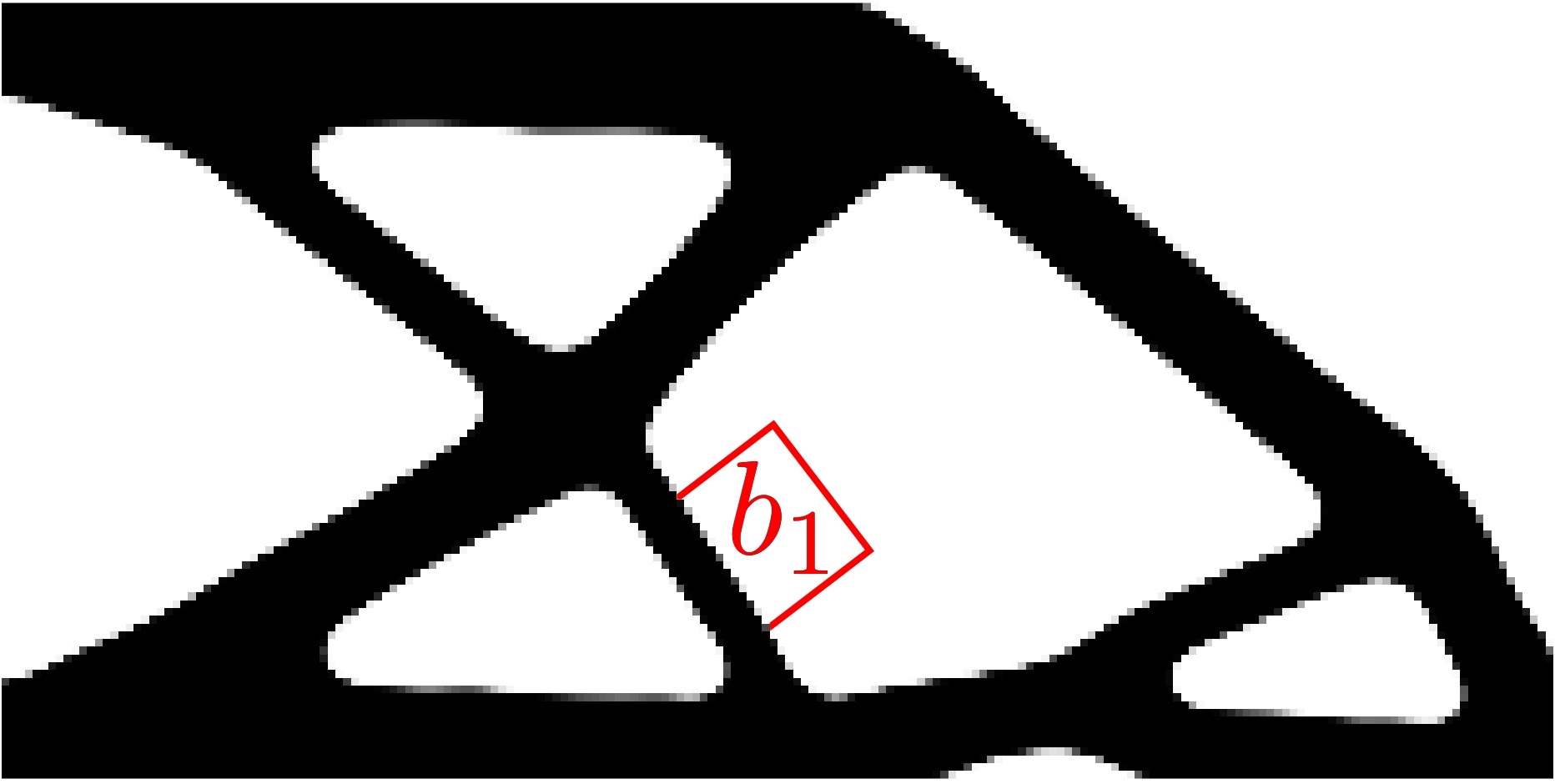} 
    		\caption{}
    	\end{subfigure}			
	&   \begin{subfigure}{1.00\linewidth}  
			\vspace{-4mm}	 	
    		\includegraphics[width=1\linewidth]{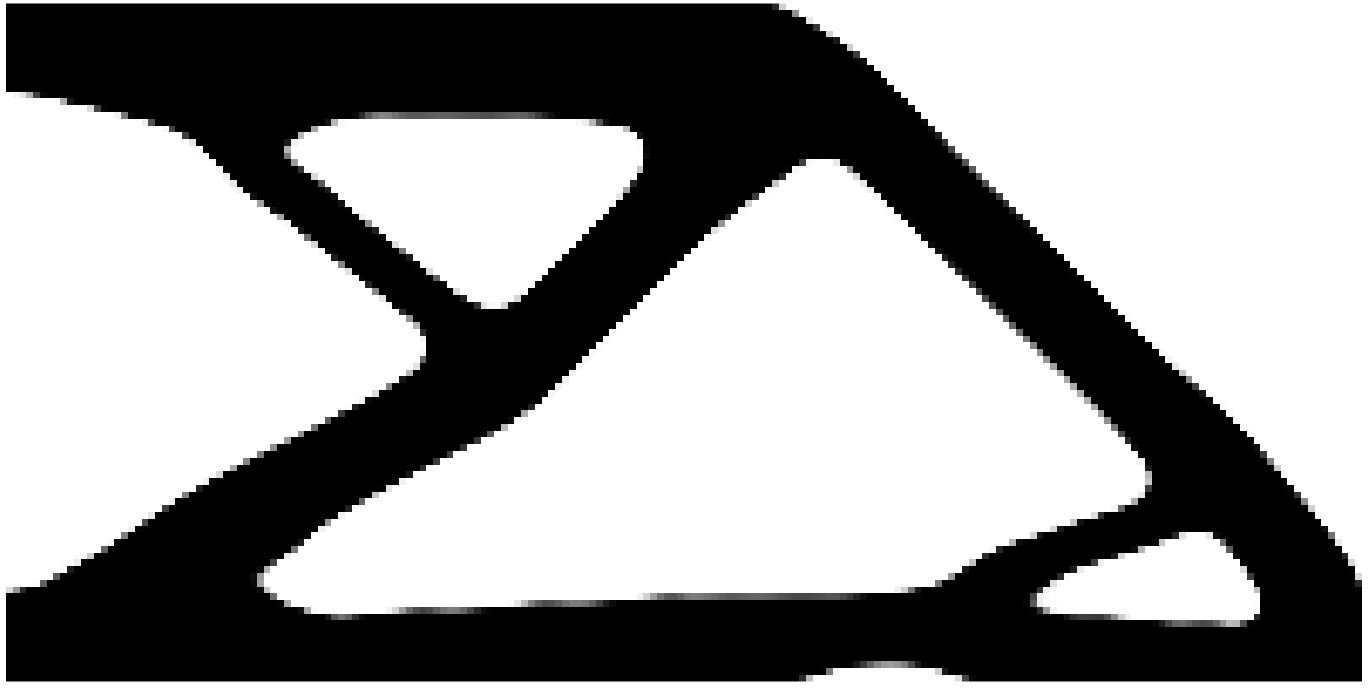} 
    		\caption{}
    	\end{subfigure}	
    &	\begin{subfigure}{1.00\linewidth}
    		\vspace{-4mm}	   	
    		\includegraphics[width=1\linewidth]{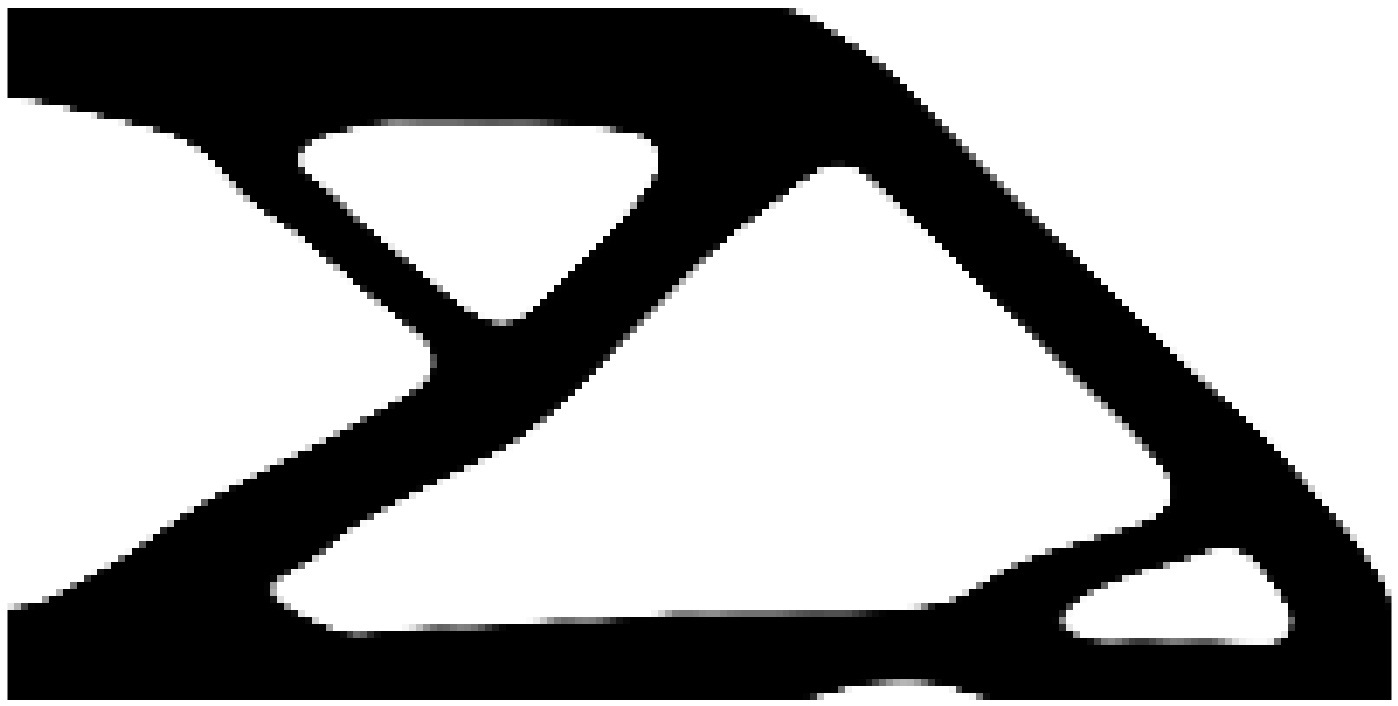} 
    		\caption{}
    	\end{subfigure}	
    &  	\begin{subfigure}{1.00\linewidth}
    		\vspace{-4mm}	   	
    		\includegraphics[width=1\linewidth]{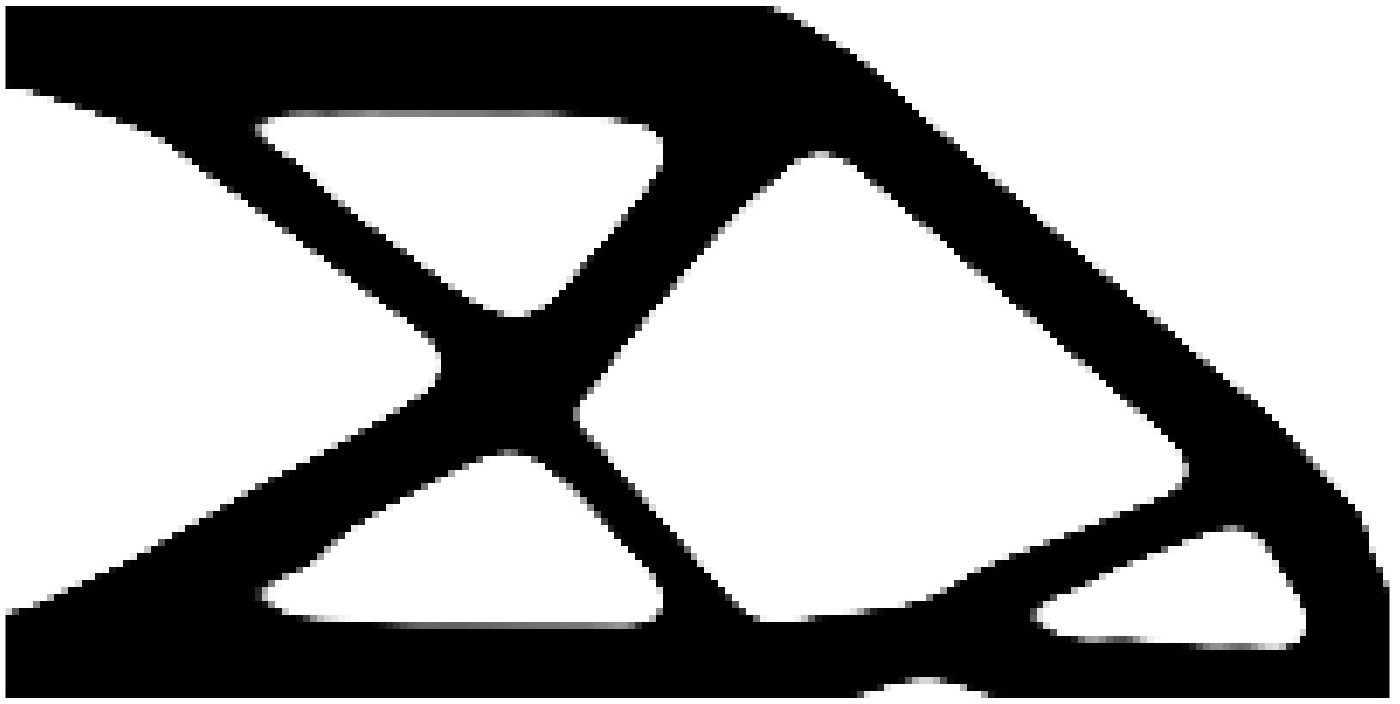} 
    		\caption{}
    	\end{subfigure}	
    \\ 
    \vspace{-7mm} (a) $C_\mathrm{int}$= $87.28$ , \; $\theta^*$= $80^\circ$ &
    \vspace{-7mm} (b) $C_\mathrm{int}$= $89.27$ , \; $\theta^*$= $90^\circ$ &
    \vspace{-7mm} (c) $C_\mathrm{int}$= $89.50$ , \; $\theta^*$= $90^\circ$ &
    \vspace{-7mm} (d) $C_\mathrm{int}$= $87.80$ , \; $\theta^*$= $92^\circ$ 		 		
\end{tabular}
    \vspace{-3mm}
	\caption{(a) Reference solution. Overhang angle constrained solutions: (b) With 36 candidate directions, (c) With 72 candidate directions, and (d) With 180 candidate directions. The constraints $\mathrm{G}_k$ are plotted on top of each solution as a function of the building direction  $\theta$.}
	\label{FIG:CB_RESULTS}	
\end{figure} 
	
	\noindent Arrays $\mathbf{a_x}$, $\mathbf{a_y}$ and $\mathbf{B}$ are defined in Eq.~\ref{EQ:AxAyB} and array ${\bm{B}_\mathbf{G}^\prime}$ is the derivative of $B_{\mathbf{G}}$ with respect to the constraints $\mathbf{G}$, determined as:
	\begin{equation}\label{EQ:Bprime}
	B_{\mathrm{G}(i)}^\prime =
	\deriv{B_\mathrm{G}}{\mathrm{G}_i} = \mathrm{t}_i^{r-1} \left( \sum_{k=1}^m (t_k)^r \right)^{1/r-1}.
	\end{equation}
	
	To compute sensitivities of $B_\mathrm{G}$ it is necessary to evaluate $m$ times the arrays $\mathbf{a_x}$, $\mathbf{a_y}$, $\mathbf{B}$, and then to evaluate only once (Eq.~\ref{EQ:DerivB_G_Compact}). This facilitates significant reduction in the computational cost of the restriction and its derivatives as matrix products involving $\mathbf{H}$, $\mathbf{D_x}$, $\mathbf{D_y}$ are performed only once at each iteration of the TO problem. For a large number of printing directions, for example $m>100$, we use an aggregation strategy that considers only a percentage of the most critical constraints in order to reduce the number of evaluations of arrays $\mathbf{a_x}$, $\mathbf{a_y}$ and $\mathbf{B}$. For this we define the set $\Psi$ that contains the indexes $k$ of the $10\%$ of the least restrictive constraints $\mathrm{G}_k$. With $\Psi$, the matrices $\mathbf{{A}_x}$ and $\mathbf{{A}_y}$ reduce their size to $N \times 0.1m$. This practice is known as aggregation with active set and it is a common practice in TO with stress constraints \citep{Paris2010}.  
	
	The free evolution and the PP steps are applied as indicated in the previous section, but the latter is performed only for the least restrictive direction of the iteration, i.e. the PP step is carried out for $\theta^*$. 
	
	Finally, the overhang-constrained optimization problem including $m$ candidate directions can be written as:
	\begin{align} \label{EQ:OPTI_Constrained_Bg}
	\begin{split}
	{\min_{\bm{\rho}}} & \quad c_\mathrm{ero}=\mathbf{f}^{\intercal} \mathbf{u}_\mathrm{ero} \\
	\mathrm{s.t.:} &\quad  \mathbf{v}^{\intercal} \bm{\bar{\rho}}_\mathrm{dil} \leq V^*_\mathrm{dil} \left( V^*_\mathrm{int} \right) 	\\
	&\quad B_\mathrm{G}(\bm{\bar{\rho}}_\mathrm{ero}) \leq 0 \\
	&\quad B_\mathrm{G}(\bm{\bar{\rho}}_\mathrm{int}) \leq 0 \\
	&\quad B_\mathrm{G}(\bm{\bar{\rho}}_\mathrm{dil}) \leq 0 \\
	&\quad 0 \leq {\rho_i} \leq1  \;\;,\;\; i=1,...\:,N.
	\end{split}
	\end{align}

	\section{Numerical examples} \label{Sec:3}
	In this section, to demonstrate the efficacy and versatility of the presented approach, we optimized various 2D benchmarks design problems with  the length scale and overhang constraints including the printing direction.  Problems consider the robust design approach \citep{Wang2011}, include the filtering treatment (Eq.~\ref{EQ:density_filter_corrected}), and are solved using the MMA optimizer \citep{Svanberg1987}. In addition, all examples use the same continuation method for $\eta$, $\beta$ and $\mathrm{m}_\mathrm{L}$ as noted earlier. The parameters of the overhang constraint and the strategy to avoid non-self-supporting parts are summarized in Table~\ref{TAB:Parameters}. 
	
	\subsection{Cantilever beam}
	We first solve cantilever beam  problem (Fig.~\ref{FIG:Cone_Example_a}). Volume fraction is set to 0.40. $r_\mathrm{min}=3s$ is set. $m$ candidate building direction is defined as $\theta_k = \frac{360^\circ(k-1)}{m}$. The aggregation exponent of $B_\mathrm{G}$ is taken $r=20$. 
	
	The reference solution is shown in Fig. \ref{FIG:CB_RESULTS_a}. The overhang-constrained solutions are shown in Figs. \ref{FIG:CB_RESULTS_b}, \ref{FIG:CB_RESULTS_c} and \ref{FIG:CB_RESULTS_d}, which include $36$, $72$ and $180$ candidate directions, respectively. On top of each solution, the restrictions $\mathrm{G}_k(\bm{\bar{\rho}}^\mathrm{int})$ for $\theta_k \in$  $[0^\circ$,\,$360^\circ]$ are plotted. The reference solution shows two orientations that minimize the unsupported surfaces, $\theta$= $80^\circ \pm 10^\circ$ and $\theta$= $260^\circ \pm 10^\circ$. However, there is no orientation $\theta_k$ that meets $\mathrm{G}_k\leq 0$, thus it is expected that the structure requires sacrificial supports for printing the designs. When $36$ and $72$ candidate directions are considered (Figs. \ref{FIG:CB_RESULTS_b} and \ref{FIG:CB_RESULTS_c}), the prevailing direction is $\theta^*=90^\circ$, which coheres with the less restrictive directions of the reference solution. The MaxOA constrained designs (Figs.~\ref{FIG:CB_RESULTS_b}-\ref{FIG:CB_RESULTS_c}) do not contain the bar $b_1$ (Fig.~\ref{FIG:CB_RESULTS_a}). Consequently, the constrained designs are relatively less stiff. With  $180$ candidate directions, the prevailing direction is $92^\circ$.  The optimized design retain the bar $b_1$, and that in turn increases stiffness the design (Fig.~\ref{FIG:CB_RESULTS_d}).
	
	\begin{table*}
\caption{Time per iteration ($t_\mathrm{iter}$) and total time ($t_\mathrm{total}$) for different optimization problems for the cantilever beam. The time is given in seconds and is obtained using a Dell Precision 5510 laptop, with an Intel Core i7-6820 HQ @2.70GHz, x64-based processor, 16GB of RAM and MATLAB R2015a.}
\begin{tabular}{p{0.03\linewidth} p{0.09\linewidth} p{0.1\linewidth} p{0.2\linewidth} p{0.09\linewidth} p{0.09\linewidth} p{0.09\linewidth}}
	\toprule	    
	 & & \multicolumn{5}{c}{Constraints included in the optimization problem} 
	 \\
	 \cmidrule(r){3-7}
     & \hfil \multirow{2}{*}{Reference} & \hfil \multirow{2}{*}{$\mathrm{G}(\bm{\bar{\rho}}^\mathrm{dil})$} & \hfil \multirow{2}{*}{$\mathrm{G}(\bm{\bar{\rho}}^\mathrm{ero}), \; \mathrm{G}(\bm{\bar{\rho}}^\mathrm{int}), \; \mathrm{G}(\bm{\bar{\rho}}^\mathrm{dil})$} & \multicolumn{3}{c}{ $B_\mathrm{G}(\bm{\bar{\rho}}^\mathrm{ero})$, $B_\mathrm{G}(\bm{\bar{\rho}}^\mathrm{int})$, $B_\mathrm{G}(\bm{\bar{\rho}}^\mathrm{dil})$}
    \\
    \cmidrule(r){5-7}
    & & & & \hfil $m=36$ & \hfil $m=72$ & \hfil $m=180$
    \vspace{1mm}\\
     & \hfil (Fig. \ref{FIG:CB_RESULTS_a}) & \hfil (Fig. \ref{FIG:Diff_MinSize_a}) & \hfil (Fig. \ref{FIG:Diff_Orientations}, $90^\circ$) & \hfil (Fig. \ref{FIG:CB_RESULTS_b}) & \hfil (Fig. \ref{FIG:CB_RESULTS_c}) & \hfil (Fig. \ref{FIG:CB_RESULTS_d})
    \\
    \cmidrule(r){2-7}
	$t_\mathrm{iter}$ & \hfil 0.60 & \hfil 0.87 & \hfil 1.32 & \hfil 1.45 & \hfil 1.51 & \hfil 1.82
	\vspace{2mm}\\
	$t_\mathrm{total}$  & \hfil 158  & \hfil 297  & \hfil 450  & \hfil 494  & \hfil 512  & \hfil 618
	\\		      
    \bottomrule
\end{tabular}
\label{TAB:time}
\end{table*}
	
	\begin{figure}
	\centering
    	\begin{subfigure}{0.45\linewidth}   	
    		\includegraphics[width=1\linewidth]{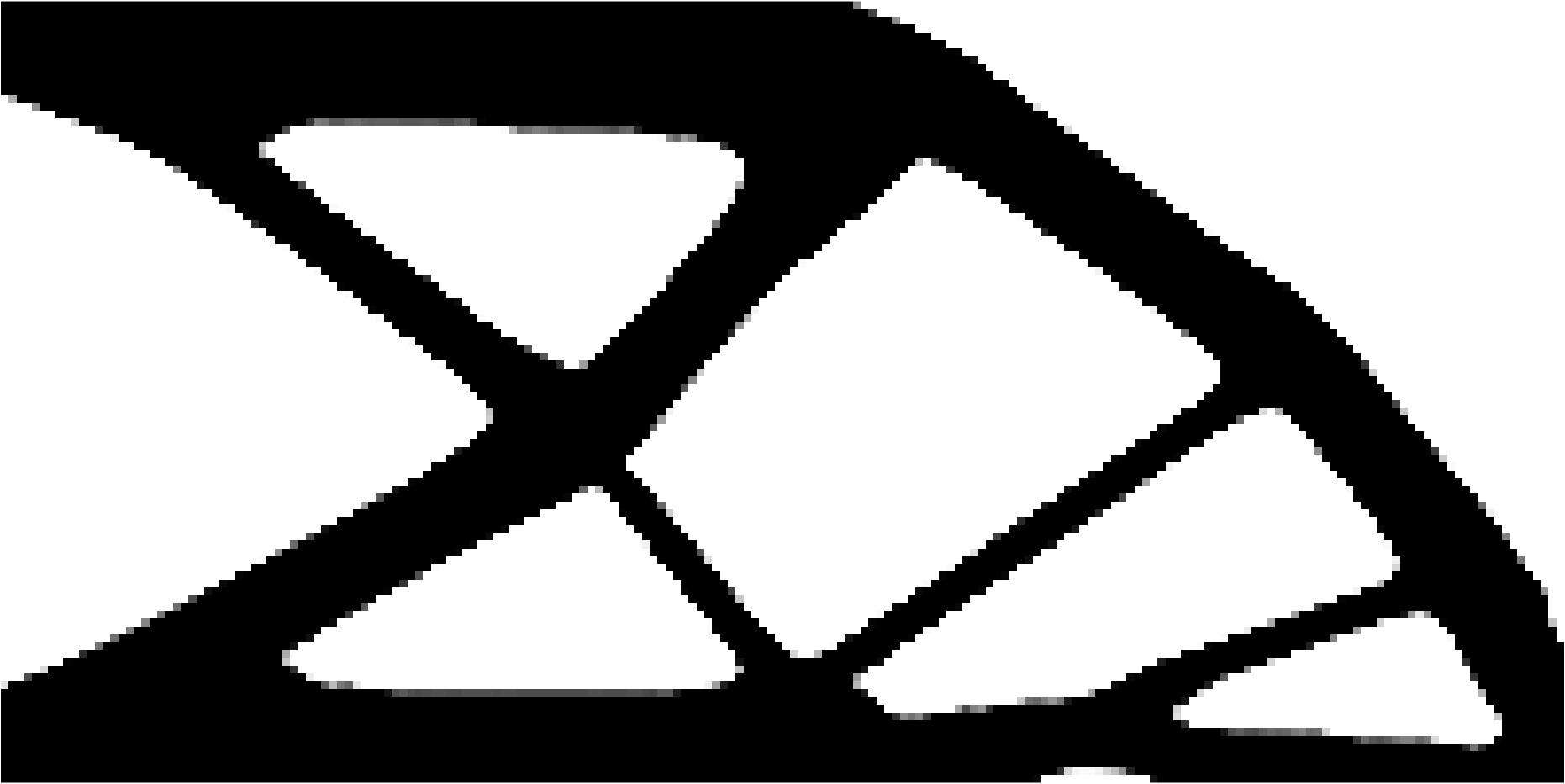} 
    		\caption{$C_\mathrm{int}$= $85.06$ , \; $\theta^*=88^\circ$}
    		\label{FIG:CB_RESULTS_2a}
    	\end{subfigure}			
         \begin{subfigure}{0.45\linewidth}   	
    		\includegraphics[width=1\linewidth]{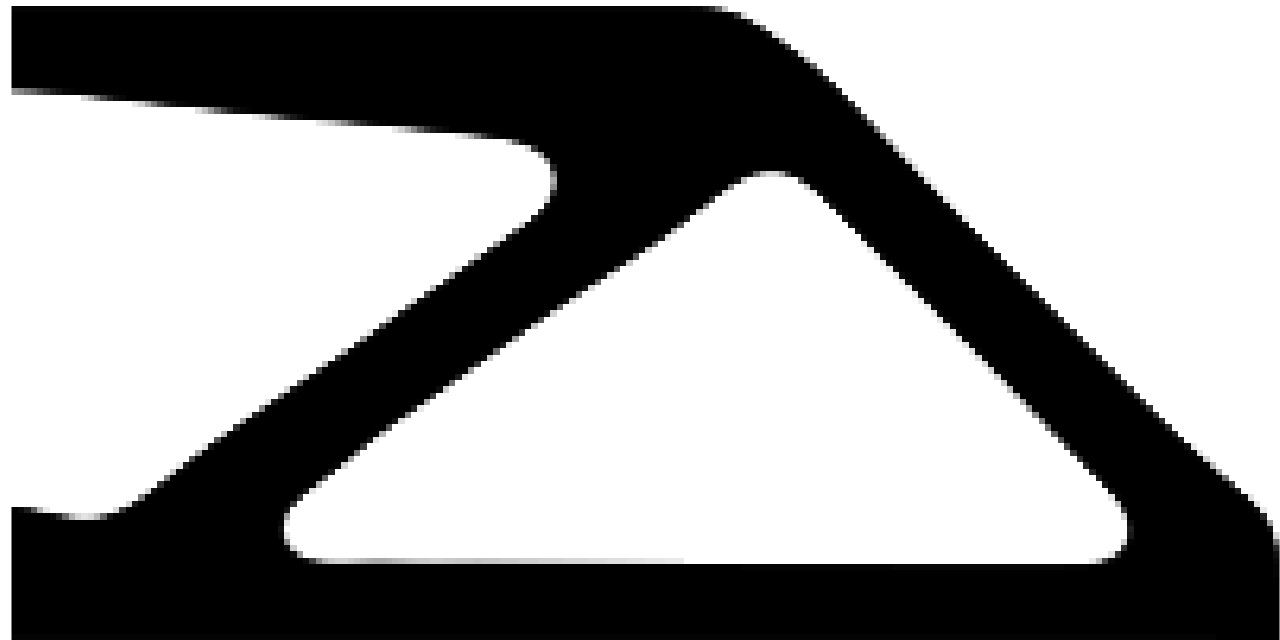} 
    		\caption{$C_\mathrm{int}$= $91.12$ , \; $\theta^*=90^\circ$}
    		\label{FIG:CB_RESULTS_2b}
    	\end{subfigure}	
	\caption{Cantilever beam with overhang constraints including 180 candidate directions. (a) $r_\mathrm{min}=2$ and (b) $r_\mathrm{min}=5$.}
	\label{FIG:CB_RESULTS_2}	
\end{figure} 
	
	\begin{figure}
	\centering
    	\begin{subfigure}{0.45\linewidth}   	
    		\includegraphics[width=1\linewidth]{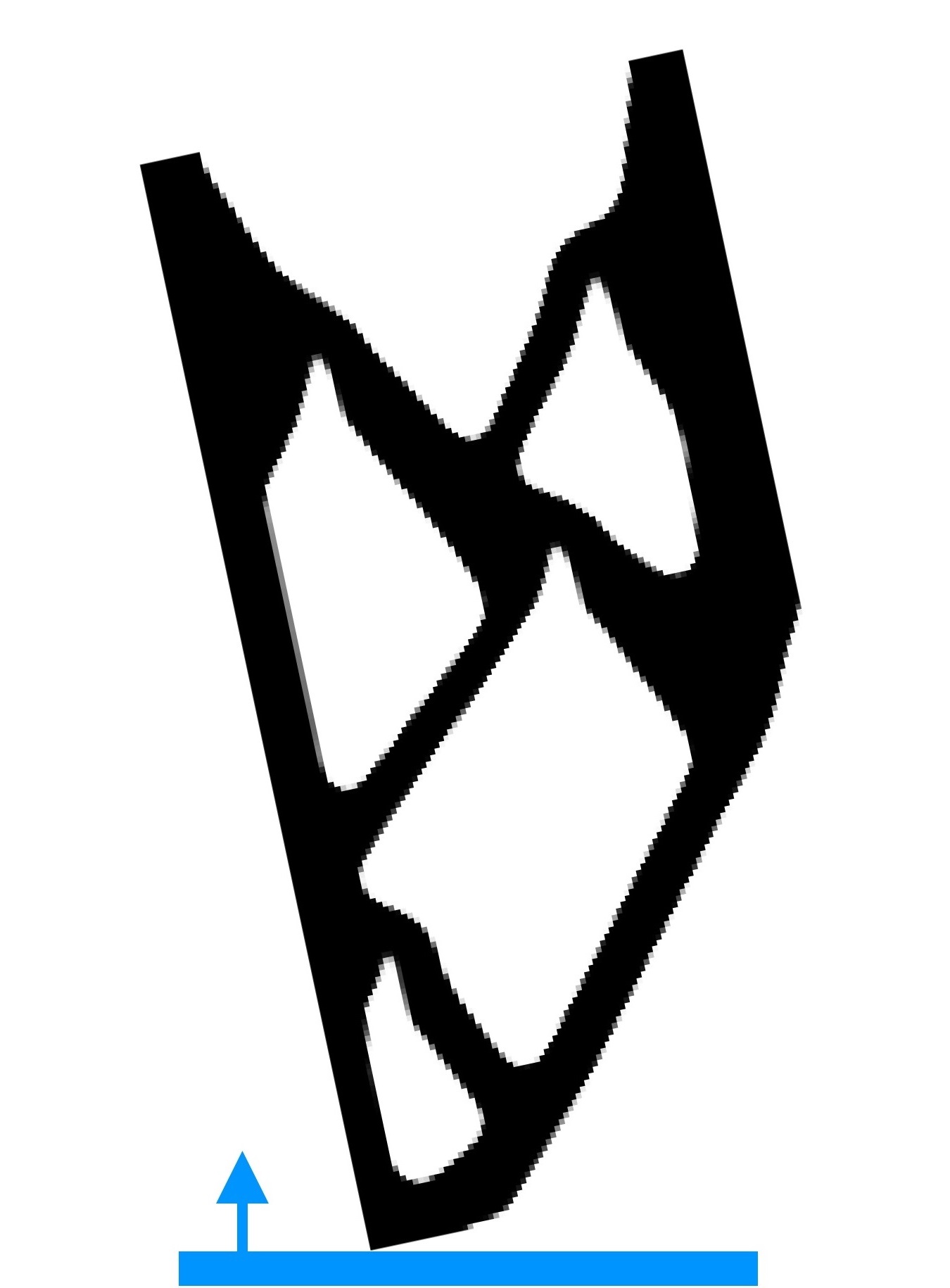} 
    		\caption{$C_\mathrm{int}$= $101.23$ , \; $\theta^*$= $78^\circ$}
    		\label{FIG:CB_RESULTS_3a}
    	\end{subfigure}			
   \begin{subfigure}{0.45\linewidth}   	
    		\includegraphics[width=1\linewidth]{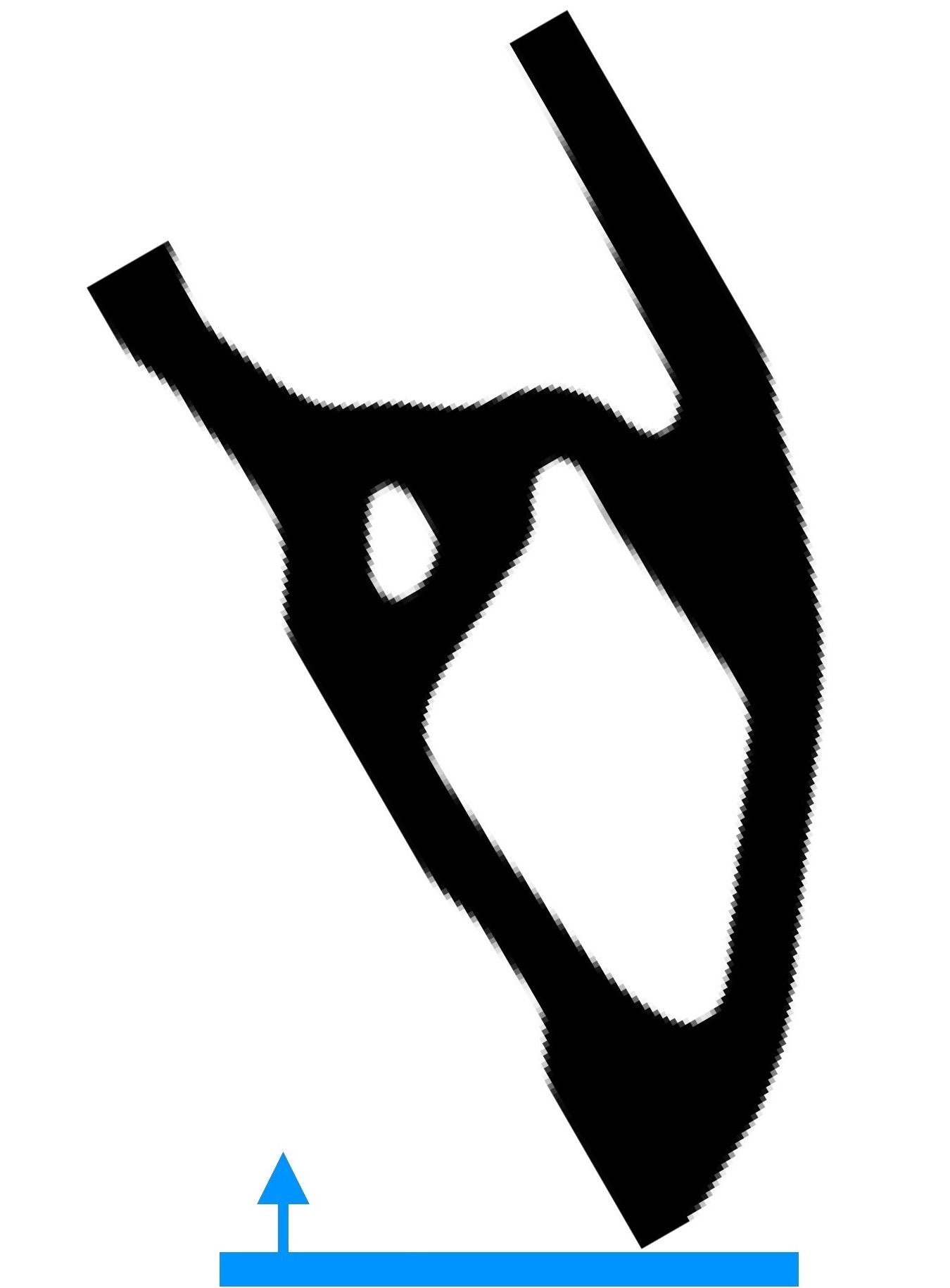} 
    		\caption{$C_\mathrm{int}$= $112.84$ , \; $\theta^*$= $60^\circ$}
    		\label{FIG:CB_RESULTS_3b}
    	\end{subfigure}	
	\caption{Cantilever beam with overhang constraints including 180 candidate directions and a minimum inclination of $60^\circ$. The minimum size is (a) $r_\mathrm{min}=3$, and (b) $r_\mathrm{min}=5$.}
	\label{FIG:CB_RESULTS_3}	
\end{figure} 

	Table~\ref{TAB:time} shows the required time to solve the above optimization problems on a machine with Intel Core i7-6820 HQ @2.70GHz, x64-based processor, 16GB of RAM and MATLAB R2015a. In addition, the time of two solutions including a single candidate direction are included. The cost of including a single restriction $\mathrm{G}(\bm{\bar{\rho}}^\mathrm{dil})$ is low, as the total computation time is increased by about 1.9 times (Table~\ref{TAB:time}). Applying the constraint on all three designs and adding several candidate directions increase the computation time significantly. This is due to  $m$ evaluations of the $\mathbf{a_x}$, $\mathbf{a_y}$ and $\mathbf{B}$ arrays. Further, it is noted that even with 180 candidate directions, the computation time is affordable on standard laptops (Table~\ref{TAB:time}).
	
	Next, the effect of the minimum size on the printing direction is evaluated. $r_\mathrm{min}=2s$ and $r_\mathrm{min}=5s$ are considered. We use the same parameters those used to generate the result displayed in Fig.~\ref{FIG:CB_RESULTS_d}. The optimized designs are shown in Fig.~\ref{FIG:CB_RESULTS_2}. In view of the prevailing directions $\theta^*$ of Figs. \ref{FIG:CB_RESULTS} and \ref{FIG:CB_RESULTS_2}, one can say that for the chosen set of parameters, the minimum size does not have significant influence on the printing direction. Note that $\theta=90^\circ \pm 2^\circ $ is a suitable candidate for the cantilever beam with control over the MinS, MinV and MaxOA. Interestingly, for the same test case \citet{Langelaar2018} gets $\theta^*=90^\circ \pm 5^\circ$ with 72 candidate directions. 
	
	Further, we assess effect of the minimum inclination angle $\alpha$ on the printing direction. The problem is now solved with $\alpha=60^\circ$. The number of candidate directions are set to 180. $r_\mathrm{min}=3$ and $r_\mathrm{min}=5$ are taken. The corresponding results are displayed in Fig.~\ref{FIG:CB_RESULTS_3}, which are orientated according to their $\theta^*$. One notices that the topology and  prevailing direction change significantly with respect to the designs shown in Fig.~\ref{FIG:CB_RESULTS}. This numerical experiment suggests that a good printing direction may not intuitively found, one needs to include the printing direction in the formulation.   
	
	\begin{figure}
	\centering
	\begin{subfigure}{0.80\linewidth}   	
    		 \includegraphics[width=1\linewidth]{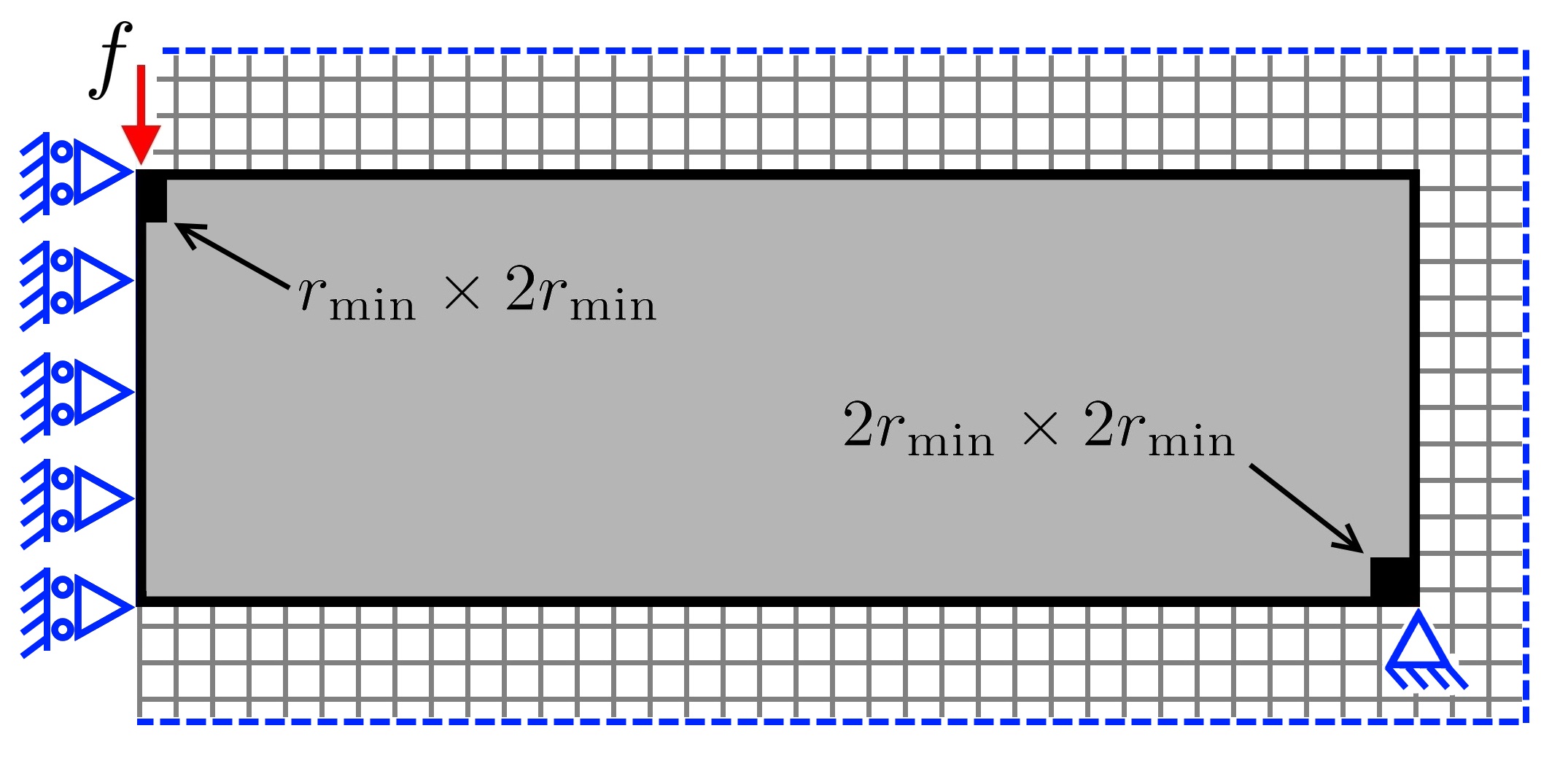} 
    		 \caption{}
    		 \label{FIG:MBB_Beam_a}
    	  \end{subfigure}	
	\begin{subfigure}{0.95\linewidth} 
			\includegraphics[width=1\linewidth]{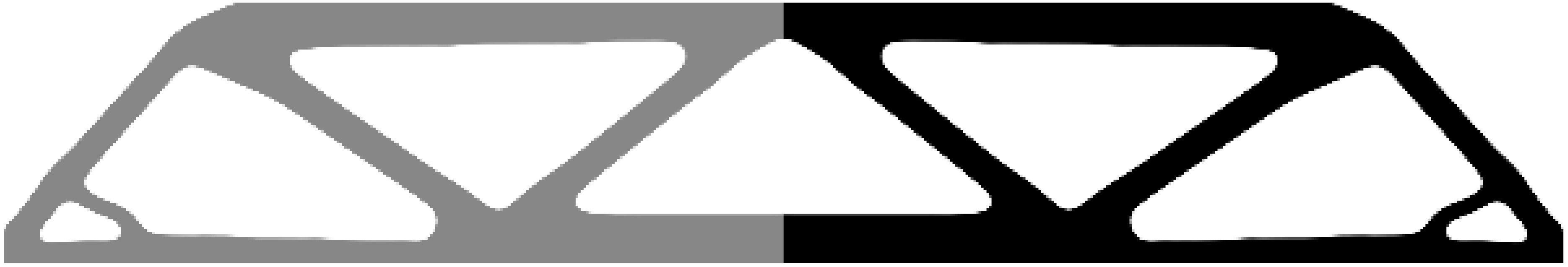} 
			\caption{$C_\mathrm{int}=236.4$ (Reference)}
			\label{FIG:MBB_Beam_b}
	\end{subfigure}
	\begin{subfigure}{0.95\linewidth} 
			\includegraphics[width=1\linewidth]{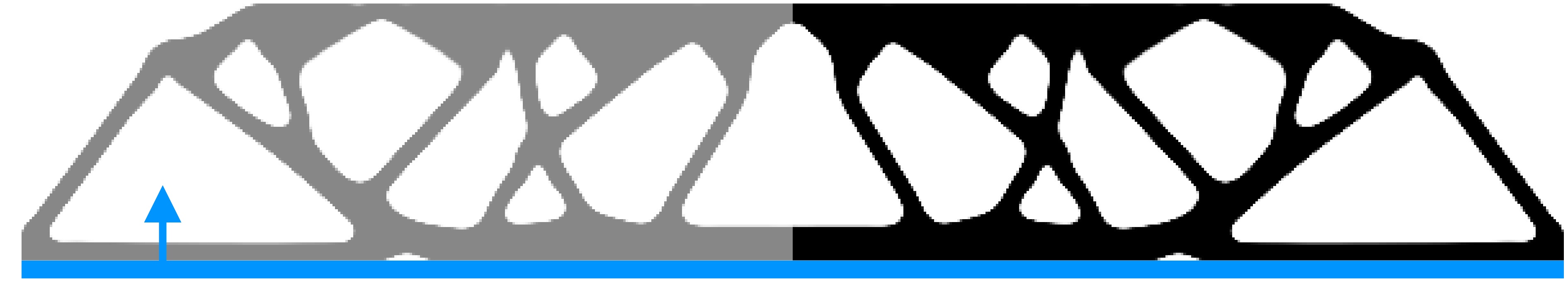} 
			\caption{$C_\mathrm{int}=265.2$ ($m=2$, $\theta^*=0^\circ$)}
			\label{FIG:MBB_Beam_c}
	\end{subfigure}	
	\caption{The symmetric half MBB beam for compliance minimization. (a) Design domain, (b) The reference solution, and (c) Overhang constrained solution including 2 candidate directions, $0^\circ$ and $180^\circ$. The symmetrical parts of the solutions are faded.}
	\label{FIG:MBB_Beam}	
\end{figure}	
	\subsection{MBB beam}
	The MBB (Messerschmitt - B\"{o}lkow - Blohm) beam is solved herein. The symmetry  design domain is as shown in Fig.~\ref{FIG:MBB_Beam_a}. As the symmetry is considered in the optimization, the overhang  constraint is corrected such that it accounts for the reflection with respect to the symmetry line. Two local constraints are considered, i.e., $g_i(\theta)$ and its mirror $g_i(360^\circ\text{-}\theta)$. These are aggregated as $G_k=G(\theta_k) + G(360^\circ\text{-}\theta_k)$. The design domain is discretized by 300$\times$100 FEs. The volume fraction is set to $40\%$. The candidate directions are defined as $\theta_k=\frac{180^\circ(k\text{-}1)}{(m\text{-}1)}$.

	
	Figures~\ref{FIG:MBB_Beam_b} and \ref{FIG:MBB_Beam_c} show the reference and the constrained solutions, respectively.  Two candidate directions are used, $0^\circ$ (upwards) and $180^\circ$ (downwards). The optimized design in Fig.~\ref{FIG:MBB_Beam_c} contains more bars or structural ramifications than the reference solution (Fig.~\ref{FIG:MBB_Beam_b}), which are required to support the upper part of the structure. As expected, the topology of  Fig.~\ref{FIG:MBB_Beam_b} and that of  Fig.~\ref{FIG:MBB_Beam_c} are entirely different from each other, wherein the performance of the latter design is lower. 
	
\begin{figure}
	\centering
	\begin{tabular}{p{0.55\linewidth}p{0.15\linewidth}p{0.15\linewidth}}
	\begin{subfigure}{0.85\linewidth}   	
    		 \includegraphics[width=1\linewidth]{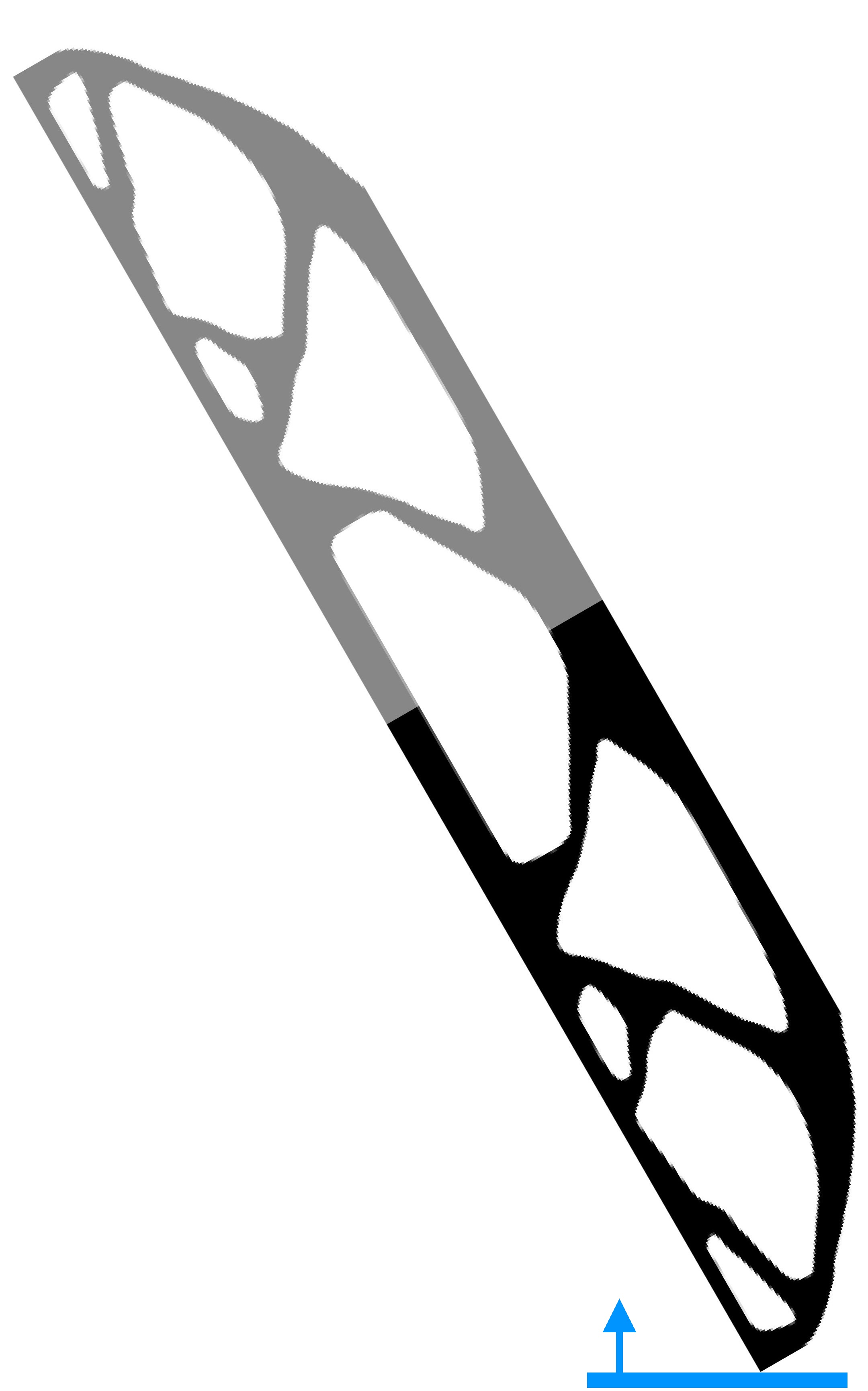} 
    		 \caption{$281.6$}
    		 \label{FIG:MBB_Beam_2_a}
    	  \end{subfigure}	
&
	\begin{subfigure}{1.00\linewidth} 
			\includegraphics[width=1\linewidth]{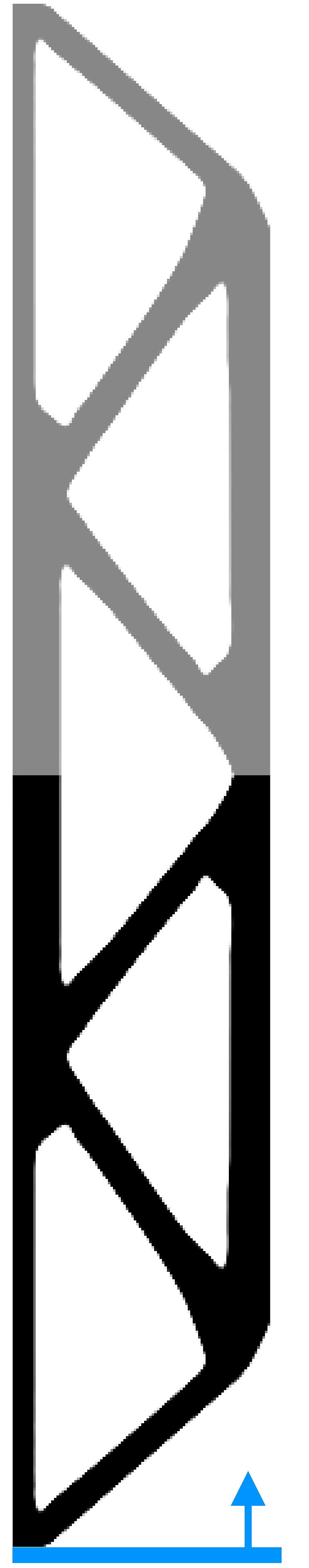} 
			\caption{$237.0$}
			\label{FIG:MBB_Beam_2_b}
	\end{subfigure}
&
	\begin{subfigure}{1.00\linewidth} 
			\includegraphics[width=1\linewidth]{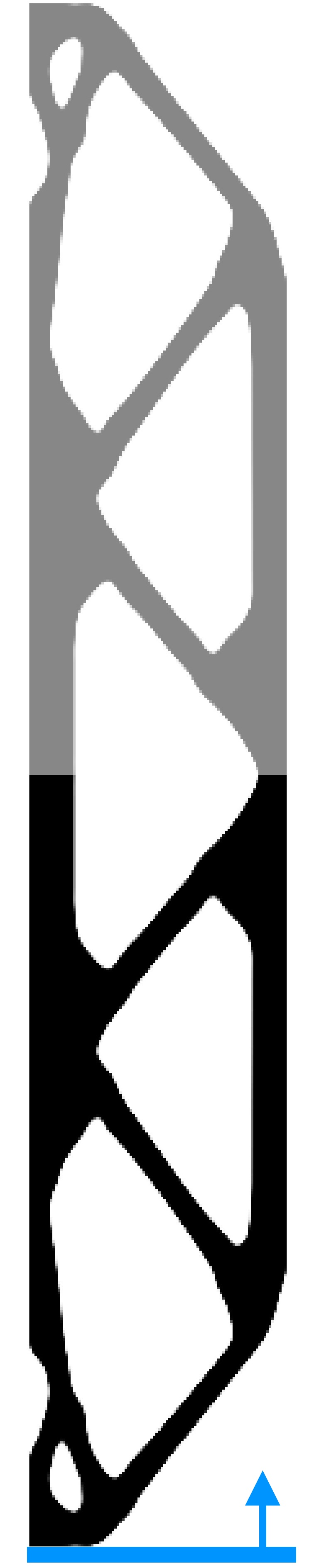} 
			\caption{$253.1$}
			\label{FIG:MBB_Beam_2_c}
	\end{subfigure}	
\end{tabular}
	\caption{Optimized MBB beam with overhang constraints. (a) $m=4$ and $\alpha=45^\circ$, (b) $m=37$ and $\alpha=45^\circ$, (c) $m=37$ and $\alpha=60^\circ$.}
	\label{FIG:MBB_Beam_2}	
\end{figure}

	Figure~\ref{FIG:MBB_Beam_2_a} shows a solution considering 4 printing directions, $0^\circ$, $60^\circ$, $120^\circ$ and $180^\circ$, wherein $\theta^*=60^\circ$ is the prevailing direction. It can be seen that the mirrored surfaces meet the desired MaxOA since the symmetry condition imposed on the restriction $\mathrm{G}_k$. The optimized design (Fig.~\ref{FIG:MBB_Beam_2_a}) exhibits a lower performance than that of Fig.~\ref{FIG:MBB_Beam_c}. This suggests that the proposed strategy for considering building orientation leads optimization to the direction that modifies the reference design  least (Fig.~\ref{FIG:MBB_Beam_b}) and it does not lead to the direction that maximizes structural performance. 
	
	The optimized design in Fig.~\ref{FIG:MBB_Beam_2_b} is obtained with 37 candidate directions and $\alpha=45^\circ$, while Fig. \ref{FIG:MBB_Beam_2_c} with  37 candidate directions and $\alpha=60^\circ$. In both cases the prevailing direction is determined to be $90^\circ$, which is not surprising as the reference design contains bars inclined at approximately $45^\circ$ with respect to the horizontal bars located at the top and bottom of the domain. Therefore, positioning the long horizontal bars vertically is a logical solution for reducing the amount of overhanging surfaces.
	
	\subsection{Compliant mechanism}
\begin{figure}
	\centering 	
    \includegraphics[width=0.75\linewidth]{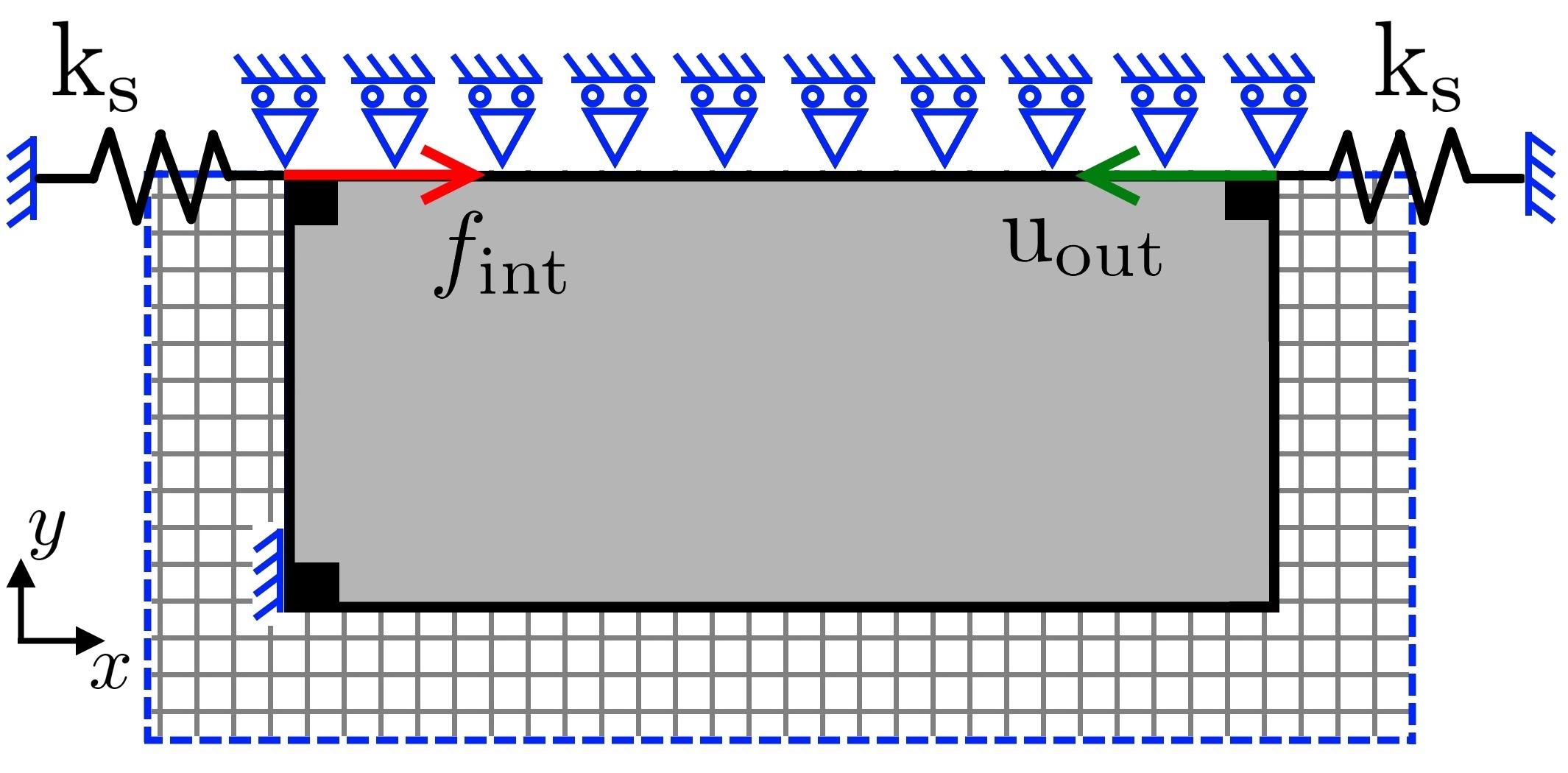} 
    \vspace{-1mm}
	\caption{Force inverter problem considering symmetry boundary conditions. Design domain of size $\mathrm{L} \times \mathrm{L/2}$. The material parameters are $\mathrm{E}_0=1$, $\mathrm{E}_\mathrm{min}=10^{-4}$, $f_\mathrm{int}$ is the input force, and $\mathrm{k}_s=1$.}
	\label{FIG:CM_Design_Domain}	
\end{figure}

	We choose the force inverter compliant mechanism in this section. The symmetric design domain with pertinent boundary conditions is shown in Fig.~\ref{FIG:CM_Design_Domain}. The optimization problem maximizes the output displacement $\mathrm{u}_\mathrm{out}$ for an input force $f_\mathrm{int}$. The optimization problem solved is:
	\begin{align} \label{EQ:OPTI_Compliant}
	\begin{split}
	{\min_{\bm{\rho}}} & \quad \mathrm{max} \left( \mathbf{L}^{\intercal} \mathbf{u}_\mathrm{ero},\mathbf{L}^{\intercal} \mathbf{u}_\mathrm{dil} \right) \\
	\mathrm{s.t.:} &\quad  \mathbf{v}^{\intercal} \bm{\bar{\rho}}_\mathrm{dil} \leq V^*_\mathrm{dil} \left( V^*_\mathrm{int} \right) 	\\
	&\quad B_\mathrm{G}(\bm{\bar{\rho}}_\mathrm{ero}) \leq 0 \\
	&\quad B_\mathrm{G}(\bm{\bar{\rho}}_\mathrm{int}) \leq 0 \\
	&\quad B_\mathrm{G}(\bm{\bar{\rho}}_\mathrm{dil}) \leq 0 \\
	&\quad 0 \leq {\rho_i} \leq1  \;\;,\;\; i=1,...\:,N \;\;, 
	\end{split}
	\end{align}
	\noindent where $\mathbf{L}$ is an array with a value of 1 at the output degree of freedom and 0 otherwise. In most of the cases, worst performer is either the eroded or dilated design. Thus, the objective function considers the worst case among the eroded and dilated designs. The volume fraction is set to $30\%$. The  minimum size is $r_\mathrm{min}=3s$. The domain is parameterized via $200\times100$ FEs. As the problem includes symmetry condition, the overhang restriction and boundary filter are treated as done for the MBB beam.

	\begin{figure}
	\centering
       \begin{subfigure}{0.45\linewidth}   	
    		\includegraphics[width=1\linewidth]{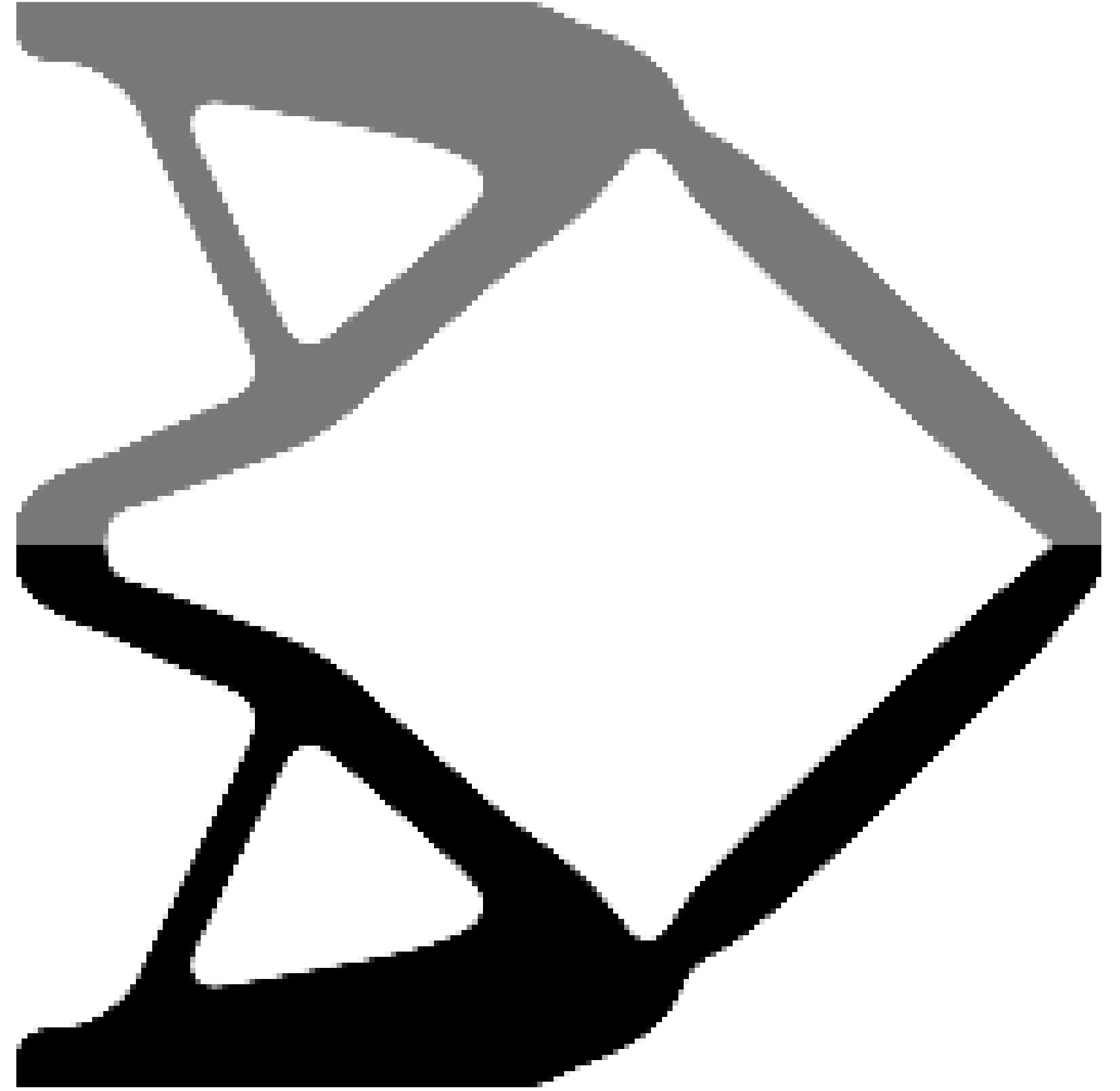} 
    		\caption{$u^\mathrm{out}_\mathrm{int}=-0.0092$}
    		\label{FIG:CM_a}
    	\end{subfigure}	
        \begin{subfigure}{0.45\linewidth}   	
    		\includegraphics[width=1\linewidth]{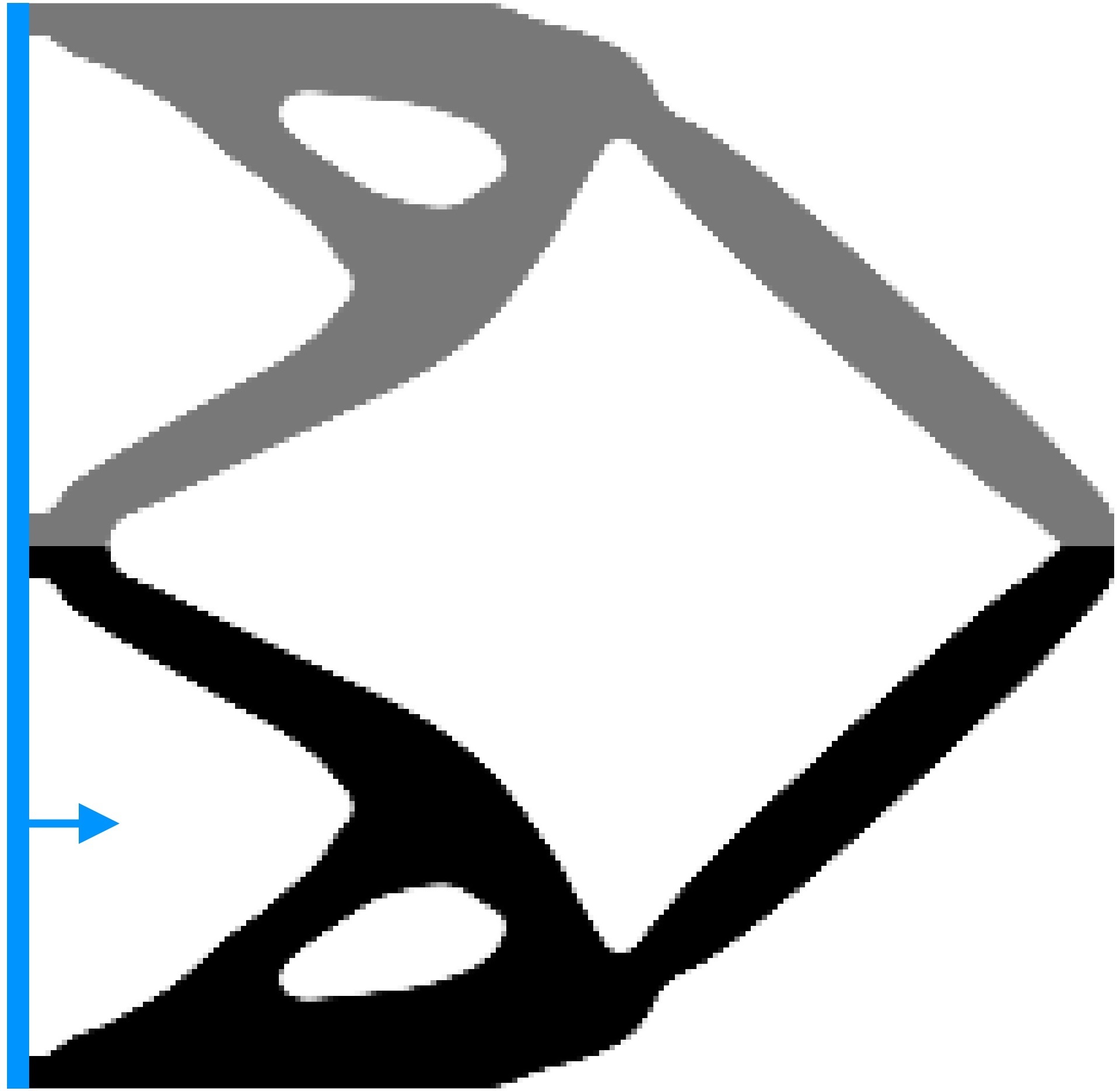} 
    		\caption{$u^\mathrm{out}_\mathrm{int}=-0.0084$}
    		\label{FIG:CM_b}
    	\end{subfigure}	
		\begin{subfigure}{0.45\linewidth}   	
    		\includegraphics[width=1\linewidth]{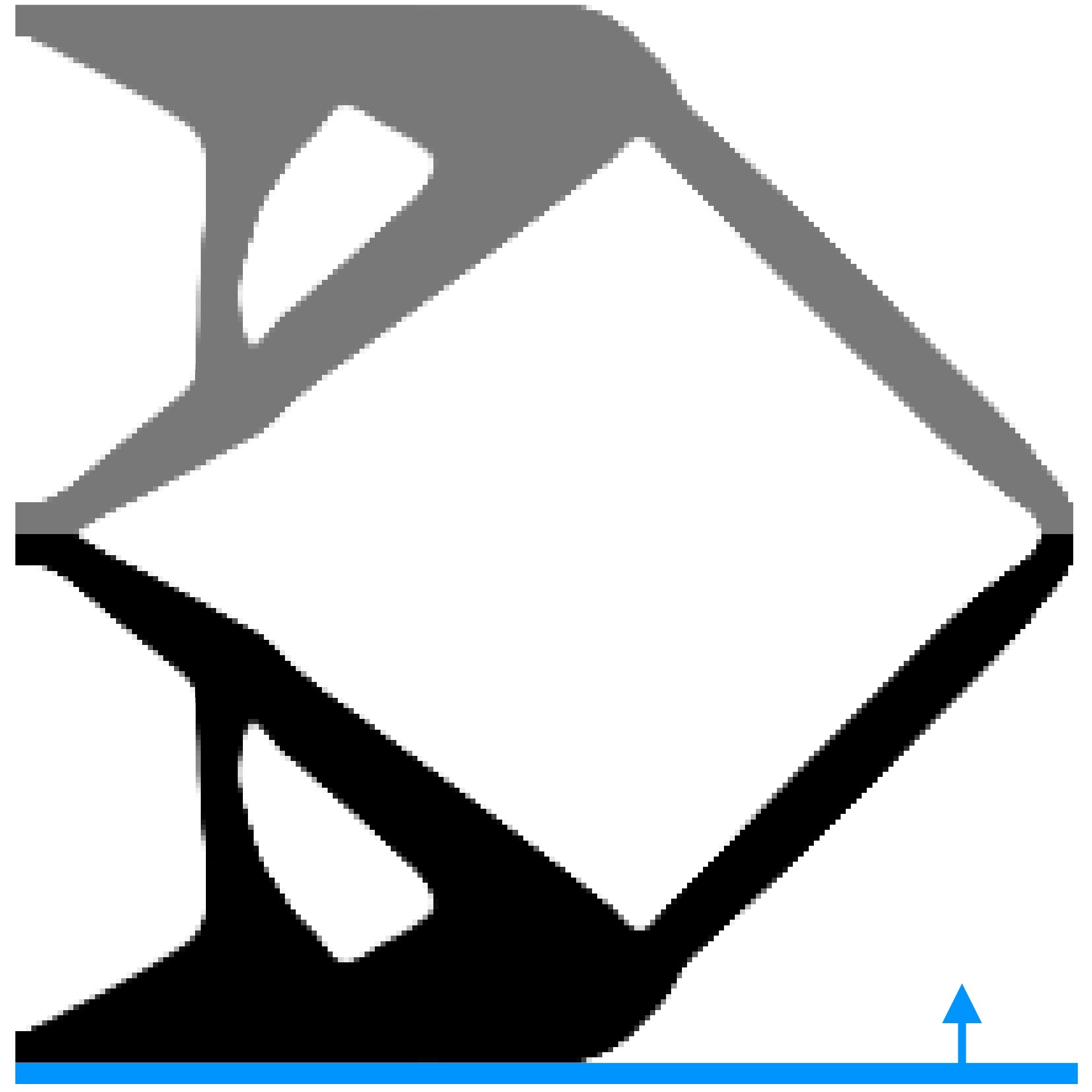} 
    		\caption{$u^\mathrm{out}_\mathrm{int}=-0.0066$}
    		\label{FIG:CM_c}
    	\end{subfigure}	
    	\begin{subfigure}{0.45\linewidth}   	
    		\includegraphics[width=1\linewidth]{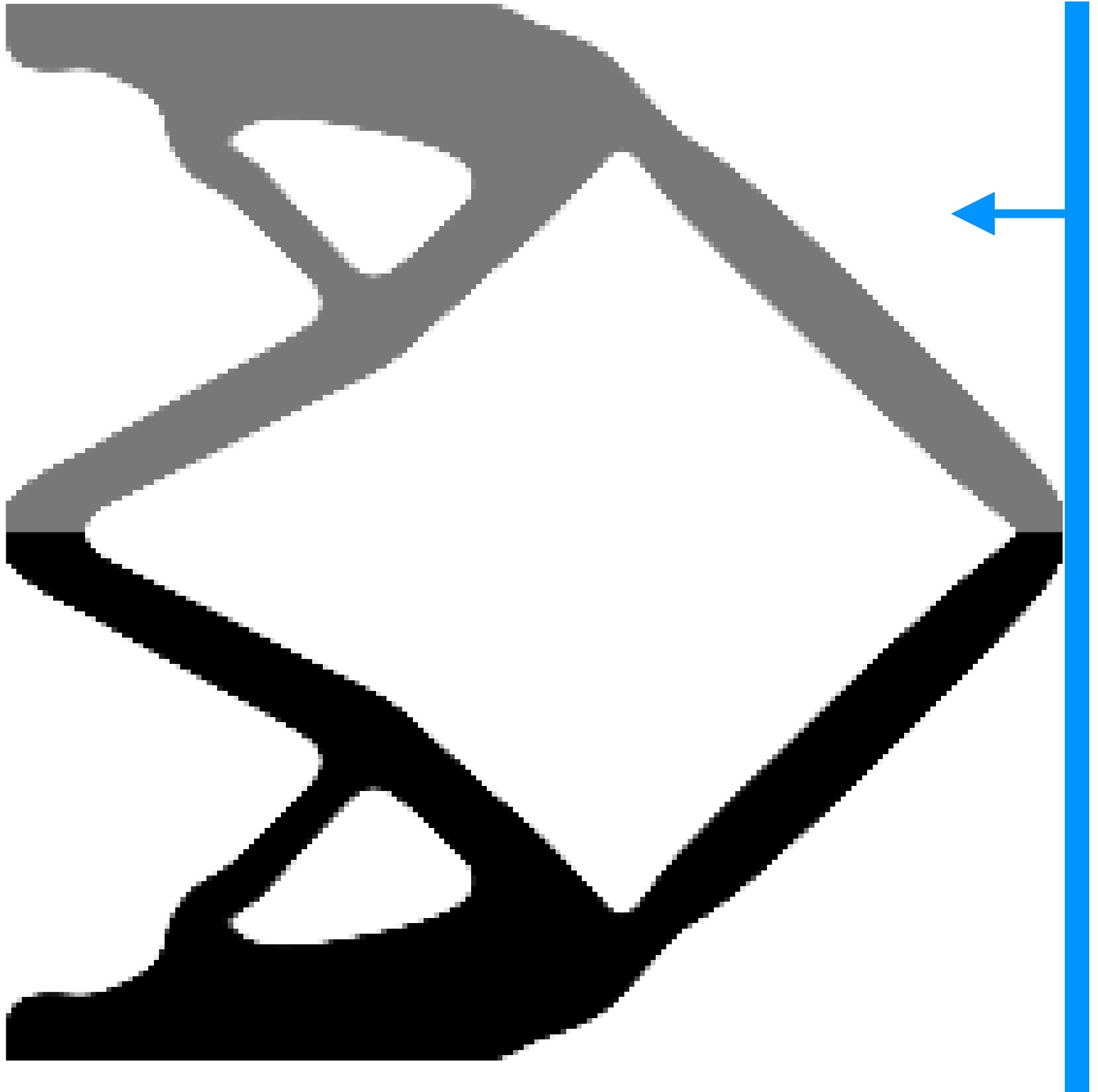} 
    		\caption{$u^\mathrm{out}_\mathrm{int}=-0.0082$}
    		\label{FIG:CM_d}
    	\end{subfigure}		
	\caption{Compliant mechanism designs. (a) The reference solution, (b) The MaxOA constrained solution considering $m$=36. (c) and (d), considering one candidate direction.}
	\label{FIG:CM}	
\end{figure}	
	The reference optimized solution including the symmetric part is shown in Fig.~\ref{FIG:CM_a}. The overhang constrained solution considering 36 candidate directions is displayed in Fig.~\ref{FIG:CM_b}. The directions $\theta_k$ are defined from $-90^\circ$ to $90^\circ$ with separations of $\Delta \theta=5^\circ$. The problem considering 36 candidate directions converges to $\theta^*=-90^\circ$. We also solve the problem using candidate direction as $\theta=0^\circ$ and $\theta=90^\circ$. The corresponding results are displayed in Figs.~\ref{FIG:CM_c} and \ref{FIG:CM_d}. It can be noticed that both solutions have a smaller output displacement (in magnitude) than the solution including 36 candidate directions. Performance of the design obtained with $\theta=90^\circ$ is very similar to that of Fig. \ref{FIG:CM_b}. This suggests that the optimal printing direction is likely to be aligned with the axis of symmetry i.e. $\theta^*$ may be $90^\circ$ or $-90^\circ$. This observation is also intuitively correct as constituting branches of the reference design are incline close to $45^\circ$ with respect to the axis of symmetry.

	\subsection{Maximum size, minimum size, maximum overhang angle, and building orientation}
	
	In this section, we present efficacy of the proposed approach on more complex problems by also including designs restricted on maximum size. Therefore, the optimization problem deals with 5 geometric features: the minimum size of solid phase, the minimum size of void phase, the maximum size of solid phase (MaxS), the maximum overhang angle, and the building direction. 
	
	The MaxS restriction used in this work is a local volume constraint. This is applied on the neighborhood elements around each element. The local formulation of such restriction is presented in \cite{Guest2009}, the aggregation strategy is introduced in \cite{fernandez2019aggregation}, and the implementation using the robust design approach is detailed in \cite{fernandez2020imposing}. We indicate the maximum size restriction by $G_\mathrm{ms}$. We provide a Matlab code \texttt{MaxSize} that implements maximum size constraint. For a detailed overview on this topic, the readers are referred to \citet{fernandez2019aggregation,fernandez2020imposing}. 
	
	\begin{figure}
\captionsetup[subfigure]{labelformat=empty}
	\centering
	\begin{tabular}{p{0.15\linewidth}p{0.15\linewidth}p{0.3\linewidth}p{0.3\linewidth}}
	\vspace{-23mm}	
	\multirow{3}{*}{  	
		\begin{subfigure}{0.54\linewidth}   	
    		 \includegraphics[width=1\linewidth]{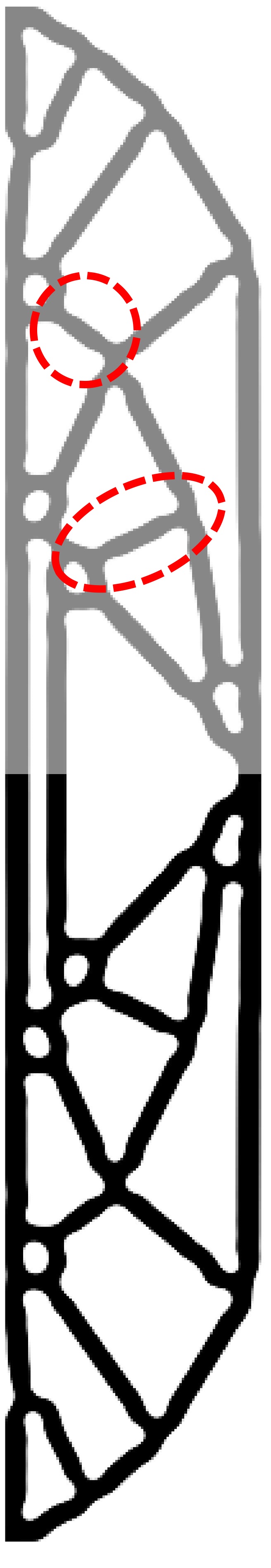} 
    		 \caption{}
    		 \label{FIG:MAXSIZE_a}
    	\end{subfigure}	
		}	
	&
	\vspace{-23mm}
	\multirow{3}{*}{
		\begin{subfigure}{0.54\linewidth}   	
    		 \includegraphics[width=1\linewidth]{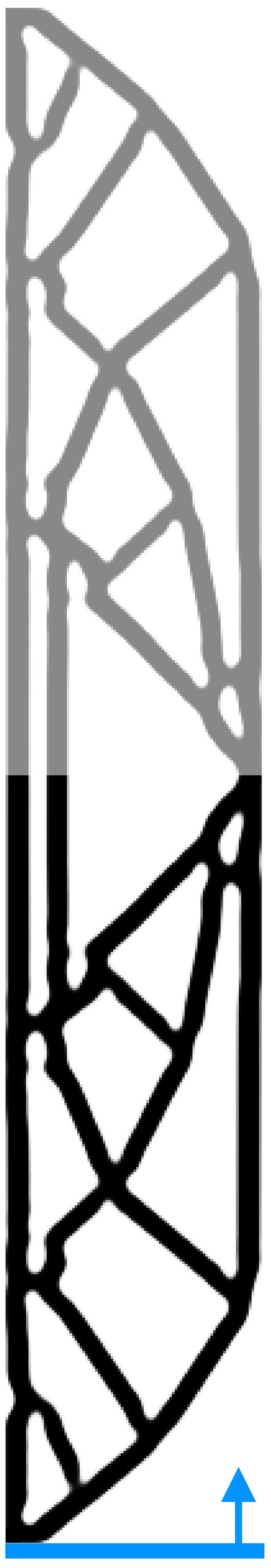} 
    		 \caption{}
    		 \label{FIG:MAXSIZE_b}
    	\end{subfigure}	
		}
	&	
	\begin{subfigure}{0.73\linewidth}   	
    		 \includegraphics[width=1\linewidth]{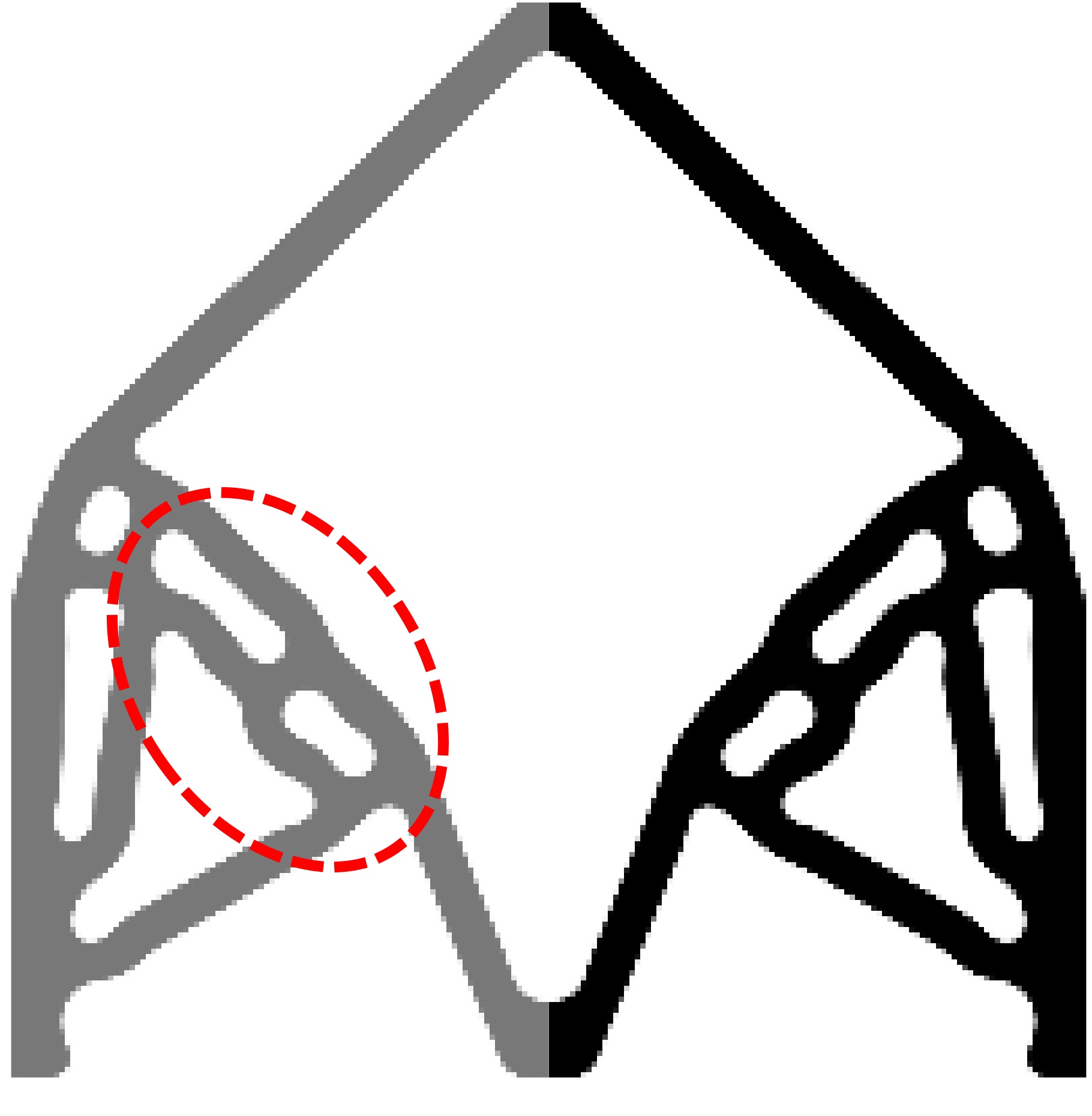} 
    		 \caption{}
    		 \label{FIG:MAXSIZE_c}
    \end{subfigure}	
    &
	\begin{subfigure}{0.73\linewidth}   	
    		 \includegraphics[width=1\linewidth]{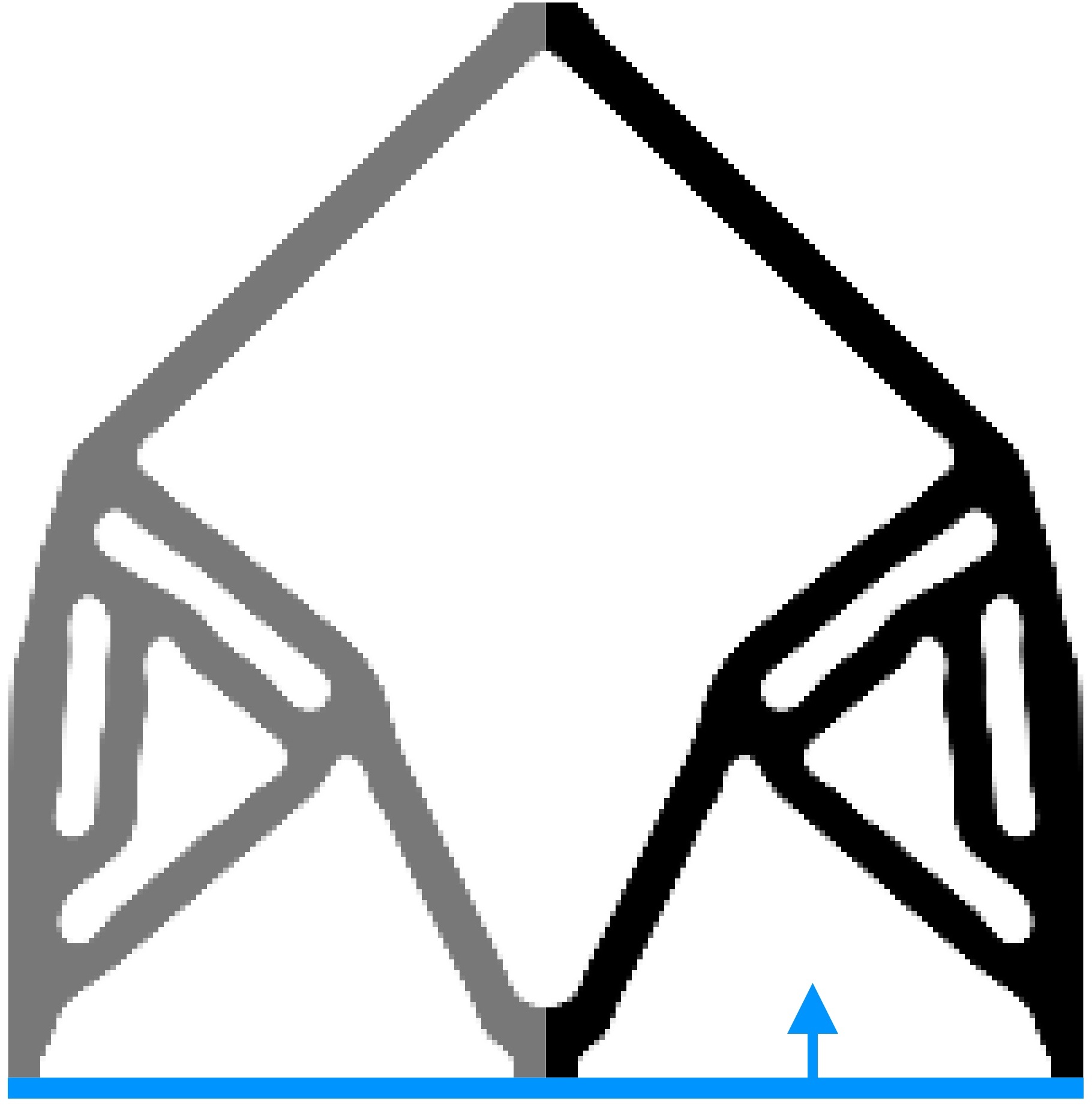} 
    		 \caption{}
    		 \label{FIG:MAXSIZE_d}
    \end{subfigure}
    \\ 
    & & \vspace{-8mm}(a) $u_\mathrm{int}^\mathrm{out}=-0.0054$ & \vspace{-8mm} (b) $u_\mathrm{int}^\mathrm{out}=-0.0047$ $(-90^\circ)$
    \\
    & &
	\hspace{13mm}    
    \begin{subfigure}{0.42\linewidth}   
			 \vspace{-2mm}    		 
    		 \includegraphics[width=1\linewidth]{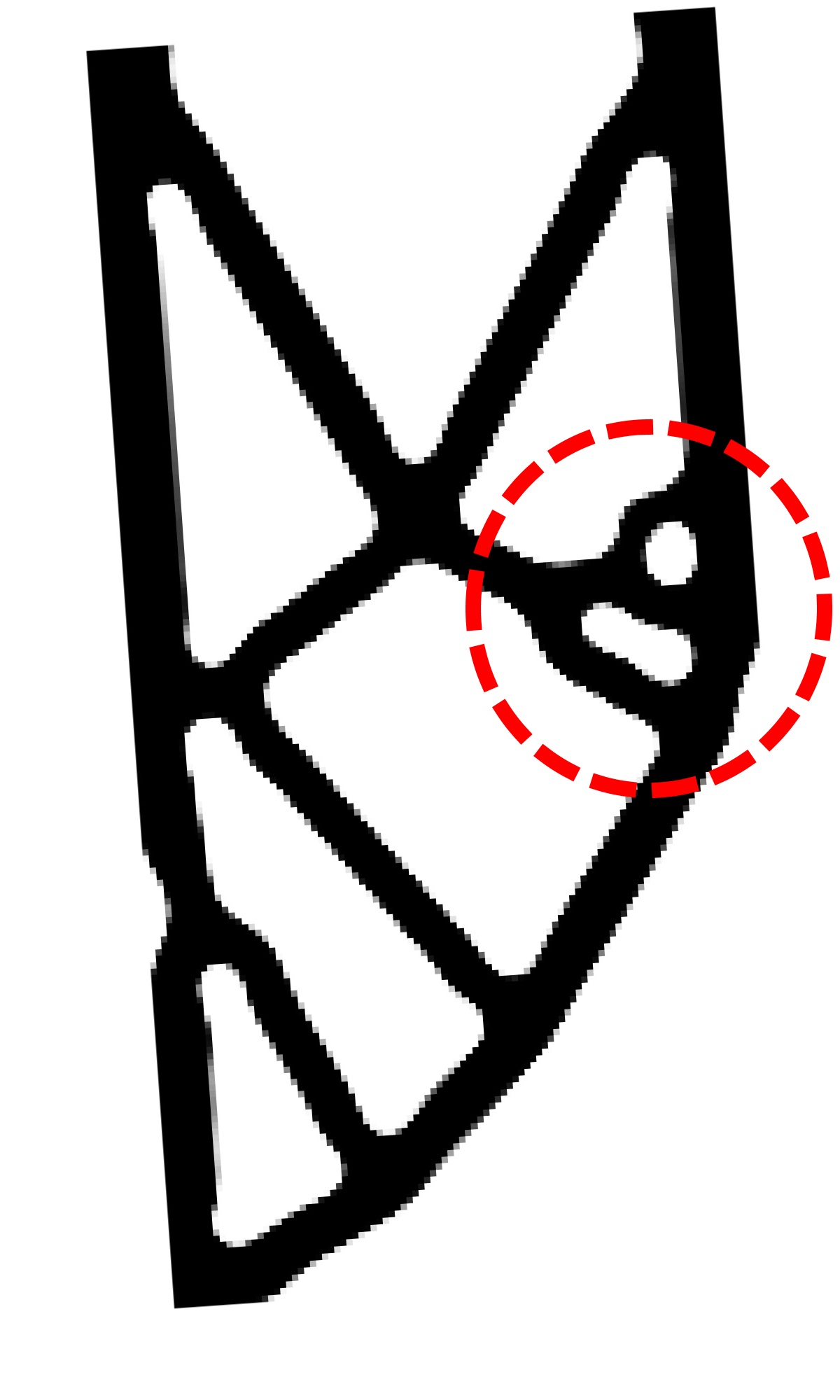} 
    		 \caption{}
    		 \label{FIG:MAXSIZE_e}
    \end{subfigure}	
    &
    \hspace{13mm}\begin{subfigure}{0.42\linewidth}   	
    		 \vspace{-2mm}
    		 \includegraphics[width=1\linewidth]{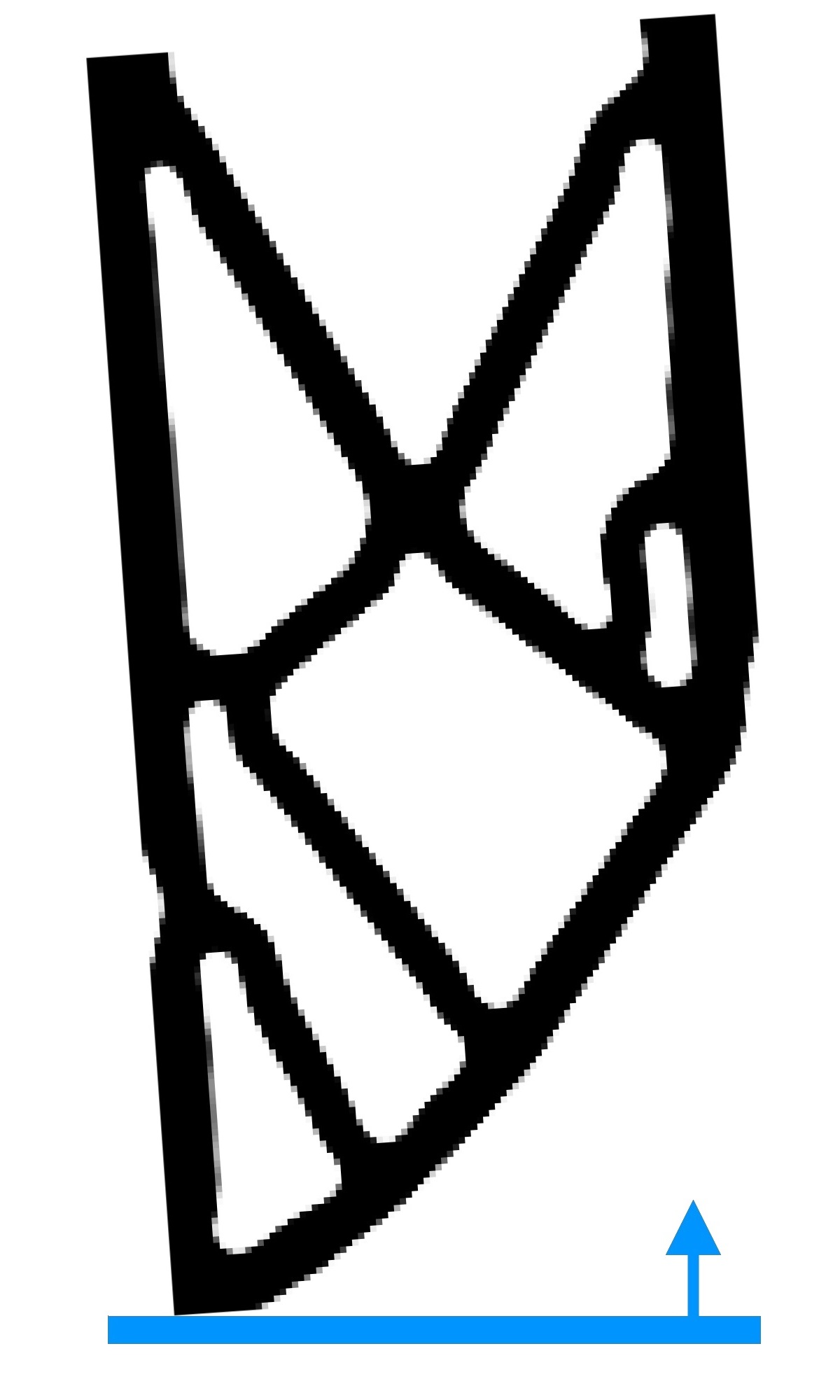} 
    		 \caption{}
    		 \label{FIG:MAXSIZE_f}
    \end{subfigure}
    \\	
    \vspace{-8.5mm}(c) $C_\mathrm{int}$=$300.86$
    &
    \vspace{-8.5mm}\hspace{-3mm}(d) $C_\mathrm{int}$=$316.35$ $(90^\circ)$
    &
    \vspace{-8.5mm}\hspace{10mm}(e) $C_\mathrm{int}=105.83$
    &
    \vspace{-8.5mm}\hspace{10mm}(f) $C_\mathrm{int}=108.32$ $(86^\circ)$
    \end{tabular}   
    \vspace{-6mm}
    \caption{Optimized designs with maximum size constraints. (a), (c) and (e), the reference solutions without Maximum Overhang Angle control. (b), (d) and (f), the Overhang-constrained solutions. Arrows indicate the building orientation.}
	\label{FIG:MAXSIZE}	
\end{figure}

	The optimization problem including the MinS, MinV, MaxS, MaxOA and building orientation can be written as:
	\begin{align} \label{EQ:OPTI_Constrained_Bg_Gms}
	\begin{split}
	{\min_{\bm{\rho}}} & \quad O_\mathrm{bj} \\
	\mathrm{s.t.:} &\quad \mathbf{v}^{\intercal} \bm{\bar{\rho}}_\mathrm{dil} \leq V^*_\mathrm{dil} \left( V^*_\mathrm{int} \right)  \\
	&\quad B_\mathrm{G}(\bm{\bar{\rho}}_\mathrm{int}) \leq 0 \\
	&\quad B_\mathrm{G}(\bm{\bar{\rho}}_\mathrm{dil}) \leq 0 \\
	&\quad G_\mathrm{ms}(\bm{\bar{\rho}}_\mathrm{int}) \leq 0 \\
	&\quad G_\mathrm{ms}(\bm{\bar{\rho}}_\mathrm{dil}) \leq 0 \\
	&\quad 0 \leq {\rho_i} \leq1  \;\;,\;\; i=1,...\:,N. 
	\end{split}
	\end{align}
	
	\noindent The problem (Eq.~\ref{EQ:OPTI_Constrained_Bg_Gms}) is solved for the cantilever beam, the MBB beam and the force inverter. For the former two test cases, $O_\mathrm{bj}$ represents compliance of the eroded design, i.e. $C_\mathrm{ero}$, while for the compliant mechanism test case, $O_\mathrm{bj}$ represents the output displacement as reported in Eq.~\ref{EQ:OPTI_Compliant}. We use the same $G_\mathrm{ms}$ for all the three test cases. The provided code \texttt{MaxSize} also indicates the same.
	
	The volume fractions for the force inverter, MBB beam and cantilever beam are set $30\%$, $40\%$ and $40\%$, respectively. MinS and MinV is set to $r_\mathrm{min}=3s$ elements, and  MaxS is taken $9s$. The reference results, i.e. without $B_\mathrm{G}$ restrictions, are shown in Figs.~\ref{FIG:MAXSIZE_a}, \ref{FIG:MAXSIZE_c} and \ref{FIG:MAXSIZE_e}.

	The MaxOA constrained results are shown in Figs.~\ref{FIG:MAXSIZE_b}, \ref{FIG:MAXSIZE_d} and \ref{FIG:MAXSIZE_f}. These are obtained for $45^\circ$ minimum inclination angle. 36 candidate directions for the MBB beam and force inverter, and 180 candidate directions for the cantilever beam are considered. The prevailing directions are indicated in parentheses next to the objective function are $-90^\circ$ $90^\circ$ and $86^\circ$  (Fig.~\ref{FIG:MAXSIZE}). In addition, the main topological differences with respect to their reference designs are highlighted by red dashed curves.  
	
	We note that in the previous examples, the MaxOA constrained designs do not undergo significant modifications with respect to the reference designs. The prevailing directions are similar to those of previous examples. Thus, even with maximum size restrictions, the proposed strategy allows to obtain a result similar to the reference one but with improved manufacturability. In addition, the designs satisfy the desired MinS, MinV, MaxS, and most of parts feature the desired MaxOA. The parameters needed to solve the problems are provided in Table \ref{TAB:Parameters}. One can realize that adjustment of parameters for a 3D problem setting will be more demanding and that forms our one of the future research directions.

	\begin{table*}[width=0.82\textwidth,cols=4,pos=h] 
		\caption{List of parameters used to obtain the reported solutions.}
		\begin{tabular}{c c c c c c c c c c c c c c c}
			\toprule	    
			& \multicolumn{13}{c}{Figure}
			\\
			\cmidrule(r){2-15}
			Parameter & \ref{FIG:CB_RESULTS} & \ref{FIG:CB_RESULTS_2} & \ref{FIG:CB_RESULTS_3a} & \ref{FIG:CB_RESULTS_3b} & \ref{FIG:MBB_Beam_c} & \ref{FIG:MBB_Beam_2_a} & \ref{FIG:MBB_Beam_2_b} & \ref{FIG:MBB_Beam_2_c} & \ref{FIG:CM_b} & \ref{FIG:CM_c} & \ref{FIG:CM_d} & \ref{FIG:MAXSIZE_b} & \ref{FIG:MAXSIZE_d} & \ref{FIG:MAXSIZE_f}
			\\ 
			\cmidrule(r){1-15}
			\vspace{1mm} $p$    
			& 60  & 60  & 80  & 80  & 60  & 60  & 60  & 60  & 80  & 80  & 80  & 80  & 60 & 60
			\\ \vspace{1mm} $\varepsilon_\mathrm{n}^\mathrm{ini}$ 
			& 0.2 & 0.2 & 0.2 & 0.1 & 0.3 & 0.3 & 0.2 & 0.2 & 0.2 & 0.2 & 0.2 & 0.4 & 0.5 & 0.4
			\\ \vspace{1mm} $\varepsilon_\mathrm{n}^\mathrm{end}$ 
			& 2.0 & 2.0 & 1.5 & 1.5 & 1.0 & 1.0 & 1.5 & 1.5 & 1.0 & 1.0 & 1.0 & 1.0 & 2.0 & 1.0
			\\ \vspace{1mm} $it_\mathrm{free}$
			& 10  & 10  & 2   & 2   & 4   & 4   & 5   & 5   & 5   & 5   & 5   & 10  & 10 & 5
			\\ $r$ 
			& 20  & 20  & 40  & 40  & 10  & 10  & 30  & 30  & 30  & -   & -   & 15  & 20 & 15
			\\		      
			\bottomrule
		\end{tabular}
		\label{TAB:Parameters}
	\end{table*}

	\section{Conclusions} \label{Sec:4}
	
	This paper presents a design approach for topology optimization that includes various limitations of the additive manufacturing, e.g., maximum and minimum size of the parts, maximum overhang angle, and building orientation. The efficacy and versatility of the presented approach are demonstrated by designing various 2D benchmark problems including stiff  structure and compliant mechanism designs. 
	
	The work focuses on additive manufacturing and in the inability to build overhanging parts without the use of sacrificial support structures. This limitation is geometrically addressed using a constraint that restricts the maximum overhang angle of the parts. The constraint is formulate using structural gradient, which is computationally cheap to evaluate. This facilitates us to determine the constraint in several building directions such that optimization can be driven towards the least restrictive one. To achieve solution close to 0-1 and to impose the minimum size of solid and void states, we use the robust design approach based on eroded, intermediate and dilated designs. Particularly, based on performed numerical experiments, we note the following observations:
	
	\begin{itemize}
		\item The maximum overhang angle constraint may come into conflict with the eroded and intermediate designs, as these fields feature minimum cavity size i.e. the maximum overhang angle constraint may not be met at the re-entrant corners. The dilated design does not come into conflict as it does not feature minimum cavity size.
		\item The gradient-based overhang constraint is prone to introduce undesirable triangular hanging parts. These parts are not self-supporting, however meet the local slope criterion. They do not transmit loads, therefore they are detected using the deformation energy in the proposed approach.
		\item The maximum overhang angle constraint improves printability of the optimized designs, however, it does not guarantee self-supporting designs. Such designs are not exempt from collapsing during printing due to, for example, their self-weight.
		
		\item Evaluating the maximum overhang angle constraint at several building orientations and including  the least restrictive one in the optimization problem, may lead the optimization towards the direction that modifies the unconstrained design least, but not to the direction that maximizes structural performance. 
	\end{itemize}
	
	Results are consistent with the desired minimum size of solid and void states, and most of the parts feature the desired maximum overhang angle. As other TO approaches, the obtained optimized designs by the presented method also depends upon different parameters. We also provide the associated MATLAB codes for the cantilever beam design. Extension of the approach for 3D topology designs forms the future research direction.  
	\section*{Declaration of interest}
	The authors declare that there is no conflict of interest.
	
	\section*{Acknowledgement}
	The authors are grateful to Prof. Krister Svanberg for providing the MATLAB implementation of the Method of Moving Asymptotes, which is used in this work. P. Kumar acknowledges financial support from the Science \& Engineering
	research board, Department of Science and Technology, Government of India under the project
	file number RJF/2020/000023.
	

	\bibliographystyle{model6-num-names}
	\bibliography{myreference}

	\appendix
%
	\section{Reference code (\texttt{topCbeam.m})} \label{APP:1}
%
	\section{Code for gradient-based overhang constraint (\texttt{GBOHC.m})} \label{APP:2}
%
	\section{Code for post-Processing triangular features (\texttt{PPTri.m})} \label{APP:3}
%
	\section{Code for maximum size constraint (\texttt{MaxSize.m})} \label{APP:4}
\end{document}